\documentclass[]{imag-ms-template}
\usepackage{xcolor}
\providecommand{\rev}[1]{#1}
\title{Geometric Modeling of Hippocampal Tau Deposition: A Surface-Based Framework for Covariate Analysis and Off-Target Contamination Detection}

\author{Liangkang Wang$^{1}$, Akhil Ambekar$^{2}$, Ani Eloyan$^{1,2\ast,\ast\ast}$\\
{\small $^{1}$Department of Biostatistics, Brown University}\\
{\small $^{1}$Center for Biostatistics and Health Data Science, Brown University}\\
{\small $\ast$Correspondence:  ani\_eloyan@brown.edu} \\
{\small  $\ast\ast$ Data used in preparation of this article were obtained from the Alzheimer’s Disease } \\ {\small Neuroimaging Initiative
(ADNI) database (adni.loni.usc.edu).  As such, the investigators } \\ {\small within the ADNI contributed to the  design
and implementation of ADNI and/or  } \\ {\small provided data but did not participate in analysis or writing of this report. } \\ {\small
A complete listing of ADNI investigators can be found at:
} \\
{\small
http://adni.loni.usc.edu/wp\-content/uploads/how\_to\_apply/ADNI\_Acknowledgement\_List.pdf}
}

\begin{document} 

\maketitle

\begin{abstract}
We introduce a novel framework combining geometric modeling with disease progression analysis to investigate tau deposition in Alzheimer's disease (AD) using positron emission tomography (PET) data. We focus on the hippocampus—a region of the brain affected early in AD—and construct a principal surface such that the projection of voxels onto this surface captures the spatial distribution and morphological changes of tau pathology. We use bidirectional projection distances and interpolated standardized uptake value ratio (SUVR) values to map tau PET data onto this surface, enabling quantification of coverage, intensity, and thickness. The projection provides a low-dimensional embedding space, reducing potential issues related to multiple comparisons while preserving spatial specificity. To analyze covariate effects, we apply a two-stage regression model with inverse probability weighting to adjust for signal sparsity and selection bias. Using the SuStaIn model, we identify subtypes and stages of AD and investigate tau dynamics related to AD subtypes. We find that the limbic-predominant subtype exhibits age-associated, nonlinear accumulation patterns in coverage and thickness, while the posterior subtype shows more uniform SUVR increases across progression. Model-based predictions further reveal that tau deposition within the hippocampus follows a structured spatial trajectory, originating centrally and expanding bidirectionally with increasing thickness in one subtype, while our second identified subtype exhibits earlier tau involvement in the posterior hippocampus—consistent with their respective whole-brain subtype definitions. Finally, we demonstrate that directional signal patterns on the principal surface can be used to detect potential contamination from the adjacent choroid plexus, highlighting the broader applicability of our framework.
\end{abstract}


\section{Introduction}

Alzheimer’s disease (AD) is a progressive neurodegenerative disorder and the leading cause of dementia worldwide, accounting for approximately 60–80\% of all cases \citep{better2023alzheimer}. AD is characterized by the accumulation of tau protein and beta-amyloid plaques in the brain \citep{RN230}. Tau protein deposition plays a critical role in neuronal damage and disease progression. Current methods quantify tau deposition using standardized uptake value ratio (SUVR) from positron emission tomography (PET) scans \citep{RN238, RN232}, but primarily focus on regional or global SUVR comparisons \citep{RN241, RN277}. However, these approaches often overlook the spatial distribution and morphological variations of tau protein deposition, limiting their ability to characterize disease progression comprehensively. Prior studies have also highlighted variability in tau biomarker levels across different populations, raising concerns about the applicability of global SUVR thresholds \citep{lah2024lower}. These findings underscore the need for a more localized and structurally informed quantification approach for fine-grained tau analysis.

Voxel-wise analysis of tau PET data has been widely adopted to map spatial patterns of pathology \citep{waragai2009comparison,RN320}. While such methods offer fine-grained resolution, they face key limitations: multiple comparisons burden, spatial noise, and lack of anatomical interpretability. In particular, voxel-wise approaches often ignore the intrinsic geometry of curved brain structures, such as the hippocampus, making it difficult to capture smooth and consistent patterns of tau deposition. From a biological perspective, tau accumulation is unlikely to increase linearly with disease progression; rather, pathological effects are believed to emerge only once local SUVR exceeds a critical threshold~\citep{RN336}. Yet, most voxel- and region-based analyses assume a linear relationship between SUVR and disease severity, overlooking this threshold-driven transition. How spatially localized tau accumulation emerges and expands along disease progression therefore remains poorly understood, motivating our geometry-aware analysis of its spatial distribution and morphological trajectory.

In this study, we propose a manifold representation (m-rep) framework (\textbf{see Section~\ref{sec: workflow}} for methodological details) to quantify the spatial distribution and morphological changes in tau deposition. Specifically, we represent each subject’s tau PET signal on an anatomically aligned principal surface using features that encode suprathreshold coverage, signal intensity, and local geometry. To enable biologically meaningful and geometrically consistent surface construction, we adapt the outward-flux skeletonization idea by using flux values to identify medial candidate points within the atlas-defined hippocampal template, rather than using the original flux-based procedure to construct a fully connected skeleton topology. We then fit a principal surface to these flux-derived medial points, providing a smooth and computationally efficient medial representation for downstream PET signal projection. Covariate effects on the derived spatial features are subsequently assessed using two-stage regression with inverse probability weighting ~\citep{RN331}, though these biomarkers can be used in other settings depending on the specific analysis objectives.

AD is known to exhibit substantial clinical and pathological heterogeneity, particularly in how tau deposition patterns evolve across individuals. Emerging evidence suggests that AD comprises multiple subtypes that are characterized by different spatial trajectories of tau spread \citep{RN277}. Traditional approaches, such as group-level averaging or voxel-wise analysis, primarily compare tau accumulation across broad diagnostic categories (e.g., cognitively normal, mild cognitive impairment (MCI), AD) and implicitly assume a single, homogeneous disease trajectory. Such methods fail to capture subtype-specific differences in both the spatial distribution and progression dynamics of tau pathology. To explicitly account for this population-level heterogeneity, we applied the Subtype and Stage Inference (SuStaIn) algorithm \citep{RN289} to estimate each participant’s disease subtype and stage based on regional SUVR profiles. Incorporating these latent variables enables our geometric framework to distinguish subtype-specific spatial patterns and to relate morphological trajectories to disease progression.

Our surface representation and spatial feature extraction approach enables spatially localized modeling of hippocampal tau pathology. Stage and subtype covariates derived using SuStaIn are incorporated in subsequent analyses. Our primary objective in this study is to characterize subtype-specific tau deposition patterns and their progression dynamics, shedding light on the heterogeneity of AD. To this end, our analyses focus on modeling spatial deposition patterns—specifically, signal intensity and surface morphology—and their relationships with disease stage and subtype. While developed in the context of hippocampal tau pathology, the approach is readily extendable to other imaging applications.

In addition to capturing biologically meaningful patterns of tau deposition, the proposed framework also enables the identification of off-target contamination from the choroid plexus (CP), a well-known confound in hippocampal \textsuperscript{18}F-AV1451/flortaucipir PET imaging~\citep{RN334,RN335}.
The CP is a highly vascularized structure adjacent to the hippocampus that can exhibit strong nonspecific binding in \textsuperscript{18}F-AV1451/flortaucipir PET.
Because of its high signal intensity and close anatomical proximity, CP activity can leak into the hippocampal region during image preprocessing and registration, leading to apparent tau elevation or suppression unrelated to true pathology. Such misregistration effects are exacerbated in the elderly and populations with neurodegenerative disease, where ventricular enlargement increases spatial variability of the CP~\citep{RN323}. By analyzing directional signal asymmetries and spatial gradients on the hippocampal principal surface, our framework localizes CP spill-in effects and quantifies their dependence on clinical cognitive status, illustrating its broader utility for artifact detection and quality assessment in PET imaging.

\section{Materials and Methods}

In this section, we describe the data source, the overall workflow of our proposed framework, and the methodological framework for surface representation and feature extraction, followed by the derivation of disease subgroup and stage using SuStaIn and statistical analyses used in downstream modeling. For the analysis described in this manuscript, we used data from the Alzheimer's Disease NeuroImaging Initiative (ADNI), which is a landmark study launched in 2003 with the goal of collecting and sharing data to have the means of developing biomarkers for disease progression evaluation, characterization of AD, and treatment development. The study data collection protocols, documentation, and data dictionaries are available at \texttt{adni.loni.usc.edu}. We briefly describe our selection of variables and imaging used in this paper, further details can be found in ADNI data dictionaries. 

\subsection{Data Source and Preprocessing}

\paragraph{Sample characteristics}  
In the analysis, we included 628 PET--MRI pairs from 392 participants from the ADNI study \citep{RN328,RN270}, each with T1-weighted MRI acquired within two years of the corresponding tau PET scan. Clinical cognitive status was obtained directly from ADNI diagnostic records and categorized as cognitively normal (CN), mild cognitive impairment (MCI), or Alzheimer's disease dementia (AD). These classifications follow ADNI diagnostic criteria and are based on standardized clinical and neuropsychological assessments, including Clinical Dementia Rating (CDR), Mini-Mental State Examination (MMSE), Wechsler Logical Memory II delayed recall, assessment of functional status, and clinician judgment \citep{RN328}. Of these PET--MRI pairs, 304 were classified as CN, 218 as MCI, and 106 as AD.

The sample included 292 males and 336 females, with a mean age of 75.1 years (SD = 7.6, range = 55.2--94.4 years). Amyloid PET Centiloid values were obtained from the UC Berkeley ADNI amyloid PET analysis outputs and matched to the tau PET scans by participant ID using the tau PET \texttt{Study.Date} and amyloid PET \texttt{SCANDATE}, allowing a maximum scan-date difference of 730 days. Centiloid values, which provide a standardized scale for comparing amyloid burden across tracers \citep{RN236}, were available for 616 of 628 tau PET scans. The median absolute time difference between matched tau and amyloid PET scans was 13 days. Table \ref{table:demographics} describes the general demographic characteristics by ADNI clinical cognitive status for the PET--MRI pairs included in the final sample.

\begin{table}[htbp]
\centering
\caption{Baseline characteristics by ADNI clinical cognitive status}
\label{table:demographics}
\begin{tabular}{lcccc}
\toprule
 & \textbf{CN (N=304)} & \textbf{MCI (N=218)} & \textbf{AD (N=106)} & \textbf{Total (N=628)} \\
\midrule
\multicolumn{5}{l}{\textbf{Age (years)}} \\
\quad Mean (SD)         & 74.6 (7.20) & 74.9 (7.68) & 76.7 (8.16) & 75.1 (7.56) \\
\quad Median [Min, Max] & 74.3 [55.2, 94.4] & 74.8 [55.5, 93.2] & 76.6 [55.7, 92.0] & 75.2 [55.2, 94.4] \\
\addlinespace
\multicolumn{5}{l}{\textbf{Sex}} \\
\quad Female            & 158 (52.0\%) & 77 (35.3\%) & 53 (50.0\%) & 288 (45.9\%) \\
\quad Male              & 146 (48.0\%) & 141 (64.7\%) & 53 (50.0\%) & 340 (54.1\%) \\
\addlinespace
\multicolumn{5}{l}{\textbf{APOE $\epsilon$4 Carrier}} \\
\quad Non-carrier       & 182 (59.9\%) & 95 (43.6\%) & 26 (24.5\%) & 303 (48.2\%) \\
\quad Carrier           & 92 (30.3\%)  & 62 (28.4\%) & 41 (38.7\%) & 195 (31.1\%) \\
\quad Missing           & 30 (9.9\%)   & 61 (28.0\%) & 39 (36.8\%) & 130 (20.7\%) \\
\addlinespace
\multicolumn{5}{l}{\textbf{Amyloid PET Centiloid}} \\
\quad Available, n & 300 & 215 & 101 & 616 \\
\quad Mean (SD) & 23.4 (38.8) & 37.0 (45.1) & 79.0 (49.0) & 37.3 (47.0) \\
\quad Median [Min, Max] & 8.0 [-26, 157] & 23.0 [-39, 142] & 82.0 [-30, 207] & 18.5 [-39, 207] \\
\addlinespace
\multicolumn{5}{l}{\textbf{Amyloid status among scans with matched Centiloid}} \\
\quad Amyloid negative & 189 (63.0\%) & 105 (48.8\%) & 17 (16.8\%) & 311 (50.5\%) \\
\quad Amyloid positive & 111 (37.0\%) & 110 (51.2\%) & 84 (83.2\%) & 305 (49.5\%) \\
\quad Missing Centiloid & 4 (1.3\%) & 3 (1.4\%) & 5 (4.7\%) & 12 (1.9\%) \\
\bottomrule
\end{tabular}
\end{table}

\paragraph{Image acquisition and preprocessing}

Details on tau PET image acquisition and preprocessing are described elsewhere \citep{ADNI3PETTechnicalManual}. Briefly, tau PET imaging was performed using \textsuperscript{18}F-AV1451, also known as flortaucipir. Following ADNI PET acquisition protocols, participants received a nominal injected dose of 370 MBq (10.0 mCi) \(\pm 10\%\). Dynamic PET acquisition was performed approximately 75--105 minutes post-injection as a 30-minute scan consisting of six 5-minute frames.
All scans were pre-processed following the standard ADNI pipeline. The ADNI MRI and PET data used in our implementation had undergone standardized ADNI quality-control procedures. For MRI, ADNI performs automated protocol checks and manual review by trained analysts at the Mayo ADIR Laboratory, with scans graded for artifacts, motion, anatomical coverage, completeness, and overall image quality. For PET, the ADNI PET Core standardizes acquisition, quality control, preprocessing, and analysis, including pre-check and full QC at the University of Michigan and output QC at UC Berkeley \citep{ADNIPETQC}. We used images that passed MRI and PET QC based on ADNI protocols. Tau PET images were first co-registered to the corresponding native-space T1-weighted MRI, and then non-linearly warped to the MNI152 template space to ensure consistent anatomical alignment across subjects. SUVR images were computed using the cerebellar gray matter as the reference region, and partial volume correction was applied to reduce spill-over effects. Because the analysis implemented in this manuscript required MRI--PET co-registration and template-space alignment for hippocampal surface projection, we used FSL FLIRT affine registration with 12 degrees of freedom for the registration steps and FreeSurfer for MRI preprocessing and segmentation. To account for potential site-related variability, we applied the neuroCombat algorithm \citep{neuroCombat} to correct for scanner- and protocol-induced batch effects. Scans were grouped by scanner model, with rare models (fewer than five scans) merged by manufacturer. Age, sex, and cognitive status were included as biological covariates in the harmonization model to preserve meaningful variance while removing non-biological noise.

\paragraph{Hippocampal Structure and Tau Signal Definition}  \label{sec: atlas}

For spatial localization, hippocampal geometry was defined using the MNI152-space T1-weighted MRI automated anatomical labeling (AAL) atlas \citep{RN313}. The hippocampal region defined in this atlas was used to construct a medial surface representation, which served as a shared spatial surface for projecting tau PET signals. Since all PET images were nonlinearly registered to the MNI152 template, this surface-based framework enabled consistent, anatomically grounded comparison of tau deposition patterns across study participants. To identify regions with significant tau deposition, we applied a fixed thresholding procedure to each participant’s PET SUVR image. Voxels with SUVR values exceeding 2.0—approximately two standard deviations above the mean regional SUVR—were considered positive for tau signal. This subject-specific binary mask was then projected onto the medial hippocampal surface for downstream shape-based analysis. 
This threshold was adopted from the SuStaIn model (see Section~\ref{sec:sustain_method}) and provided a biologically meaningful cutoff for downstream analysis.

\subsection{Workflow Overview} \label{sec: workflow}

\begin{figure}[ht]
    \centering
    \includegraphics[width=0.95\textwidth]{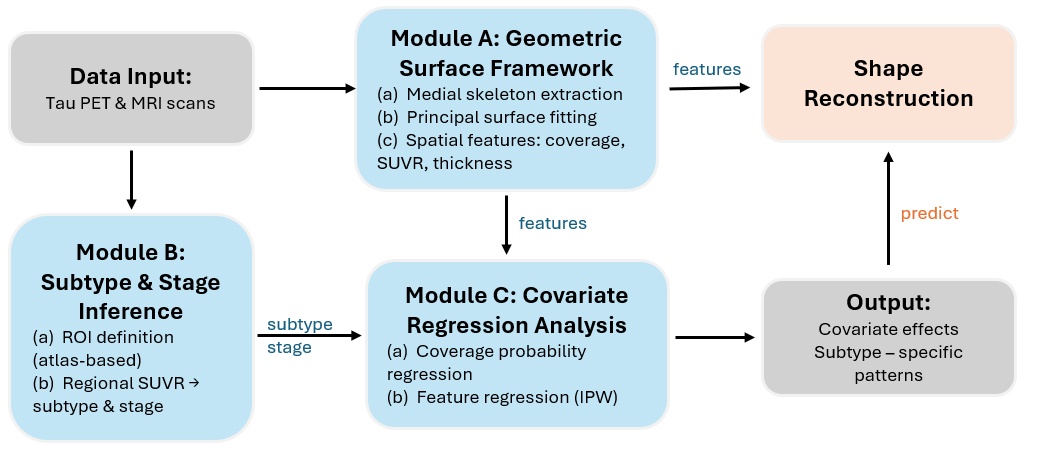} 
    \caption{
Workflow for spatial tau feature extraction, subtype modeling, and regression analysis.  
Module A constructs a shared medial surface representation and extracts surface-based tau features, including suprathreshold coverage, interpolated SUVR intensity, and bidirectional projection-distance measures.
Module B: Subtype \& Stage Inference applies SuStaIn to atlas-defined regional SUVR values to derive disease subtype and stage labels.  
Module C: Regression Analysis evaluates the effects of demographic and clinical variables on spatial tau features via (a) coverage probability modeling and (b) inverse probability weighted (IPW) regression.  
This workflow supports both shape reconstruction from geometric features and subtype-specific covariate effect analysis, enabling detailed investigation of heterogeneous spatial tau deposition patterns.
}
    \label{fig:workflow}
\end{figure}

In this study, we present a surface-based workflow for analyzing subject-specific tau deposition patterns and structural features in both the left and right hippocampus. The workflow presented in Figure~\ref{fig:workflow} integrates standard image pre-processing, geometric modeling and feature extraction, subtype and stage inference, and statistical analysis.
We process individual PET scans to extract tau signal distributions projected onto a shared medial surface, enabling consistent per-subject shape characterization. By anchoring the analysis in this anatomically consistent coordinate system, we are able to aggregate subject-level representations and conduct group-wise comparisons.
This design supports biologically interpretable feature extraction, which serves as the basis for downstream regression modeling. The modules and steps of the workflow are illustrated in Figure~\ref{fig:workflow}.

To clarify the role of each component, this workflow combines established methods with study-specific adaptations and modeling components. ADNI image preprocessing, atlas-based hippocampal definition, SuStaIn subtype and stage inference, outward-flux skeletonization, principal surface fitting, and conventional voxel-wise regression are used for data preparation, disease stratification, geometric representation, and group comparisons. In particular, we use outward flux to identify medial candidate points in the hippocampal template, but replace the computationally intensive step of constructing a fully connected flux-based skeleton topology with principal surface fitting adapted for this setting. This yields a smooth shared medial surface for projecting tau PET signal and deriving coverage, smoothed SUVR intensity, and bidirectional projection-distance features, which are then analyzed using two-stage surface-based regression models for coverage probability and IPW-adjusted tau intensity and thickness. The main contribution of this study is the integration and adaptation of these components into a surface-based PET analysis framework, including the optimized flux-based medial skeleton construction for hippocampal PET surface representation, the shared medial-surface projection framework for deriving suprathreshold coverage, interpolated SUVR intensity, and bidirectional projection-distance features, and the application of surface-based regression analyses, including IPW-adjusted analyses, to evaluate SUVR intensity and projection-based thickness.

\paragraph{Module A: Geometric Surface Modeling and Feature Extraction:} The core methodological framework consists of geometric surface modeling and spatial feature extraction. We transform the 3D voxel-wise tau PET signal within the hippocampus into a structured 2D representation on a medial principal surface. Surface construction involves two steps: (1) computing outward-flux values on the atlas-defined hippocampal structure to identify medial candidate points, and (2) fitting a smooth principal surface to these flux-derived points via thin-plate spline (TPS) regularization. Directly applying principal surface fitting to the complex hippocampal geometry often produced convoluted, self-folding surfaces; thus, our two-step approach ensured biologically meaningful and geometrically consistent surface modeling. After aligning subject-level tau PET data to the MNI152 space, suprathreshold tau signals are projected and interpolated onto this shared surface, enabling consistent per-subject shape characterization and group-wise comparisons. This transformation reduces data dimensionality, while preserving the anatomically grounded spatial organization of tau signals. From the projected data, we derive standardized spatial features including suprathreshold coverage, SUVR intensity, and projection-based thickness.

\paragraph{Module B: Subtype and Stage Classification with SuStaIn:} To stratify heterogeneity in tau progression, we applied the Subtype and Stage Inference (SuStaIn) algorithm to derive disease subtype and stage for each participant. SuStaIn, a widely adopted method in AD research, uses z-score thresholds and regional severity scores to model disease progression trajectories and has been instrumental in identifying biologically meaningful AD subtypes. Although not part of the core geometric framework, we compared our estimated tau deposition patterns between the SuStaIn subtypes and stages.

\paragraph{Module C: Modeling of Spatial Tau Deposition:} Following feature extraction, we performed a two-stage regression analysis to examine differences of tau deposition between AD subtypes and stages on tau deposition patterns. Covariate-by-subtype interaction terms were included to enable formal comparison of subtype-specific effects. In the first stage, we estimated the probability of suprathreshold signal coverage at each uniformly sampled location on the principal surface, while in the second stage we modeled the effects of demographic and disease-related covariates—including age, sex, cognitive status, and SuStaIn-derived stage—on SUVR intensity and projection-based thickness. To mitigate selection bias from conditioning on signal presence, inverse probability weighting (IPW) was applied. This approach allows for robust estimation of both signal intensity and morphological variation, accounting for spatial sparsity and subtype-specific heterogeneity across spatially consistent surface points.

This workflow was applied to both left and right hippocampi as a case study, demonstrating its utility in quantifying and comparing tau pathology across subtypes and stages. While developed for hippocampal tau analysis, the workflow is generalizable and can be extended to other brain regions or disease types, providing a robust approach for analyzing structural and pathological heterogeneity.

\subsection{Notation and Definitions} \label{sec: notation}
To ensure consistency throughout this paper, we adopt the following notations, with descriptive definitions to clarify their roles in the study:

\begin{itemize}

    \item \(R_0 \subset \mathbb{R}^3\) denotes the fixed atlas-defined hippocampal template region used in this study, and \(S_0=\partial R_0\) denotes its boundary surface. Flux-derived medial points and the fitted principal surface are obtained from \(R_0\), hence \(R_0\) provides the shared anatomical reference geometry onto which subject-level tau PET signal is projected.

    \item For subject $i=1,\ldots,n$, $R_i \subset R_0$ denotes the set of suprathreshold tau-positive voxels within the hippocampal template region after PET registration and thresholding. The set $R_i$ is not used to construct a subject-specific principal surface; rather, its voxels are projected onto the shared surface derived from $R_0$ for feature extraction.

    \item For subject $i$, $\mathbf{q}_{iv}=(x_{iv},y_{iv},z_{iv})\in R_i$, $v=1,\ldots,V_i$, denotes the coordinate of the $v$th suprathreshold tau-positive voxel, where $V_i$ is the total number of suprathreshold voxels for subject $i$. Each voxel is associated with an SUVR value $s_{iv}$.

    \item $d_0(\mathbf q)$ denotes the distance-transform function of the fixed hippocampal template region $R_0$. For any point $\mathbf q\in R_0$, $d_0(\mathbf q)$ is defined as the shortest Euclidean distance from $\mathbf q$ to the template boundary $S_0=\partial R_0$:
    \[
    d_0(\mathbf q)=\min_{\mathbf p\in S_0}\|\mathbf q-\mathbf p\|_2.
    \]
    Equivalently, if $\pi_0(\mathbf q)$ denotes a closest boundary point on $S_0$, then $d_0(\mathbf q)=\|\mathbf q-\pi_0(\mathbf q)\|_2$. This distance-transform function is used to compute the gradient field for outward-flux-based identification of medial points.

    \item $\dot{\mathbf{q}}=\nabla d_0(\mathbf{q})=(\partial_x d_0,\partial_y d_0,\partial_z d_0)$ denotes the gradient vector field of the distance function, indicating the local direction of maximal increase of distance, which in this context points inward from the boundary toward the interior. The vector field is normalized such that $\|\dot{\mathbf{q}}\|=1$ wherever the gradient is defined.

    \item $\Phi(\mathbf{q})$ denotes the average outward flux of the normalized distance-gradient field $\dot{\mathbf{q}}$ at location $\mathbf{q}\in R_0$. It is defined in the continuous setting as the limit of the average directional projection over a vanishing spherical neighborhood:

    \begin{equation}\label{eq:Phi}
    \Phi(\mathbf{q}) = \lim_{\varepsilon \to 0} \frac{1}{|\partial B_\varepsilon(\mathbf{q})|} \int_{\mathbf{p} \in \partial B_\varepsilon(\mathbf{q})} \left\langle \dot{\mathbf{q}}(\mathbf{p}), \mathbf{n}_{\mathbf{p}} \right\rangle \, dS(\mathbf{p}),
    \end{equation}
    
    where $B_\varepsilon(\mathbf{q})$ denotes a closed ball of radius $\varepsilon$ centered at $\mathbf{q}$, $\partial B_\varepsilon(\mathbf{q})$ is its boundary surface, $\mathbf{n}_{\mathbf{p}}$ is the unit outward normal vector at $\mathbf{p}\in\partial B_\varepsilon(\mathbf{q})$, and $dS$ is the surface area element. The flux quantifies the local directional consistency of the distance-gradient field $\dot{\mathbf{q}}$. Values of $\Phi(\mathbf{q})$ near zero correspond to homogeneous regions, while strictly negative values are indicative of medial points—locations equidistant from multiple boundary regions where the vector field is highly divergent.

    \item In this study, we distinguish between two types of unit normal vectors. \(\mathcal{N}_{\text{MS}}\) represents the local medial normal direction associated with the flux-derived medial structure, defined by the vector from the center of the corresponding maximal inscribed sphere to its tangent point on the boundary surface \(S_0\). \(\mathcal{N}_{\text{PS}}\) is the normal vector of the principal surface, defined as the unit vector orthogonal to the local tangent plane at a point on the surface. Although \(\mathcal{N}_{\text{MS}}\) and \(\mathcal{N}_{\text{PS}}\) originate from distinct geometric definitions, they are closely aligned in practice because the principal surface is fitted as a smooth approximation to the flux-derived medial point cloud. Accordingly, we treat them as directionally equivalent for projection-distance calculation and use \(\mathcal{N}\) to denote the normal direction unless otherwise specified; the directional-consistency assumption is discussed in Supplementary Figure~\ref{fig:validation_reconstruction}.

    \item \( \mathbf{t} = (t_1, t_2) \in [0, 1] \times [0, 1] \) is used as a generic 2D parameterization coordinate for the principal surface, representing a location in the normalized 2D parameter domain.
    
    \item \( f: [0,1]^2 \to \mathbb{R}^3 \) represents a smooth mapping function for the principal surface in 3D space. It maps a 2D parameterization coordinate \( \mathbf{t} \) to its corresponding point \( f(\mathbf{t}) \) on the principal surface.
    
    \item \( \lambda_f: \mathbb{R}^3 \to [0,1]^2 \) is a projection function that maps a 3D point \( \mathbf{q} \) onto the parameterization space, identifying the closest corresponding point on the principal surface. For a subject-specific tau-positive voxel $\mathbf{q}_{iv}$, we write $\mathbf{t}_{iv}=\lambda_f(\mathbf{q}_{iv})$ for its projected 2D coordinate.

    \item $D^{+}_{ik}$ and $D^{-}_{ik}$ denote the positive- and negative-direction projection-distance features for subject $i$ at grid point $k$. These features measure the local extent of suprathreshold tau signal from the principal surface along the two opposite directions of the surface normal.

\end{itemize}

\subsection{Geometric Surface Modeling and Feature Extraction}\label{sec: geometric rep}

\begin{figure}[htbp]
    \centering
    \includegraphics[width=0.95\textwidth]{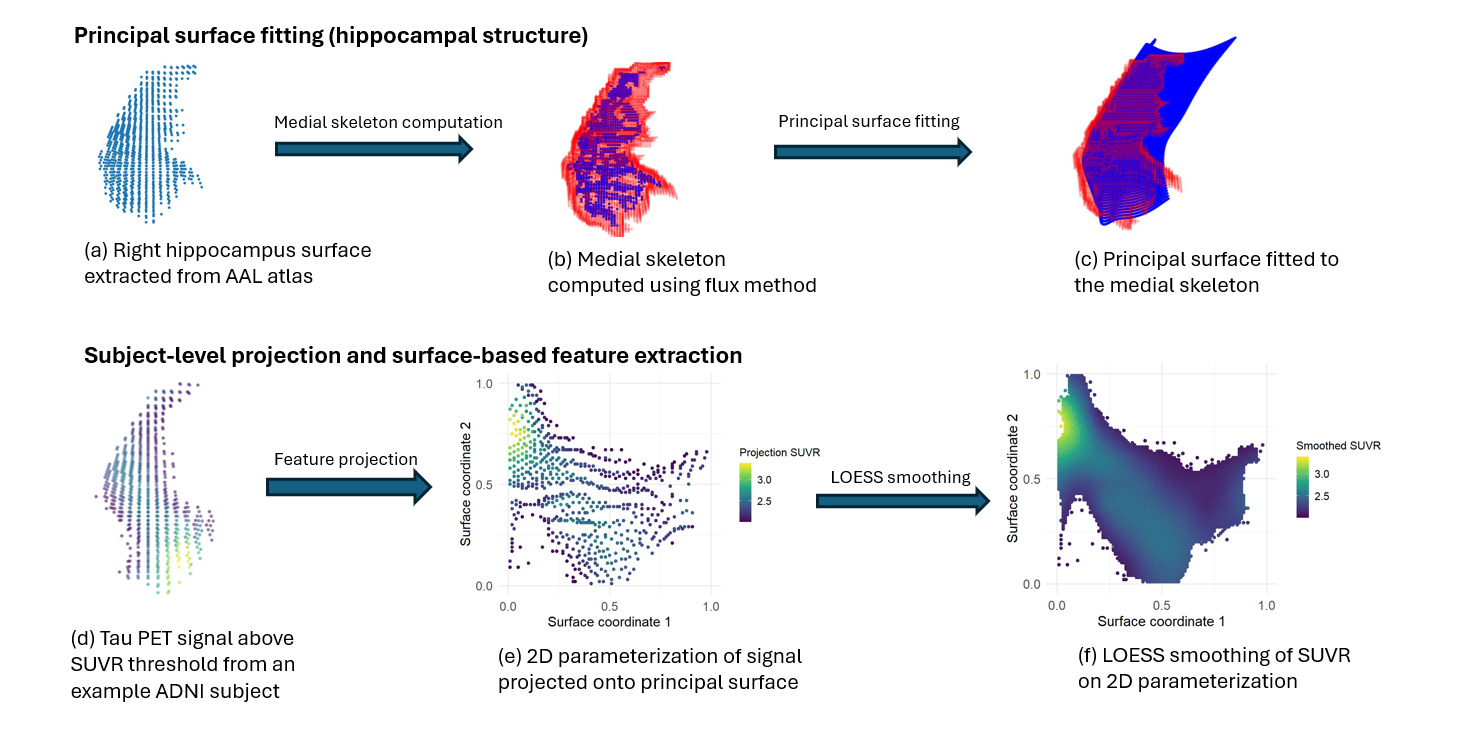} 
    \caption{Framework for medial surface-based feature extraction from hippocampal tau PET data, corresponding to Module A in Figure~\ref{fig:workflow}.
(a) The right hippocampus structure was extracted from the AAL atlas and displayed from the anterior-facing view used throughout the workflow.
(b) Outward flux was used to identify medial points capturing the intrinsic shape geometry.
(c) A principal surface was fitted to the flux-derived medial points to provide a smooth 2D representation.
(d) Tau PET SUVR signals above threshold from an example ADNI participant were identified within the hippocampal structure. 
(e) These signals were projected onto the principal surface and parameterized using normalized surface coordinates.
(f) LOESS smoothing was applied to the projected SUVR values to visualize spatial patterns of tau accumulation across the 2D surface.
Voxel-coordinate axes are omitted in the 3D panels because they represent image-space locations rather than workflow-specific quantities; surface-coordinate axes are retained in the 2D projection and smoothing panels. The arrows indicate the sequential processing steps from anatomical surface extraction to signal projection and smoothing. }
    \label{fig:mapping_interpolation_suvr}
\end{figure}

We developed a geometric surface modeling framework that integrates outward-flux-derived medial points with principal surface fitting to represent subject-specific tau deposition within the hippocampus. This approach, corresponding to Module A of our overall workflow shown in Figure~\ref{fig:workflow} and illustrated in detail in Figure~\ref{fig:mapping_interpolation_suvr}, integrates outward-flux-based identification of medial points and principal surface (PS) fitting to create an anatomically aligned, low-dimensional representation. The resulting medial representation (m-rep) enables standardized spatial feature extraction across individuals, including suprathreshold coverage, SUVR intensity, and projection-based thickness, forming the basis for subsequent analyses.

\subsubsection{Medial Point Identification via Outward Flux} \label{sec: medial skeleton}

Accurately capturing the intrinsic geometry of the hippocampus is essential for modeling spatial patterns of tau deposition. As the first component of the geometric modeling workflow (Module A-a in Figure~\ref{fig:workflow}; Figure~\ref{fig:mapping_interpolation_suvr}a--b), we used outward flux to identify medial candidate points within the atlas-defined hippocampal template \(R_0\). 
These flux-derived medial points were computed from the fixed atlas-defined hippocampal template region \(R_0\), rather than separately from each subject-specific tau-positive voxel set. Thus, they provide a shared anatomical reference geometry for subsequent projection of subject-level PET signal.

The medial candidate points are motivated by the classical grassfire flow model~\citep{blum_transformation_1967}, which conceptualizes the inward propagation of a shape's boundary at constant speed along the inward normal direction, with medial locations emerging where inward-propagating fronts meet. This process is commonly formalized using Hamilton-Jacobi equations~\citep{RN264,RN262}. For implementation on the voxelized hippocampal template region, we computed the average outward flux of the distance-transform gradient field following prior flux-based shape analysis work~\citep{RN263}. 

Specifically, we compute the average outward flux \( \Phi(\mathbf{q}) \) at each voxel location \( \mathbf{q}\in R_0 \) by evaluating the average directional relationship between the local distance gradient field \( \dot{\mathbf{q}}(\mathbf{p}) \) and the outward normal vector \( \mathbf{n}_{\mathbf{p}} \) over a small spherical neighborhood using equation~(\ref{eq:Phi}). 

In homogeneous interior regions, where the distance-gradient field varies smoothly, flux values approach zero. In contrast, near medial locations, where distance-gradient directions from opposing boundaries meet, the average outward flux becomes negative, indicating local symmetry of the surrounding shape~\citep{RN263}. 
In the original flux-based skeletonization setting, these flux-derived points can be further processed to construct a connected skeleton topology. In our framework, we use the flux computation only to obtain a medial point cloud and do not rely on the original topology-construction procedure. Instead, the flux-derived medial points are used as the input for principal surface fitting, which provides a smooth, continuous, and computationally efficient medial representation for subsequent PET signal projection.

\subsubsection{Principal Surface Fitting} \label{sec: principal surface}

To convert the flux-derived medial point cloud into a smooth surface representation, we applied the principal surface fitting method proposed by \cite{RN221}. This step (Module A-b in Figure~\ref{fig:workflow}; Figure~\ref{fig:mapping_interpolation_suvr}c) regularizes the geometry and establishes a continuous two-dimensional parameterization of the hippocampal structure. The resulting parameter space facilitates spatial alignment, dimensionality reduction, and consistent representation of tau deposition patterns across individuals. We first centered and projected the flux-derived medial points in \( \mathbb{R}^3 \) onto their first two principal components using principal component analysis (PCA). The resulting coordinates captured the dominant geometric variation and were standardized to the domain \([0, 1] \times [0, 1]\), providing an initial two-dimensional parameterization consistent with downstream interpolation and analysis. We then applied a bivariate thin-plate spline (TPS) to fit a smooth principal surface, minimizing penalized least-squares error to achieve geometric regularization~\citep{mgcv}. 

The TPS produced a smooth mapping \( f: [0,1]^2 \to \mathbb{R}^3 \) from the standardized 2D parameter domain to the fitted principal surface:
\[
f(\mathbf{t}) = 
\begin{bmatrix}
    \hat{f}_{11}(t_1) + \hat{f}_{12}(t_2) + \hat{f}_{13}(t_1,t_2) \\
    \hat{f}_{21}(t_1) + \hat{f}_{22}(t_2) + \hat{f}_{23}(t_1,t_2) \\
    \hat{f}_{31}(t_1) + \hat{f}_{32}(t_2) + \hat{f}_{33}(t_1,t_2)
\end{bmatrix},
\quad \mathbf{t}=(t_1,t_2)\in[0,1]^2.
\]
where \(\mathbf{t}\) denotes a location in the standardized 2D parameter domain, and \(f(\mathbf{t})\) is the corresponding 3D point on the fitted principal surface. This parameterization step provides a smooth and consistent mapping from the flux-derived medial point cloud to the principal surface.

Once the initial surface \(f(\mathbf{t})\) was fitted, each flux-derived medial point \( \mathbf{q}^{\mathrm{med}}_{\ell} \) was projected onto the current principal surface using the projection function \( \lambda_f \), which identified the closest point on the surface:
\[
\lambda_f(\mathbf{q}^{\mathrm{med}}_{\ell}) =
\sup_{\mathbf{t}} \left\{\mathbf{t} : 
\left\| \mathbf{q}^{\mathrm{med}}_{\ell} - f(\mathbf{t}) \right\|
=
\inf_{\boldsymbol{\mu}}
\left\| \mathbf{q}^{\mathrm{med}}_{\ell} - f(\boldsymbol{\mu}) \right\|
\right\}.
\]
In cases where multiple parameterizations \( \mathbf{t} \) yielded identical projection distances, the largest \( \mathbf{t} \) value under the ordering used in the original implementation was selected to ensure consistent mapping and to resolve potential non-uniqueness in the projection, following the original implementation by~\cite{RN221}. The parameterization was then iteratively updated until changes across iterations fell below a predefined threshold. This iterative refinement ensured that the final principal surface accurately represented the flux-derived medial geometry and provided a stable parameterization for subsequent spatial feature extraction.

\subsubsection{Projection and Feature Extraction}  \label{sec: loess}

To ensure standardized feature extraction across participants (Module A-c in Figure~\ref{fig:workflow}; Figure~\ref{fig:mapping_interpolation_suvr} d-f), we projected subject-specific suprathreshold tau-PET signal onto the shared principal surface constructed from the atlas-defined hippocampal template region \(R_0\). Specifically, for subject \(i\), let \(R_i \subset R_0\) denote the set of hippocampal voxels with \(\mathrm{SUVR}>2.0\) after registration and thresholding. Each voxel \(\mathbf{q}_{iv}\in R_i\) was projected to its nearest point on the shared principal surface using Euclidean distance. The PS was derived from \(R_0\) through outward-flux-based identification of medial points and principal surface fitting (see Sections~\ref{sec: medial skeleton} and~\ref{sec: principal surface}).

\paragraph{SUVR Interpolation}

To construct reference anchors for surface-based interpolation, we uniformly sampled 235 fixed grid points on the left PS and 262 on the right to ensure coverage on both sides of the principal surface and spatial uniformity. These numbers were specific to fit the hippocampal surface and can be adjusted for other regions depending on their geometry and spatial extent. Let \(\mathbf{t}_k\in[0,1]^2\), \(k=1,\ldots,K\), denote a fixed grid point on the shared principal-surface parameter domain, and let \(f(\mathbf{t}_k)\in\mathbb{R}^3\) denote its corresponding 3D surface location. For subject \(i\), each projected tau-positive voxel \(\mathbf{q}_{iv}\) has a corresponding two-dimensional coordinate \(\mathbf{t}_{iv}\) on the principal-surface domain and an SUVR value \(s_{iv}\).

At each reference location \(\mathbf{t}_k\), SUVR values were estimated using LOESS interpolation~\citep{loess}. 
We define the interpolated SUVR value for subject \(j\) at grid point \(k\) by fitting a local linear model to the projected suprathreshold points from that subject:
\[
(\hat{\beta}_{0,ik}, \hat{\boldsymbol{\beta}}_{ik})
=
\arg\min_{\beta_0,\boldsymbol{\beta}}
\sum_{v=1}^{V_i}
a_{iv}(\mathbf{t}_k)
\left(
s_{iv}
-
\beta_0
-
\boldsymbol{\beta}^{\top}(\mathbf{t}_{iv}-\mathbf{t}_k)
\right)^2 ,
\]
where \(a_{iv}(\mathbf{t}_k)\) is a tricube LOESS weight that decreases with the distance between \(\mathbf{t}_{iv}\) and \(\mathbf{t}_k\). The interpolated SUVR value is then \(\hat{s}_{ik}=\hat{\beta}_{0,ik}\).
Interpolation is limited to regions with sufficient local data support (Figure~\ref{fig:mapping_interpolation_suvr}b–d). The same fixed grid points were used to summarize geometric projection-distance features for downstream analysis.

\paragraph{Projection-Based Shape Representation} \label{sec:reconstruction}

The spatial features—including suprathreshold coverage, interpolated SUVR intensity, and bidirectional projection distances—provide a compact yet comprehensive characterization of tau deposition. 
For subject \(i\) at grid point \(k\), we denote by \(D^{+}_{ik}\) and \(D^{-}_{ik}\) the positive- and negative-direction projection-distance features, respectively. These quantities measure the local extent of suprathreshold tau signal from the principal surface along the two opposite directions of the surface normal.
These directional measurements capture local thickness and anisotropy of deposition patterns and are interpolated at the same grid points for downstream shape analysis.

Together, these features support both feature-based statistical modeling and surface-based shape reconstruction. For example, by combining the 2D parameter coordinates \(\mathbf{t}_k\), their corresponding surface normals, and interpolated projection distances, we can reconstruct the tau deposition boundary for a subject as
\[
\hat{\mathbf{q}}^{+}_{ik}
=
f(\mathbf{t}_k)
+
D^{+}_{ik}\mathcal{N}_{\mathbf{t}_k},
\qquad
\hat{\mathbf{q}}^{-}_{ik}
=
f(\mathbf{t}_k)
-
D^{-}_{ik}\mathcal{N}_{\mathbf{t}_k}.
\]
This reconstruction represents subject-level tau deposition geometry relative to the shared principal surface. This reconstruction demonstrates that the geometry and thickness of tau accumulation can be recovered from the principal surface representation. By integrating projection-based features with the anatomical alignment of the PS, our medial representation (m-rep) preserves sufficient structural detail for both visualization and statistical modeling.

\subsection{Disease Subtype and Stage Modeling with SuStaIn} 
\label{sec:sustain_method}

To characterize heterogeneous patterns of tau progression, we applied the Subtype and Stage Inference (SuStaIn) model \citep{RN289} to derive subtype and stage classifications. SuStaIn infers disease subtypes and progression stages by simultaneously clustering and staging subjects based on regional SUVR variation. We implemented the model using its publicly available Python package \citep{RN290}. Following prior studies \citep{RN313}, we constructed a subject-by-feature matrix using mean SUVR values from 10 consolidated regions of interest (ROIs) defined by the AAL atlas: left and right frontal lobes, left and right parietal lobes, left and right occipital lobes, left and right temporal lobes, and left and right medial temporal lobes.
For each ROI, regional SUVR values among cognitively normal participants were modeled using a two-component Gaussian mixture model, and the lower-mean Gaussian component was used as the estimated normal reference distribution. Regional SUVR values were then converted to z-scores relative to this estimated normal distribution, such that a \(z=2\) event represented an SUVR value two standard deviations above the estimated normal mean. Two event thresholds (\(z>2\) and \(z>5\)) were used for each ROI, resulting in 20 SuStaIn event features per subject. The corresponding ROI-specific SUVR cutoffs are reported in Table~\ref{tab:suvr_cutoffs}.

The optimal number of subtypes (\(K_{\mathrm{sub}}=4\)) was selected via ten-fold cross-validation based on the lowest CVIC and out-of-sample likelihood. This solution aligned with previous findings \citep{RN277}, identifying two dominant patterns in the ADNI dataset \citep{RN270,RN328}: a limbic-predominant subtype and an occipitotemporal/posterior variant. Each participant was assigned a most likely subtype and a disease stage (1--20), representing sequential tau accumulation across ROIs. These SuStaIn-derived classifications were used in downstream regression analyses, enabling subtype-specific comparisons and stage-based trajectory modeling.

\begin{table}[htbp]
\centering
\caption{Region-specific SUVR cutoffs corresponding to SuStaIn z-score event thresholds in the ADNI tau PET cohort.}
\label{tab:suvr_cutoffs}
\vspace{0.6em}
\begin{tabular}{llcc}
\toprule
\textbf{ROI} & \textbf{Anatomical region} & \textbf{\(z=2\) SUVR cutoff} & \textbf{\(z=5\) SUVR cutoff} \\
\midrule
FLL  & Left frontal lobe          & 1.739 & 2.305 \\
FLR  & Right frontal lobe         & 1.698 & 2.263 \\
PLL  & Left parietal lobe         & 1.692 & 2.267 \\
PLR  & Right parietal lobe        & 1.612 & 2.159 \\
OLL  & Left occipital lobe        & 1.957 & 2.656 \\
OLR  & Right occipital lobe       & 1.934 & 2.642 \\
\textbf{TLL}  & \textbf{Left temporal lobe}         & \textbf{2.008} & \textbf{2.724} \\
\textbf{TLR}  & \textbf{Right temporal lobe}        & \textbf{1.976} & \textbf{2.700} \\
\textbf{MTLL} & \textbf{Left medial temporal lobe}  & \textbf{2.025} & \textbf{2.785} \\
\textbf{MTLR} & \textbf{Right medial temporal lobe} & \textbf{2.057} & \textbf{2.799} \\
\midrule
Average & Across 10 ROIs           & 1.870 & 2.530 \\
\bottomrule
\end{tabular}

\vspace{0.5em}
\begin{minipage}{0.92\textwidth}
\footnotesize
\textit{Note.} For each ROI, regional SUVR values among cognitively normal participants were modeled using a two-component Gaussian mixture model. The lower-mean Gaussian component was used as the estimated normal reference distribution, and the \(z=2\) and \(z=5\) event thresholds were back-transformed to the original SUVR scale. Temporal and medial temporal ROIs are highlighted because they are most relevant to hippocampal tau involvement and have \(z=2\) cutoffs close to the primary \(\mathrm{SUVR}>2.0\) threshold used to define suprathreshold hippocampal tau signal.
\end{minipage}
\end{table}

\subsection{Tau PET Threshold Selection}
\label{sec:threshold_selection}

We used a primary threshold of \(\mathrm{SUVR}>2.0\) to define suprathreshold hippocampal tau signal for surface projection and downstream modeling. This choice was derived from the SuStaIn reference-distribution framework described above. Specifically, the SuStaIn \(z=2\) event thresholds were back-transformed to the original SUVR scale, yielding temporal and medial temporal cutoffs close to 2.0: 2.008 for the left temporal lobe, 1.976 for the right temporal lobe, 2.025 for the left medial temporal lobe, and 2.057 for the right medial temporal lobe (Table \ref{tab:suvr_cutoffs}). Thus, the fixed threshold \(\mathrm{SUVR}>2.0\) closely matched the region-specific abnormality thresholds implied by the SuStaIn model while providing a single interpretable cutoff for defining suprathreshold hippocampal tau signal.

Prior work has also shown that \textsuperscript{18}F-AV1451/flortaucipir tau PET cut-points vary by ROI, reference group, and thresholding method; in particular, \citet{Weigand2022CutPoint} showed that reference-group mean-plus-two-standard-deviation cut-points in the entorhinal cortex and a temporal meta-ROI provided relatively conservative thresholds with strong concurrent validity for tau PET positivity classification. Because the choice of threshold determines which hippocampal voxels are projected onto the principal surface, we further evaluated the robustness of the surface-projection and downstream regression results across nearby SUVR cutoffs, as described below.

\subsection{Surface-Based Regression Analysis of Tau Features}
\label{sec: 2stage regression}

To investigate how demographic and clinical variables influence hippocampal tau pathology, we performed pointwise regression analyses on spatial tau features across the hippocampal surface. Analyses were conducted at 497 uniformly sampled locations on the principal surface (235 left, 262 right), corresponding to the projection grid points defined during surface reconstruction (see Section~\ref{sec:reconstruction}). Due to spatial sparsity in tau accumulation, not all subjects exhibited suprathreshold signal at every location, requiring a modeling framework that accounts for missing data and selection bias. To evaluate whether covariate effects differed significantly across disease subtypes, we employed a unified regression model incorporating covariate-by-subtype interaction terms. The interaction terms captured differences in effect magnitude between subtype 1 and subtype 2, enabling statistical inference on heterogeneity. 

\paragraph{Stage 1: Logistic Regression for Coverage} \label{sec: coverage regression method}

For each fixed surface grid point \(k\), we defined a binary coverage indicator \(C_{ik}\) for subject \(i\), where \(C_{ik}=1\) indicates the presence of suprathreshold tau signal at that surface location and \(C_{ik}=0\) otherwise. We fit a logistic regression model for \(C_{ik}\) using age, sex, cognitive status (CN, MCI, AD), SuStaIn disease stage encoded using a second-order polynomial basis, subtype, and interaction terms between subtype and each covariate. These models estimate the probability \(\hat p_{ik}=P(C_{ik}=1\mid \mathbf X_i)\) of suprathreshold tau coverage at surface point \(k\). The predicted probabilities from this stage were used to compute inverse probability weights for downstream analyses.

\paragraph{Stage 2: Inverse-Probability-Weighted Linear Regression on Tau Features}

Conditional on coverage, we modeled covariate effects on three surface-derived features: 
(1) positive projection distance (signal expansion above the surface),
(2) negative projection distance (signal expansion below the surface), 
(3) smoothed SUVR values.

As these features are only defined at covered suprathreshold locations, directly modeling them without correction may induce selection bias. To address this, we applied inverse probability weighting (IPW)~\citep{RN331}. For subject \(i\) at surface point \(k\), the inverse probability weight was defined as \(w_{ik}=1/\hat p_{ik}\), where \(\hat p_{ik}\) is the predicted probability of coverage from Stage~1. For each surface point \(k\), weighted least squares regression was then used to estimate covariate effects on each selected tau feature:
\[
\hat{\boldsymbol{\beta}}_k
=
\arg\min_{\boldsymbol{\beta}}
\sum_{i:C_{ik}=1}
w_{ik}
\left(
Y_{ik}
-
\mathbf X_i^\top \boldsymbol{\beta}
\right)^2 ,
\]
where \(Y_{ik}\) denotes one of the surface-derived tau features \(\hat{s}_{ik}\), \(D^{+}_{ik}\), or \(D^{-}_{ik}\), and \(\mathbf X_i\) includes the same covariates, subtype indicator, and interaction terms used in Stage~1.

This regression method enables formal statistical comparison of covariate effects across subtypes while correcting for signal-dependent inclusion. P-values from all models were adjusted using the Benjamini--Hochberg (BH) procedure~\citep{RN332} across all surface points and hemispheres.

\subsection{Threshold Sensitivity Analysis}
\label{sec:threshold_sensitivity}

To assess whether the main findings depend on the specific choice of \(\mathrm{SUVR}>2.0\), we repeated the surface-projection workflow using five thresholds centered around the primary cutoff: 1.8, 1.9, 2.0, 2.1, and 2.2. For each threshold, suprathreshold hippocampal voxels were projected onto the shared principal surface using the same projection procedure as in the primary analysis. We then summarized threshold-dependent changes in tau coverage at two levels: subject-level surface coverage burden, defined as the proportion of principal-surface points covered by suprathreshold signal within each scan, and surface-point coverage frequency, defined as the proportion of scans in which suprathreshold signal was mapped to each surface location.

We also repeated the IPW-adjusted smoothed SUVR regression analysis under each threshold to evaluate the stability of downstream covariate-effect findings. For each regression term and SuStaIn subtype, the primary \(\mathrm{SUVR}>2.0\) analysis was treated as the reference map. We converted the BH-corrected point-wise p-value maps into binary significance masks and compared each alternative-threshold mask with the corresponding primary-threshold mask using the Dice similarity coefficient:
\[
\mathrm{Dice}(A,B_t)=\frac{2|A\cap B_t|}{|A|+|B_t|},
\]
where \(A\) denotes the set of significant surface points under the primary \(\mathrm{SUVR}>2.0\) threshold and \(B_t\) denotes the corresponding set under alternative threshold \(t\). Dice values were calculated using pooled left-and-right hippocampal surface points. As a secondary analysis, we also compared the continuous point-wise coefficient maps using Pearson and Spearman correlations. The corresponding threshold-sensitivity results are reported in Section~\ref{sec:threshold_sensitivity_results}.
\subsection{Comparison with Voxel-wise Regression and Monte Carlo Simulation}

We evaluated the improvement of the proposed surface-based framework over conventional voxel-wise analysis at two complementary levels. First, we performed an empirical comparison on the observed ADNI tau PET data using voxel-wise linear regression, a commonly used approach for mapping spatial PET effects at high resolution~\citep{waragai2009comparison,RN320}. At each voxel within the bilateral hippocampus defined by the AAL2 atlas, SUVR was modeled using the same covariates and subtype interaction terms used in the surface-based analysis, including age, sex, cognitive status, SuStaIn stage, and subtypes. Unlike the proposed framework, this voxel-wise analysis was performed directly in the original voxel space and did not use principal-surface projection, surface-based coverage modeling, or IPW correction for signal-dependent inclusion. P-values were adjusted across voxel locations using the Benjamini--Hochberg procedure. To enable visual comparison with the surface-based maps, we extracted the voxels closest to the medial principal surface and displayed the corresponding subtype 1 significance maps.

Second, we conducted a Monte Carlo simulation to compare the performance of the two methods in recovering a known spatial pattern. We used the subtype 1 age effect for smoothed SUVR, one of the primary empirical findings from the surface-based analysis, as the known spatial pattern to construct the data-generating process. We defined the ground truth on the principal surface using the primary IPW regression results at the \(\mathrm{SUVR}>2.0\) threshold: in the simulated data surface points with significant age effects retained their empirical age-effect coefficients, whereas all other surface points were assigned an age-effect coefficient of zero. For each simulation replicate, we sampled \(n=152\) participants, where subject-level covariate profiles were sampled with replacement from the observed subtype 1 cohort. This empirical covariate resampling preserved the observed subtype 1 covariate distribution without imposing a parametric model for age, sex, cognitive status, or SuStaIn stage. Simulated smoothed SUVR values were generated using a pointwise linear model with centered age, the known age-effect map, and the empirical non-age covariate effects for sex, cognitive status, and SuStaIn stage. Independent Gaussian noise with mean 0 and standard deviation \(\sigma\) was added to represent replicate-specific measurement variation. In the real-data smoothed SUVR analysis, residual variability was spatially heterogeneous, with pointwise residual standard deviations across hippocampal surface locations having an interquartile range of approximately 0.18--0.26. To simplify the data-generating process while spanning the empirical range of residual variability, we evaluated three common-noise settings, \(\sigma=0.18\), \(0.24\), and \(0.30\), representing low, intermediate, and high measurement-variation regimes. Within each setting, the same value of \(\sigma\) was used for all simulated locations and for both the surface-based and voxel-wise simulated outcomes.

Within each simulation replicate, we analyzed the simulated data using both the proposed principal-surface IPW regression and conventional voxel-wise linear regression. Each method produced a detected age-effect region after Benjamini--Hochberg correction, which was compared with the prespecified ground-truth age-effect region used to generate the simulated data. For the voxel-wise analysis, the surface-defined age-effect truth was transferred to voxel locations by matching rounded spatial coordinates, allowing both methods to be evaluated against the same underlying spatial truth. For each Gaussian noise setting, we repeated the simulation for 100 replicates and summarized performance using sensitivity for true-region recovery, false positive rate among true-null points, false discovery proportion among detected significant points, and Dice overlap between the detected and true age-effect regions.

\subsection{Detection of Choroid Plexus Contamination}

In this subsection, we describe the procedure used to detect off-target signal contamination from the CP within the principal surface (PS) framework, motivated by the known relevance of CP spill-in for hippocampal \textsuperscript{18}F-AV1451/flortaucipir PET.
We formally define and quantify the effect of CP contamination on hippocampal tau PET measurements.
Specifically, we model how registration inaccuracy and spatial signal mixing can introduce systematic SUVR bias and develop regression-based metrics to localize these effects on the hippocampal surface.

\subsubsection{Modeling Registration-Induced CP Contamination}

To formalize this effect, we modeled registration-induced signal leakage using a two-step transformation process typical in PET--MRI alignment. 
Let \( x \) denote a voxel coordinate in the original PET space. The corresponding PET signal intensity is \( I_{\text{PET}}^{\text{personal}}(x) \), and the subject's anatomical MRI intensity is denoted by \( I_{\text{MRI}}^{\text{personal}}(x) \). The rigid transformation \( T_{\text{PET}} \) maps PET space to subject MRI space, such that \( x' = T_{\text{PET}}(x) \). Next, a non-rigid deformation field \( u(x') \) maps the MRI space to the template space as \( x'' = x' + u(x') \), where \( x'' \) is the voxel coordinate in template space and \( I_{\text{MRI}}^{\text{template}}(x'') \) is the reference template MRI. The full transformation from PET to template space is thus \( x'' = T_{\text{MRI}}(T_{\text{PET}}(x)) \), and the corresponding PET signal in template space becomes:
\begin{equation}\nonumber
    I_{\text{PET}}^{\text{template}}(x'') = I_{\text{PET}}^{\text{personal}}(T_{\text{PET}}^{-1}(T_{\text{MRI}}^{-1}(x''))).
\end{equation}

In practice, the estimated deformation \( \hat{u}(x') \) may deviate from the true deformation \( u^*(x') \), leading to a registration error \( \Delta u(x') = \hat{u}(x') - u^*(x') \). As a result, the PET signal is assigned to a misregistered location
 \(x''_{\text{misregistered}} = x' + u^*(x') + \Delta u(x')\), and the retrieved PET signal becomes 
\( I_{\text{PET}}^{\text{misregistered}}(x'') = I_{\text{PET}}^{\text{personal}} \left( T_{\text{PET}}^{-1}(T_{\text{MRI}}^{-1}(x''_{\text{misregistered}})) \right)\). 

We define the contamination error at voxel \( x'' \) as
\begin{equation}\nonumber
    \Delta_{\text{contamination}}(x'') = I_{\text{PET}}^{\text{misregistered}}(x'') - I_{\text{PET}}^{\text{template}}(x''),
\end{equation}
which captures the difference between the ideal PET signal and the one affected by registration inaccuracy. This formulation characterizes how spatial misalignments propagate into the PET signal and introduce systematic contamination, particularly in anatomically constrained and signal-sparse regions such as the hippocampus.

We quantified contamination effects from the CP on hippocampal tau PET signal using a regression-based approach within the PS framework. For each subject \(i\), hippocampal SUVR voxels \(\mathbf{q}_{iv}\) were projected onto the shared PS and assigned to their nearest selected PS point. Let \(\mathcal{V}_{ik}\) denote the set of voxels from subject \(i\) assigned to selected surface point \(k\). For each subject \(i\) and surface point \(k\), we fit the local linear regression model
\begin{equation}\nonumber
y_{ivk}
=
\beta_{0,ik}
+
\beta_{1,ik} d^{\mathrm{CP}}_{ivk}
+
\epsilon_{ivk},
\qquad v\in\mathcal{V}_{ik},
\end{equation}
where \(y_{ivk}\) is the SUVR value of voxel \(v\) from subject \(i\) assigned to surface point \(k\), and \(d^{\mathrm{CP}}_{ivk}\) is the signed projection distance of that voxel relative to the assigned principal-surface point. The sign of \(d^{\mathrm{CP}}_{ivk}\) encodes the local direction of the voxel relative to the principal surface, with directions standardized across hemispheres for CP contamination analysis.

The slope \(\beta_{1,ik}\) represents the subject-specific local association between SUVR and signed projection distance at surface point \(k\). Positive values \((\beta_{1,ik}>0)\) indicate that SUVR increases along the positive signed-distance direction, whereas negative values \((\beta_{1,ik}<0)\) indicate the opposite directional gradient.
These directional SUVR gradients were used as local measures of potential CP-related spill-in or spill-out. For each subject and surface point, one-sided hypothesis tests were performed to evaluate positive \((\beta_{1,ik}>0)\) or negative \((\beta_{1,ik}<0)\) directional effects, with false discovery rate (FDR) controlled using the Benjamini--Hochberg procedure~\citep{RN332}. To assess reproducibility across individuals, the subject-level slope estimates \(\beta_{1,jk}\) were further summarized using group-level one-sample \(t\)-tests within each cognitive-status group (CN, MCI, and AD) at each surface point \(k\). FDR correction was again applied across surface points separately for each direction. This procedure yielded spatially localized significance maps of CP-driven contamination effects, which were used in downstream visualization and interpretation of artifact-prone regions.

\section{Results}

In this section, we present the empirical findings from the surface-based analysis of hippocampal tau PET data. We first describe the SuStaIn-derived disease subtypes and stage distributions, followed by surface-based analyses of tau coverage, SUVR intensity, and projection-based thickness.
We then evaluate the robustness of the results to tau PET threshold selection, use the fitted models to generate stage-wise predictions of hippocampal tau shape, compare the proposed framework with conventional voxel-wise regression, and examine spatial patterns of choroid plexus contamination.

\subsection{SuStaIn-Derived Subtypes and Disease Staging}

Implementation of SuStaIn yielded four non-zero disease subtypes, with event-ordering patterns shown in the SuStaIn positional variance diagram (Supplementary Figure~\ref{fig: sustain_pvd}). These subtypes corresponded to previously described anatomical phenotypes~\citep{RN277}:

\begin{itemize}
  \item \textbf{Subtype 1 (Limbic-dominant)}: Tau accumulation begins in the medial temporal lobe (MTL) and spreads to the parietal and temporal cortices by Stage 11.
  \item \textbf{Subtype 2 (Posterior occipitotemporal phenotype)}: Starts in the occipital cortex and proceeds anteriorly through the temporal lobe and MTL before reaching parietal and frontal areas.
  \item \textbf{Subtype 3 (Parietal-dominant, MTL-sparing phenotype)}: Initiates in the parietal and temporal cortices, progressing to the frontal and occipital regions.
  \item \textbf{Subtype 4 (Left-temporal phenotype)}: Characterized by early tau involvement in the left temporal cortex, followed by bilateral temporal and parietal spread.
\end{itemize}

\begin{figure}[htbp]
    \centering
    \begin{tabular}{cc}
        \includegraphics[width=0.45\textwidth]{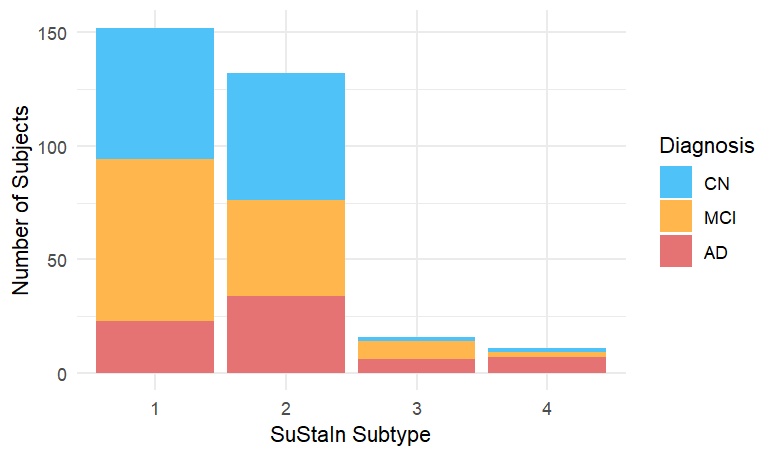}&
        \includegraphics[width=0.45\textwidth]{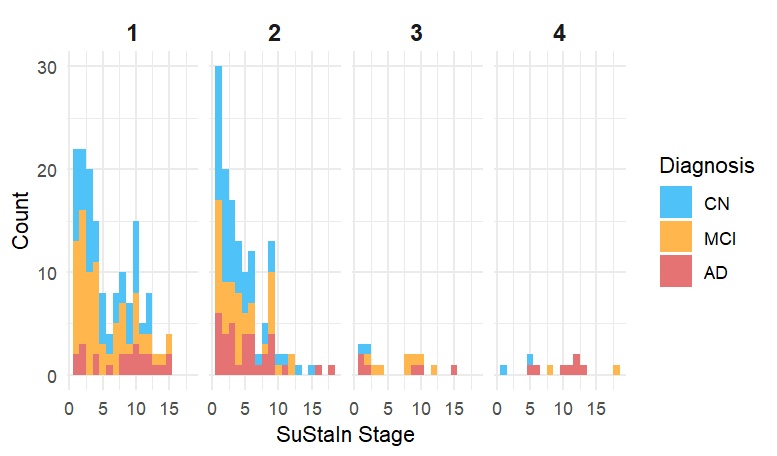} \\
        (a) Cognitive status within SuStaIn subtypes  & (b) Stage across SuStaIn subtypes  \\
        
    \end{tabular}
   \caption{Overview of SuStaIn-derived subtypes
    (a) Cognitive-status composition (CN, MCI, AD) within each subtype, shown as a stacked bar plot. 
    (b) Distribution of SuStaIn stages within each subtype, stratified by cognitive status. 
    Subtype 1 and 2 are predominantly early-stage with a higher proportion of CN and MCI cases, while subtypes 3 and 4 are later-stage with more AD cases. 
    }
    \label{fig:spatial_coverage_num}
\end{figure}

The overall subtype distribution included 317 individuals assigned to subtype 0, and 152, 132, 16, and 11 assigned to subtypes 1 through 4, respectively. Subtype 0 corresponds to subjects with no suprathreshold regional tau deposition across regions, and is interpreted as a baseline group not exhibiting any trajectory of disease progression. This group includes a large proportion of CN participants and is consistent with a SuStaIn stage of 0. The cognitive-status composition and stage distributions of the four non-zero subtypes are shown in Figure~\ref{fig:spatial_coverage_num}a--b. Subtypes 1 and 2 span a wide range of stages and include both CN and symptomatic individuals, while subtypes 3 and 4 are smaller and more skewed toward later stages and participants with AD. Given the limited sample size and reduced clinical variability of subtypes 3 and 4, all subsequent analyses focus on subtypes 1 and 2, which represent the dominant patterns of hippocampal tau accumulation in this cohort.

\subsection{Pointwise Coverage of Tau Deposition on the Hippocampal Surface} \label{sec:coverage_reg}

\begin{figure}[htbp]
    \centering
    \begin{tabular}{cc}
        \includegraphics[width=0.4\textwidth]{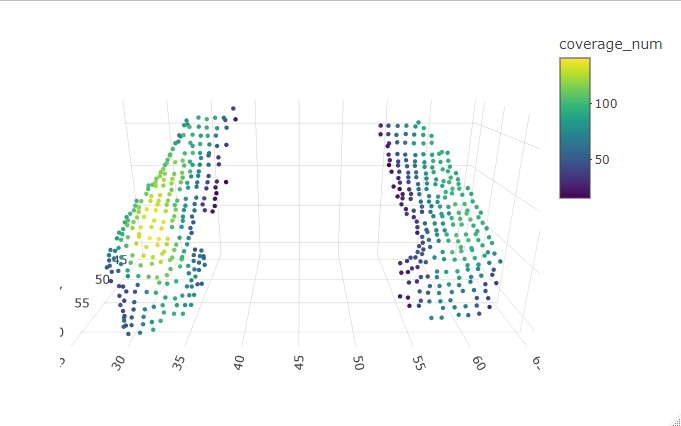}&
        \includegraphics[width=0.4\textwidth]{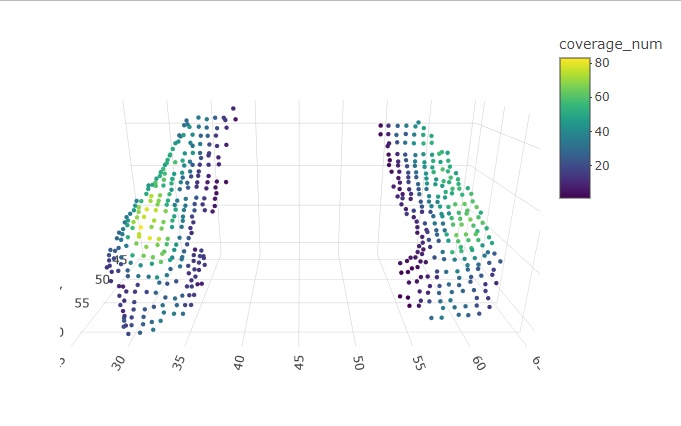} \\
        (a) Coverage map (subtype 1)  & (b) Coverage map (subtype 2) \\
        \includegraphics[width=0.4\textwidth]{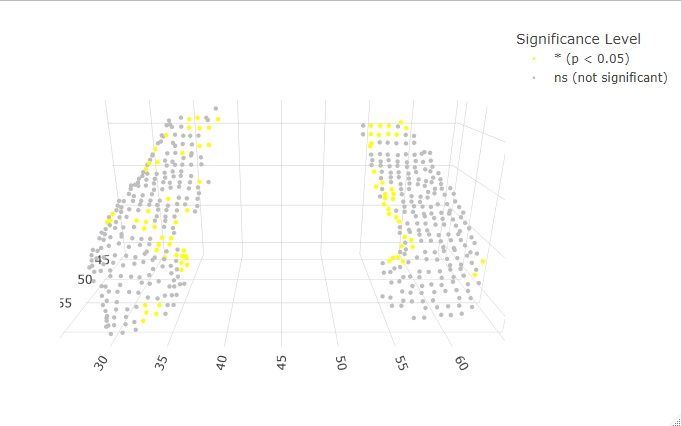}&
        \includegraphics[width=0.4\textwidth]{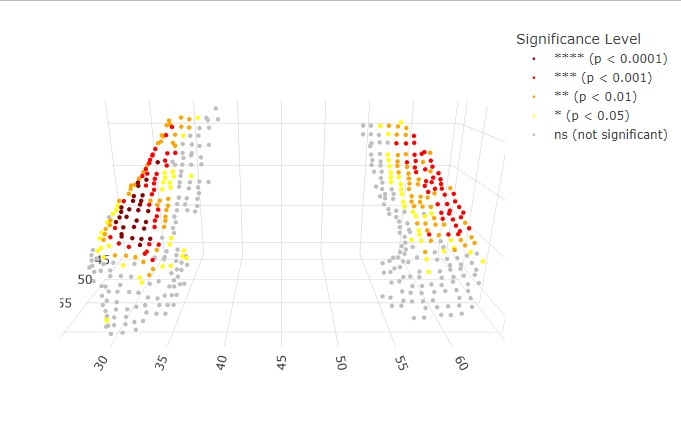} \\ 
        (c) Cognitive-status effect (subtype 1)  & (d) Cognitive-status effect (subtype 2) \\
        \includegraphics[width=0.4\textwidth]{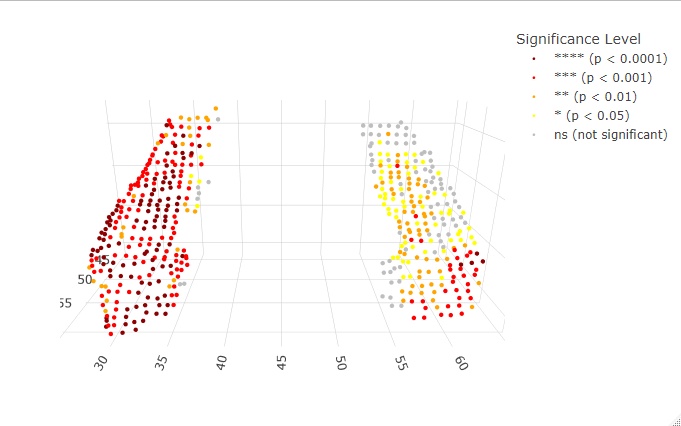}&
        \includegraphics[width=0.4\textwidth]{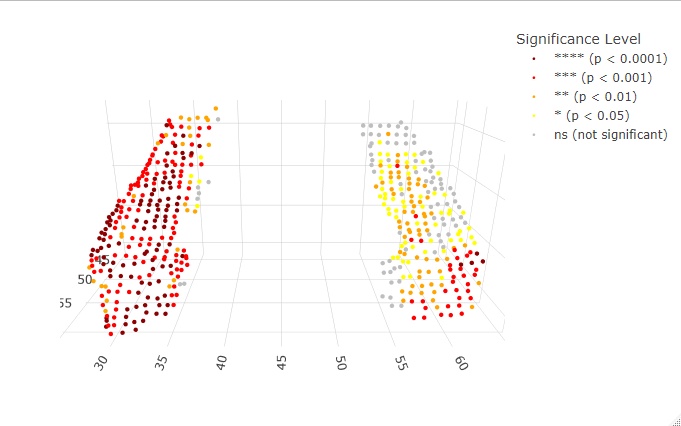} \\
        (e) Linear stage effect (subtype 1)  & (f) Linear stage effect (subtype 2) \\
        \includegraphics[width=0.4\textwidth]{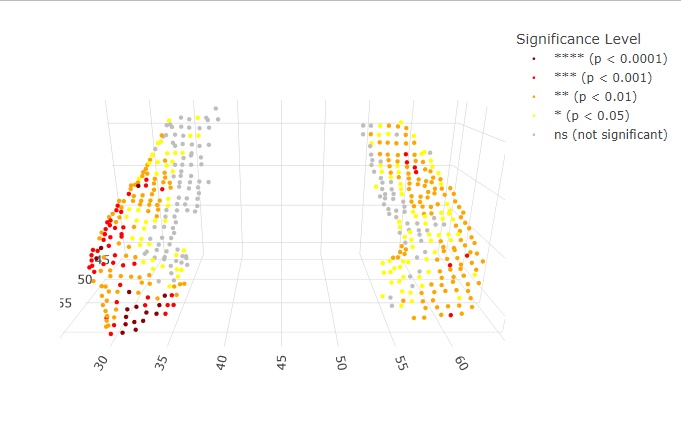} &
        \includegraphics[width=0.4\textwidth]{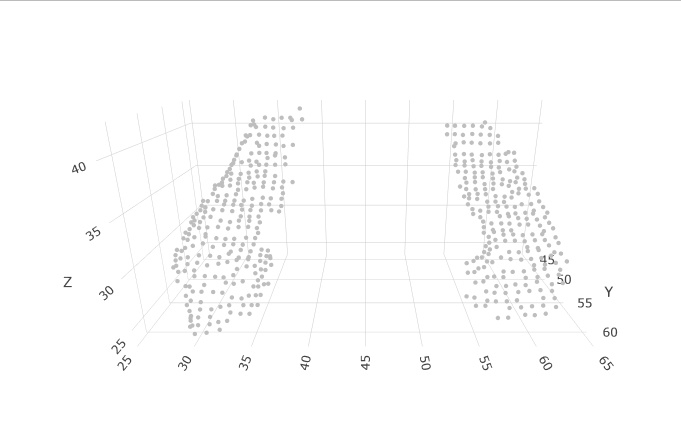} 
        \\
        (g) Quadratic stage effect (subtype 1) & (h) Quadratic stage effect (subtype 2)\
    \end{tabular}
\caption{Coverage and significance maps for covariate effects on suprathreshold tau signal across the hippocampal surface.
Panels (a–b) display the number of scans showing suprathreshold signal ($\mathrm{SUVR} > 2$) at each surface location for subtype 1 and subtype 2, respectively.
Panels (c–d) show the significance of cognitive status effects on tau coverage, estimated using IPW-adjusted logistic regression. 
Panels (e–h) depict the significance of linear and quadratic SuStaIn stage effects, where the “linear” and “quadratic” components correspond to the first two orthogonal polynomial terms.
}

    \label{fig:spatial_coverage}
\end{figure}

To understand where suprathreshold tau deposition occurs across the hippocampus, we examined spatial coverage patterns using the logistic regression model described in Section~\ref{sec: coverage regression method}. The number of scans in which each surface point was covered is visualized in Figure~\ref{fig:spatial_coverage}(a-b) for subtype 1 and subtype 2, respectively. In all plots, the surface is displayed with anterior at the bottom and posterior at the top, with the right hippocampus on the left and the left hippocampus on the right, matching a brain orientation facing the viewer.
A consistent spatial trend is observed across both subtypes: the central region of the right hippocampus exhibits substantially higher coverage than any other region, including the corresponding location in the left hippocampus. This asymmetry suggests greater tau deposition in the lateral body of the right hippocampus across subjects and supports the robustness of the medial surface alignment.

We next evaluated how coverage patterns were modulated by covariates. 
Figures~\ref{fig:spatial_coverage}(c-f) display significance maps for the logistic regression coefficients associated with cognitive status and SuStaIn stage in both subtypes. Cognitive status shows a significant and spatially distinct effect in subtype 2 (Figure~\ref{fig:spatial_coverage}d), with widespread significant regions, including many points at $p<0.001$. These regions are concentrated along the lateral surface of the mid and posterior hippocampus. In these areas, the estimated coefficients are consistently positive, indicating that more severe cognitive status is associated with an increased likelihood of suprathreshold tau signal. This pattern suggests that subtype 2 captures a progression-related accumulation process, particularly in the lateral posterior hippocampus. In subtype 1, cognitive status exhibits a weaker and more diffuse effect, with only scattered significance observed in the medial hippocampus, though the estimated coefficients remain positive.

SuStaIn stage was also included in the model to account for variation along the disease progression axis. To reflect its ordinal nature (20 levels), stage was encoded as a second-order polynomial. Both the linear and quadratic components are shown in Figure~\ref{fig:spatial_coverage} (e-h). Linear estimates were positive and broadly distributed in both subtypes, indicating that advancing stage increases the likelihood of suprathreshold tau signal.
The two subtypes showed highly similar spatial patterns: in the right hippocampus, stage effects extended across the full surface, with slightly weaker signals in the central region of subtype 1. In the left hippocampus, significant effects were concentrated medially and anteriorly, with minimal lateral involvement.
Subtype 1 showed slightly stronger effects in the mid-to-posterior medial region compared to subtype 2.
Beyond these linear effects, second-order (quadratic) components revealed marked subtype differences.
Subtype 1 exhibited widespread negative quadratic coefficients across most of the surface, except for a narrow band in the mid-posterior medial region, whereas subtype 2 showed no significant quadratic terms.
These patterns suggest a plateauing of coverage expansion in later stages for subtype 1, consistent with a rapid early accumulation phase followed by deceleration—a biologically plausible trajectory for the limbic-dominant subtype, where tau deposition typically initiates early in the medial temporal lobe.

Age and sex were also included in the model. Age showed a small but significant negative effect in the anterior medial portion of the left hippocampus in subtype 1, suggesting higher tau signal among younger individuals, whereas subtype 2 showed no significant age or sex effects. Complete results are shown in Supplementary Figure~\ref{fig:spatial_coverage_age_sex}.

\subsection{Covariate Effects on Tau Deposition Intensity (SUVR)}

We next evaluated the effects of age, cognitive status, and SuStaIn stage on hippocampal tau SUVR, separately within each subtype. The significance maps are shown in Figure~\ref{fig:spatial_SUVR}, and corresponding regression coefficient estimates are provided in Supplementary Figure~\ref{fig:spatial_SUVR_supp}. Age-related effects were observed in subtype 1, with significant regions localized to the mid-to-posterior lateral surface of the right hippocampus and the mid-lateral portion of the left hippocampus (Figure~\ref{fig:spatial_SUVR}a). These areas exhibited positive regression coefficients (Supplementary Figure~\ref{fig:spatial_SUVR_supp}), indicating that, after adjustment for sex, cognitive status, and SuStaIn stage, older individuals within this subtype tended to show greater hippocampal tau SUVR. In contrast, subtype 2 showed no significant age-related SUVR effects (Figure~\ref{fig:spatial_SUVR}b), suggesting a subtype-specific difference in the age--tau association within the hippocampus.

To assess whether this subtype-specific age effect could be explained by differences in participant age or symptom-onset timing, we compared age-related characteristics between the two dominant subtypes. Age at tau PET imaging was similar between subtype 1 (75.58 \(\pm\) 6.88 years; \(n=152\)) and subtype 2 (75.81 \(\pm\) 7.81 years; \(n=132\); Welch's \(t(263.3)=0.26\), \(p=0.796\)). 
Cognitive symptom-onset timing was available from the ADNI Participant Demographics, which records the reported calendar year of cognitive symptom onset. We used this onset year, together with the tau PET imaging year and age at imaging, to calculate cognitive symptom duration at tau PET and cognitive symptom-onset age.
Among MCI/AD scans, onset year was available for 90 of 94 scans in subtype 1 (95.7\%) and 65 of 76 scans in subtype 2 (85.5\%). The cognitive symptom-onset age was on average earlier in subtype 2 than in subtype 1, but this difference was not statistically significant (66.67 \(\pm\) 8.02 years, \(n=73\), versus 68.27 \(\pm\) 6.32 years, \(n=93\); Welch's \(t(134.4)=-1.39\), \(p=0.166\)).
Cognitive symptom duration at tau PET was also on average longer in subtype 2, but not significantly different between subtypes (7.91 \(\pm\) 4.46 versus 7.03 \(\pm\) 4.10 years; Welch's \(t(148.2)=1.32\), \(p=0.190\)).
Age at scan was also only weakly related to SuStaIn stage within each subtype, with age-stage correlations of \(r=0.030\) for subtype 1 and \(r=0.001\) for subtype 2. These results suggest that the absence of significant age-related hippocampal SUVR effects in subtype 2 is unlikely to be explained by a gross difference in age distribution, available clinical symptom-onset timing, symptom duration, or by age acting as a proxy for inferred SuStaIn stage.

Cognitive status effects were also predominantly observed in subtype 1, with spatial distributions closely resembling the age-related patterns (Figure~\ref{fig:spatial_SUVR}c). Significant regions were again localized to the right mid-to-posterior lateral surface and the left mid-lateral hippocampus. In subtype 2, the right hippocampus exhibited a similar pattern to subtype 1, while the left hippocampus showed a more spatially restricted cluster centered in the mid-region.
Interestingly, the regression coefficients for cognitive status were negative in both subtypes (Supplementary Figure~\ref{fig:spatial_SUVR_supp}), a finding that may reflect limitations of categorical clinical cognitive-status labels for capturing continuous tau progression. This apparent inversion underscores the value of SuStaIn stage as a more continuous and data-driven marker of disease progression, potentially more robust than categorical cognitive status for capturing spatial patterns of tau deposition.

Linear SuStaIn stage effects showed strong and widespread significance in both subtypes (Figure~\ref{fig:spatial_SUVR}e--f), with consistently positive regression coefficients across the hippocampal surface. The spatial patterns were highly similar between subtypes, with subtype 1 showing weaker effects in the left posterior hippocampus. Notably, anterior regions exhibited minimal significance in both subtypes, supporting a spatially constrained progression pattern that may reflect early-stage sparing of the anterior hippocampus.
Beyond these broadly similar linear effects, the quadratic stage component revealed an opposite trend to that observed in coverage (Figure~\ref{fig:spatial_SUVR}g--h). Subtype 2 exhibited significant positive effects for the quadratic stage component in the right middle and left anterior hippocampus, whereas subtype 1 showed no such effects. Because SuStaIn stage was modeled using the first two orthogonal polynomial terms, these coefficients should be interpreted as linear and nonlinear stage-related components rather than raw stage and stage-squared effects. This pattern suggests that subtype 2 shows nonlinear SUVR increases at later inferred stages, possibly reflecting a slower but more persistent accumulation trajectory, while subtype 1 shows more stable, approximately linear intensity changes across progression.

Interaction analyses further confirmed these subtype-specific divergences. 
Significant age-by-subtype interactions were observed along a continuous band in the right posterior-to-mid hippocampus, with sparser yet spatially corresponding effects in the left hemisphere (Figure~\ref{fig:interaction}a). 
A spatially confined but significant linear stage-by-subtype interaction was also detected in the posterior portion of the left hippocampus (Figure~\ref{fig:interaction}b), indicating meaningful heterogeneity in SUVR progression patterns between subtypes. 

The effect of sex was negligible in both subtypes, with only a few scattered regions showing weak association in subtype 2. Complete p-value and coefficient maps for sex are provided in Supplementary Figure~\ref{fig:sex_SUVR}.

\begin{figure}[htbp]
    \centering
    \begin{tabular}{cc}
        \includegraphics[width=0.4\textwidth]{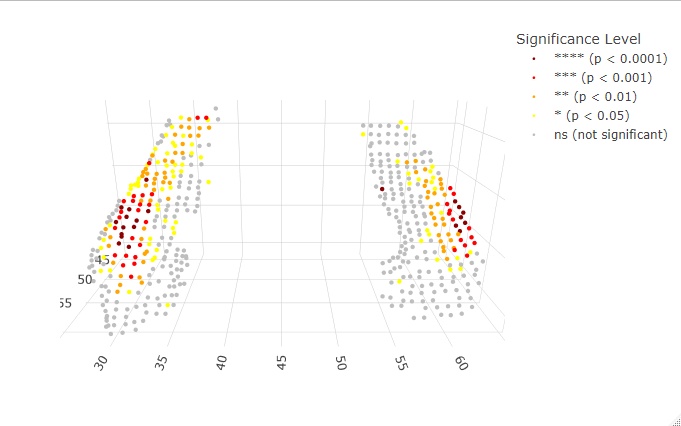}&
        \includegraphics[width=0.4\textwidth]{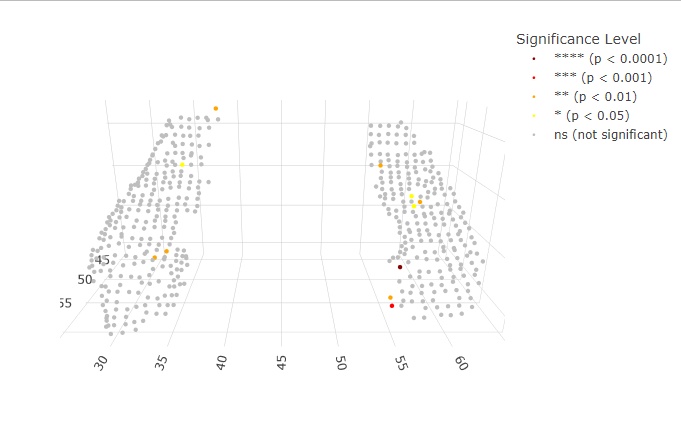} \\ 
        (a) Age effect (subtype 1)  & (b) Age effect (subtype 2)\\
        \includegraphics[width=0.4\textwidth]{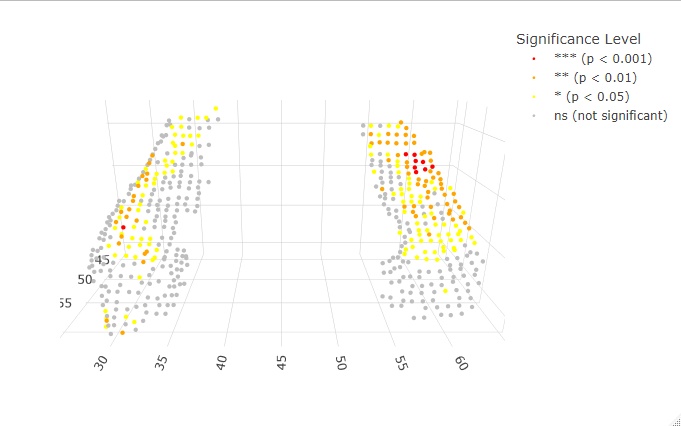}&
        \includegraphics[width=0.4\textwidth]{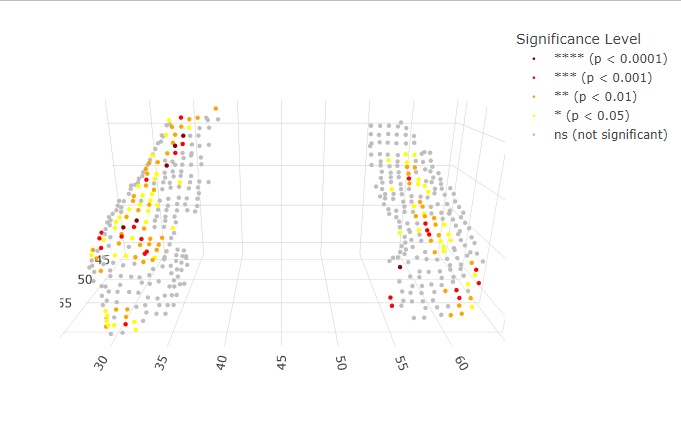} \\
        (c) Cognitive-status effect (subtype 1) & (d)  Cognitive-status effect (subtype 2)\\
        \includegraphics[width=0.4\textwidth]{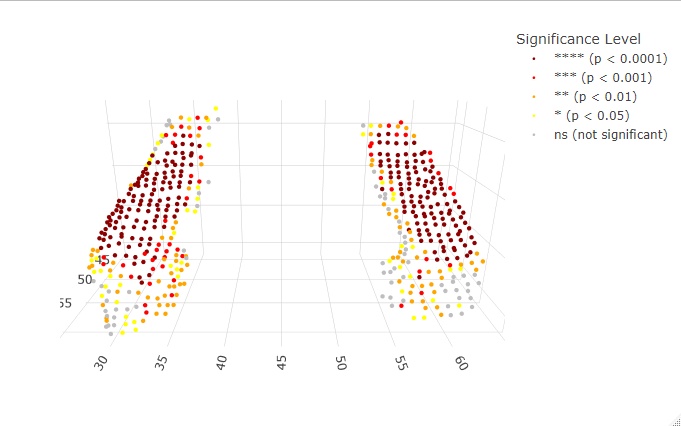}&
        \includegraphics[width=0.4\textwidth]{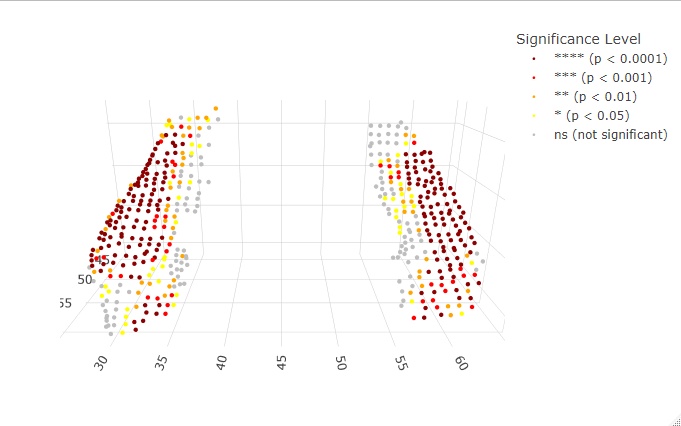} \\
        (e) Linear stage effect (subtype 1) & (f) Linear stage effect (subtype 2)\\
        \includegraphics[width=0.4\textwidth]{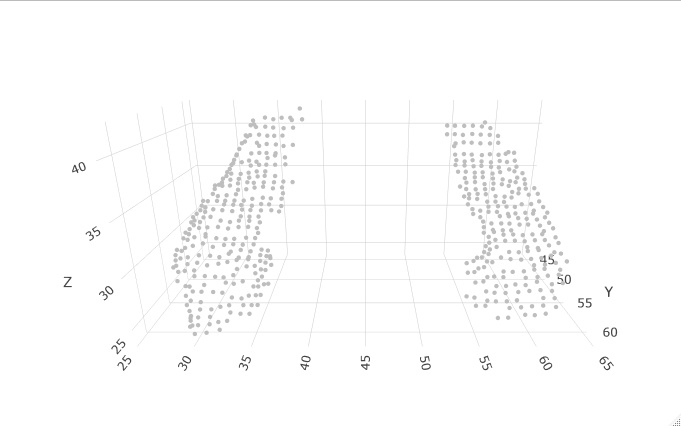}&
        \includegraphics[width=0.4\textwidth]{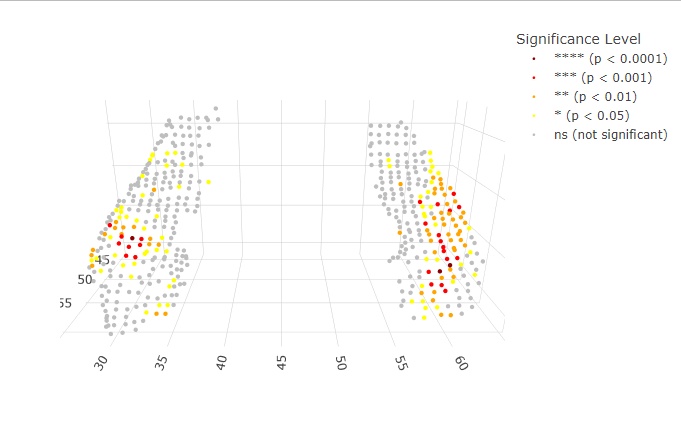} \\
        (g) Quadratic stage effect (subtype 1) & (h) Quadratic stage effect (subtype 2)\\
    \end{tabular}
    \caption{Significance maps of covariate effects on hippocampal tau SUVR. 
    (a--b) Age-related effects show a small significant cluster in the middle lateral hippocampus in subtype 1, with no significant age-related SUVR effects in subtype 2. 
    (c--d) Cognitive-status effects are stronger in subtype 1, with multiple significant regions concentrated along the mid-to-posterior lateral surface. 
    (e--h) SuStaIn stage effects are shown as the first two orthogonal polynomial components. The linear stage component shows widespread positive effects in both subtypes, whereas the nonlinear stage component is more prominent in subtype 2.}
    \label{fig:spatial_SUVR}
\end{figure}

\subsection{Stage-Dependent Effects on Tau Deposition Thickness}

To capture shape-level changes in tau pathology beyond intensity or coverage, we analyzed a geometric measure of deposition thickness based on projection distance. Specifically, we considered two complementary directional components: positive and negative projection distance, defined by projecting suprathreshold signal along opposite directions of the local surface normal from the medial principal surface. These features reflect signal expansion relative to hippocampal geometry, and together quantify bidirectional changes in local thickness centered on the surface. Covariate effects on thickness were estimated using IPW-adjusted linear regression. Age, sex, and cognitive status did not show significant effects in either projection direction across subtypes; full results are provided in Supplementary Figure~\ref{fig:thickness_age_sex_diagnosis}. We therefore focus on the effects of SuStaIn stage.

\begin{figure}[htbp]
    \centering
    \begin{tabular}{cc}
        \includegraphics[width=0.4\textwidth]{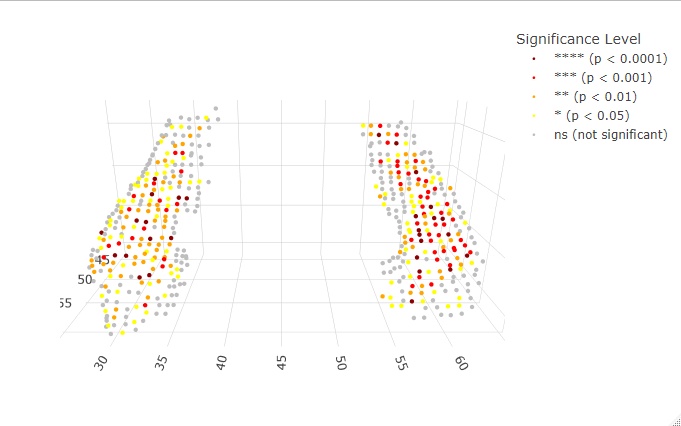}&
        \includegraphics[width=0.4\textwidth]{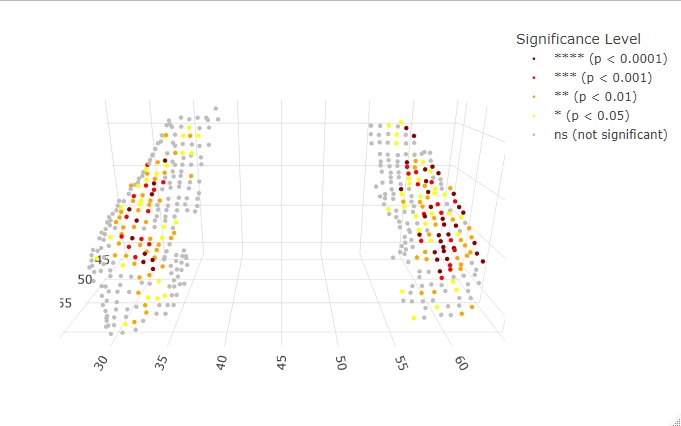} \\ 
        (a) Negative linear effect (subtype 1) & (b) Negative linear effect (subtype 2)\\
        \includegraphics[width=0.4\textwidth]{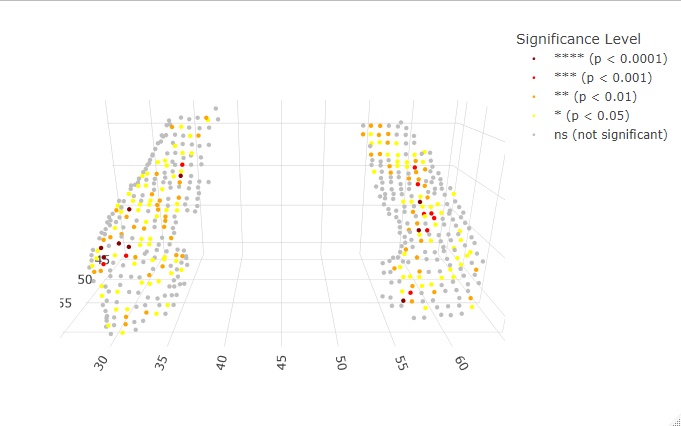}&
        \includegraphics[width=0.4\textwidth]{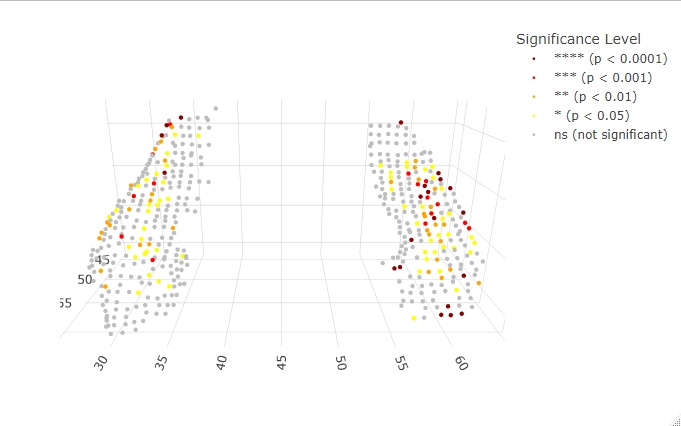} \\ 
        (c) Positive linear effect (subtype 1) & (d) Positive linear effect (subtype 2)\\
        \includegraphics[width=0.4\textwidth]{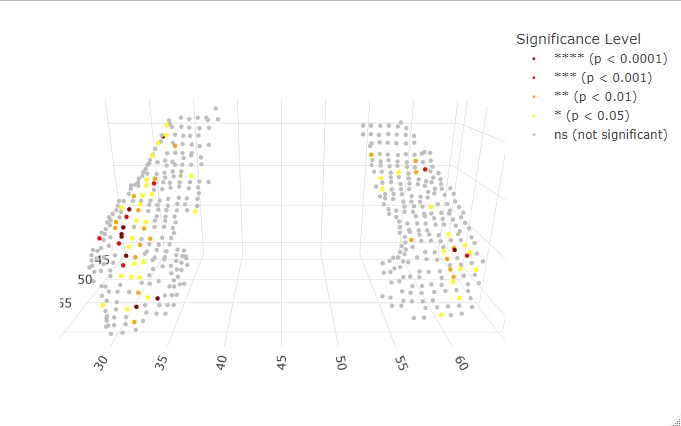}&
        \includegraphics[width=0.4\textwidth]{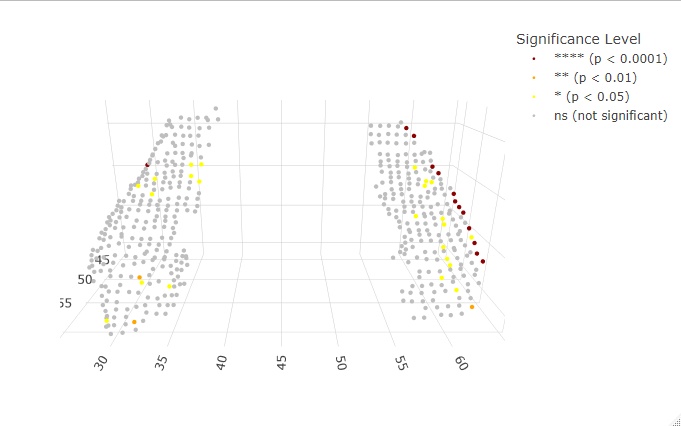} \\
        (e) Negative quadratic effect (subtype 1) & (f) Negative quadratic effect (subtype 2)\\
        
        \includegraphics[width=0.4\textwidth]{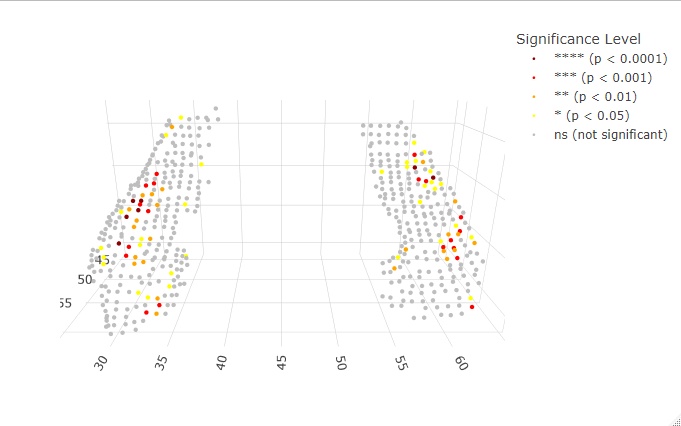}&
        \includegraphics[width=0.4\textwidth]{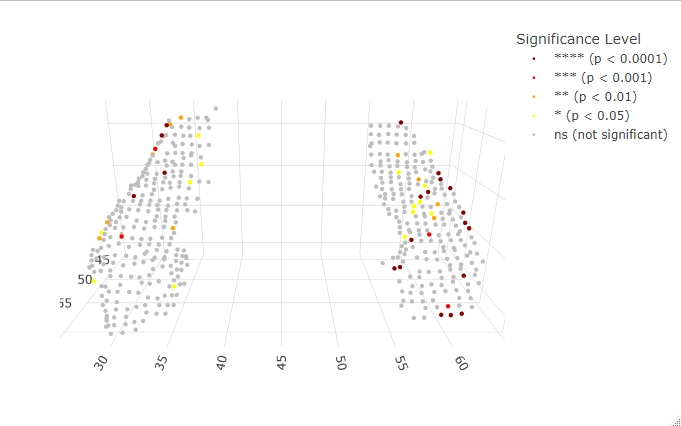} \\
        (g) Positive quadratic effect (subtype 1) & (h) Positive quadratic effect (subtype 2)\\
    \end{tabular}
\caption{Significance maps of SuStaIn stage effects on hippocampal tau deposition thickness, separated by subtype, projection direction, and polynomial term. Each panel shows results from an IPW-adjusted regression model fit within a specific subtype. Panels (a–d) correspond to linear stage effect for projection distances; (e–h) show quadratic stage effect for projection thickness. All maps display $-{\log_{10}}(p)$ values thresholded at multiple significance levels after BH correction for multiple comparisons. Subtype labels are indicated in parentheses for consistency with other figures.}
\label{fig:thickness}
\end{figure}

Linear stage effects were robust in both subtypes and across both projection directions (Figure~\ref{fig:thickness}a-d). In each subtype, the spatial pattern of significance was nearly identical between the positive and negative features, indicating that tau thickness increases occur symmetrically around the principal surface. Subtype 1 showed a broader region with smaller p-values in the negative direction, particularly in the right hippocampus, where the spatial extent of significant vertices exceeded that of the left. Subtype 2, in contrast, showed similar significance levels across directions, but with spatial localization primarily to the lateral surface, while medial regions remained non-significant. These results reinforce the effectiveness of the principal surface representation in separating hippocampal geometry into spatially meaningful upper and lower regions, while also revealing asymmetric expansion rates across directions in subtype 1 and more uniform expansion in subtype 2.

Quadratic stage effects were significant only in subtype 1 and are visualized in Figure~\ref{fig:thickness}e--h. In subtype 1, both negative and positive projection-distance features showed overlapping significant regions (Figure~\ref{fig:thickness}e,g), with consistently negative quadratic-stage coefficient estimates (Supplementary Figure~\ref{fig:thickness_sup}e,f), indicating a deceleration of thickness increase at later stages. Subtype 2 showed no significant quadratic effects in either projection direction (Figure~\ref{fig:thickness}f,h; Supplementary Figure~\ref{fig:thickness_sup}g,h).

\begin{figure}[htbp]
    \centering
    \begin{tabular}{cc}
        \includegraphics[width=0.45\textwidth]{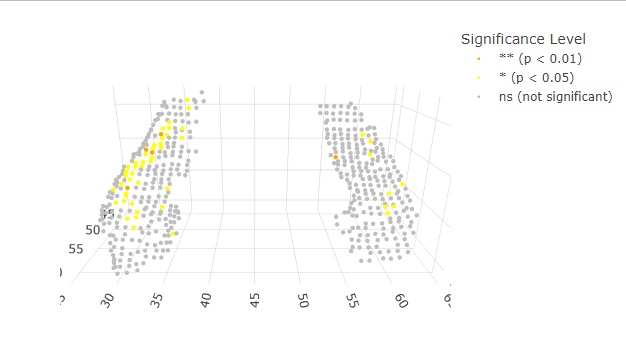}&
        \includegraphics[width=0.45\textwidth]{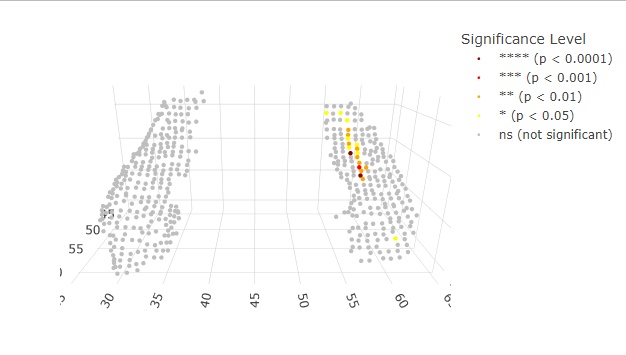} \\ 
        (a) Age × Subtype (SUVR)   & (b) Linear stage × Subtype (SUVR)\\
        \includegraphics[width=0.45\textwidth]{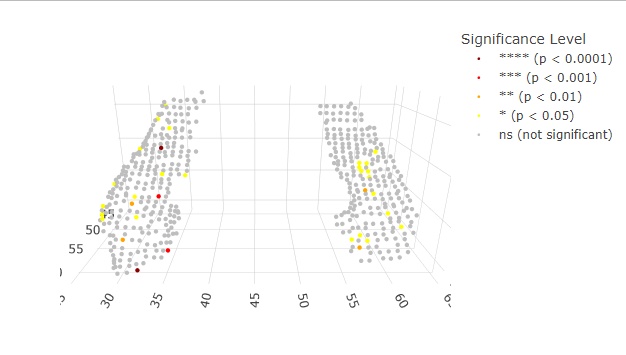}&
        \includegraphics[width=0.45\textwidth]{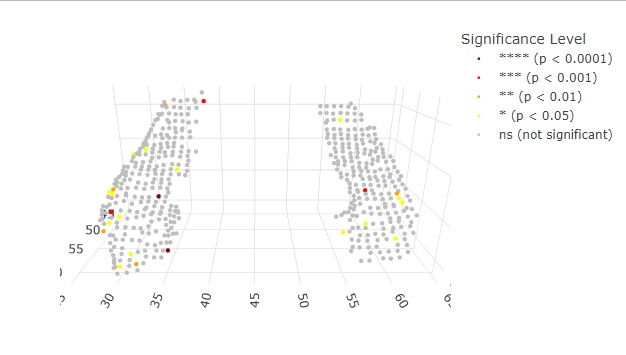} \\
        (c) Age × Subtype (Thickness)  & (d) Sex × Subtype (Thickness)\\
        \includegraphics[width=0.45\textwidth]{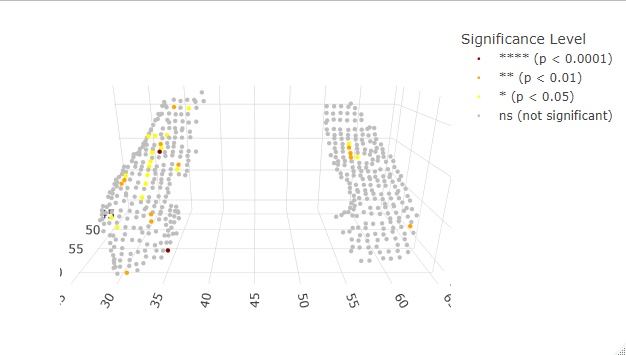}&
        \includegraphics[width=0.45\textwidth]{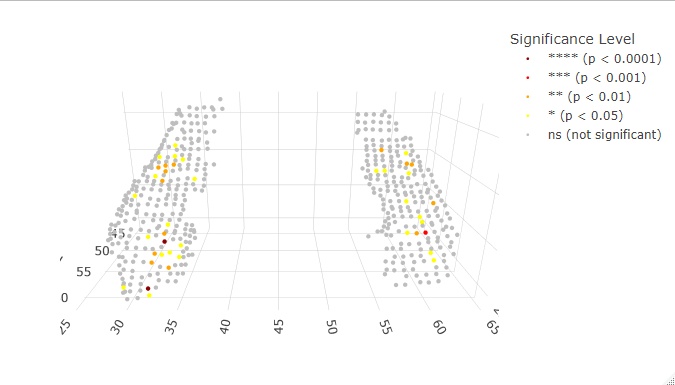} \\
        (e)  Cognitive status \(\times\) Subtype (Thickness)  & (f) Linear stage \(\times\) Subtype (Thickness)\\
    \end{tabular}
\caption{
Significance maps for covariate-by-subtype interaction terms in the unified regression model. 
Each map displays surface regions where the effect of a covariate on tau deposition differs significantly between subtypes. 
(a)~Significance of the age effect interaction on SUVR intensity; 
(b)~Interaction significance for SuStaIn stage (linear term) on SUVR intensity; 
(c)~Interaction significance of age on negative projection thickness; 
(d)~Interaction significance of sex on negative projection thickness; 
(e)~Interaction significance of cognitive status on negative projection thickness; 
(f)~Interaction significance of SuStaIn stage (linear term) on negative projection thickness. All maps display $-{\log_{10}}(p)$ values from IPW-adjusted regression models, with multiple significance levels indicated by color intensity. 
Benjamini–Hochberg correction was applied across all vertices to control for multiple comparisons.}
\label{fig:interaction}
\end{figure}

Interaction analyses yielded only sparse but spatially consistent covariate-by-subtype differences in thickness (Figure~\ref{fig:interaction}c–f). Most significant effects were localized near the central body of the hippocampus—regions known to exhibit greater morphological thickness and higher susceptibility to tau aggregation.

Taken together, these results highlight a consistent subtype divergence across all feature domains. The limbic subtype (subtype 1) displays early, aggressive expansion of tau signal in both coverage and thickness, followed by a deceleration phase—while the posterior subtype (subtype 2) exhibits slower but more persistent accumulation. Notably, while coverage and thickness capture spatial and shape-related changes, SUVR directly reflects signal intensity, and the reversal in quadratic effects between these modalities underscores the multidimensional nature of subtype-specific progression.

\subsection{Stage-Wise Prediction of Hippocampal Tau Shape via Two-Stage Regression}

Building upon pointwise regression analyses that quantified how age, sex, cognitive status, and disease stage influence tau deposition features across the hippocampal surface, we used the trained two-stage regression models to generate predicted hippocampal tau shapes across SuStaIn stages. This predictive framework combines a logistic regression for tau coverage with stage-weighted linear models for intensity and deposition thickness, and allows for model-based simulation of tau progression patterns. To simulate a representative disease trajectory, we fixed the covariates at cohort-typical values: age was set to 75.5 (the overall sample mean), sex to male, and cognitive status to MCI, which was the most prevalent and evenly distributed cognitive-status category across disease stages. Using these fixed covariates, we varied the SuStaIn stage from 1 to 15 and applied the model to each previously selected principal surface point to predict the probability of tau coverage and the associated deposition geometry.

We performed this simulation separately for subtype 1 and subtype 2, using their respective pointwise models. Since the first-stage logistic regression is central to determining whether a point is covered by tau, we evaluated the accuracy of the logistic models via ROC analysis. The area under the curve (AUC) was 0.834 for subtype 1 and 0.875 for subtype 2 (Supplementary Figure~\ref{fig:prediction_ROC}). These values indicate strong discriminative ability of the logistic models in predicting pointwise tau presence from subject-level covariates. Based on Youden’s index \citep{RN333}, we selected the optimal predicted probability threshold for tau coverage—0.498 for subtype 1 and 0.254 for subtype 2—and applied these thresholds to filter predicted coverage points when reconstructing tau deposition shapes. 

\begin{figure}[htbp]
    \centering
    \includegraphics[width=0.95\textwidth]{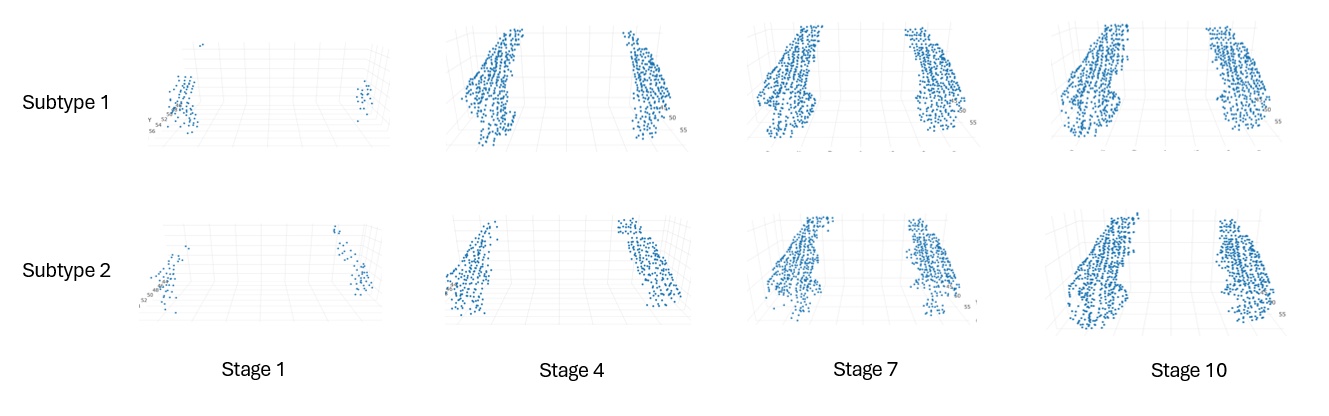} 
    \caption{
        Model-based prediction of tau deposition shapes across SuStaIn stages using two-stage regression. 
The first row shows predicted hippocampal tau deposition shapes for subtype 1 at SuStaIn stages 1, 4, 7, and 10 (left to right). 
The second row shows corresponding predictions for subtype 2. 
Predictions were generated by fixing covariates (Age = 75.5, Sex = Male, Cognitive status = MCI) and varying stage input from 1 to 10. 
Each shape represents the reconstructed hippocampal surface based on predicted pointwise tau coverage and deposition thickness.}
    \label{fig:prediction_shapes}
\end{figure}

Figure~\ref{fig:prediction_shapes} illustrates the predicted tau deposition shapes for subtype 1 and subtype 2 at SuStaIn stages 1, 4, 7, and 10. 
For subtype 1, the results demonstrate a smooth and biologically plausible progression of tau pathology across the hippocampal surface. 
Over-threshold tau accumulation (SUVR $>$ 2)  emerges initially in the middle portion of the hippocampus at stage 1, and progressively extends both anteriorly and posteriorly along the longitudinal axis, as well as medially and laterally across the medial surface. This spatial expansion is accompanied by a gradual increase in predicted deposition thickness. These patterns indicate that high-intensity tau signal does not appear diffusely from the outset, but rather follows a structured, stage-wise trajectory of propagation within the hippocampus.

For subtype 2, tau deposition exhibits an earlier onset in the posterior hippocampus, consistent with the posterior progression pattern identified by whole-brain SuStaIn modeling. 
Although the SuStaIn stages were derived from global tau patterns, the predicted spatial expansion within the hippocampus aligns with the subtype-specific whole-brain trajectory. 
Similar to subtype 1, tau deposition in subtype 2 extends from initial focal regions toward anterior, posterior, and lateral areas, with increasing thickness and eventual involvement of the entire hippocampus. 
Compared to empirical averages—which are particularly unstable in late stages due to limited sample size—these model-based predictions provide a biologically plausible visualization of disease progression.

\subsection{Robustness to Tau PET Threshold Selection}
\label{sec:threshold_sensitivity_results}

Using the threshold-sensitivity analysis described in Section~\ref{sec:threshold_sensitivity}, we evaluated whether the main surface-based findings were preserved across nearby SUVR cutoffs. Increasing the threshold from 1.8 to 2.2 reduced the overall amount of projected tau coverage, as expected, but the relative subject-level coverage patterns remained highly consistent across thresholds (Supplementary Figure~\ref{fig:threshold_sensitivity_coverage}). Compared with the primary \(\mathrm{SUVR}>2.0\) threshold, subject-level coverage proportions were strongly correlated at alternative thresholds, with correlations of 0.950, 0.978, 0.978, and 0.941 for thresholds 1.8, 1.9, 2.1, and 2.2, respectively.

The spatial organization of tau coverage on the principal surface was even more stable. Surface-point coverage-frequency maps showed highly similar anatomical patterns across thresholds in both hemispheres (Supplementary Figure~\ref{fig:threshold_sensitivity_coverage}b). Correlations of point-wise coverage frequency relative to the primary \(\mathrm{SUVR}>2.0\) threshold ranged from 0.993 to 0.996 in the left hippocampus and from 0.992 to 0.997 in the right hippocampus. Thus, changing the threshold altered the amount of suprathreshold tau signal but did not substantially change where tau signal was projected on the hippocampal principal surface.

For downstream IPW-adjusted smoothed SUVR regression, we evaluated threshold robustness primarily using Dice overlap of BH-corrected significant surface-point masks relative to the primary \(\mathrm{SUVR}>2.0\) analysis (Table~\ref{tab:threshold_ipw_dice}). The main spatial findings showed strong overlap across nearby thresholds. In subtype 1, Dice overlap remained high for age effects (0.769--0.853) and linear SuStaIn stage effects (0.793--0.958). Cognitive-status effects in subtype 1 also showed moderate-to-high overlap for thresholds 1.8, 1.9, and 2.1 (0.696--0.758), with lower overlap at the most stringent threshold of 2.2 (0.470). In subtype 2, the linear SuStaIn stage effect was similarly stable across thresholds (0.755--0.893).

\begin{table}[htbp]
\centering
\caption{Dice overlap of IPW-adjusted smoothed SUVR significant surface-point masks across SUVR thresholds.}
\label{tab:threshold_ipw_dice}
\vspace{0.5em}
\begin{tabular}{llccccc}
\toprule
\textbf{Subtype} & \textbf{Regression term} & \textbf{Primary \(n\)} & \textbf{1.8 vs 2.0} & \textbf{1.9 vs 2.0} & \textbf{2.1 vs 2.0} & \textbf{2.2 vs 2.0} \\
\midrule
1 & Age              & 241 & 0.769 & 0.853 & 0.829 & 0.803 \\
1 & Cognitive status & 172 & 0.705 & 0.758 & 0.696 & 0.470 \\
1 & Linear stage     & 451 & 0.948 & 0.958 & 0.911 & 0.793 \\
1 & Sex              & 0   & N.R.  & N.R.  & N.R.  & N.R.  \\
1 & Quadratic stage  & 0   & N.R.  & N.R.  & N.R.  & N.R.  \\
\midrule
2 & Age              & 9   & 0.182 & 0.000 & 0.138 & 0.000 \\
2 & Cognitive status & 146 & 0.553 & 0.624 & 0.517 & 0.026 \\
2 & Linear stage     & 370 & 0.868 & 0.893 & 0.890 & 0.755 \\
2 & Sex              & 37  & 0.179 & 0.066 & 0.286 & 0.044 \\
2 & Quadratic stage  & 90  & 0.021 & 0.428 & 0.469 & 0.383 \\
\bottomrule
\end{tabular}

\vspace{0.5em}
\begin{minipage}{0.95\textwidth}
\footnotesize
\textit{Note.} Dice coefficients compare BH-corrected significant surface-point masks from each alternative threshold with the primary \(\mathrm{SUVR}>2.0\) analysis. Primary \(n\) denotes the number of significant surface points under the primary threshold. N.R. indicates that no reference significant region was detected under the primary \(\mathrm{SUVR}>2.0\) threshold, so Dice overlap with the primary map was not evaluated.
\end{minipage}
\end{table}

Lower Dice overlap was concentrated in effects with small or weak primary significant regions. For example, subtype 2 age had only nine significant surface points under the primary threshold and showed low Dice overlap across alternative thresholds. Subtype 2 sex and quadratic-stage effects also showed lower spatial overlap, consistent with weaker or less spatially coherent effects in the primary analysis. For subtype 1 sex and quadratic-stage effects, no reference significant region was detected under the primary \(\mathrm{SUVR}>2.0\) threshold; therefore, Dice overlap with the primary map was not evaluated. As a secondary analysis, Pearson and Spearman correlations of continuous coefficient maps showed broadly consistent patterns (Supplementary Table~\ref{tab:threshold_ipw_correlation}), supporting the same conclusion that the principal spatial findings were robust across reasonable SUVR thresholds.

\subsection{Comparison with Voxel-wise Regression and Monte Carlo Simulation}
\label{sec: compare and validation}

We compared the surface-based results with conventional voxel-wise regression to evaluate whether the principal-surface representation improved the detection of spatially organized hippocampal tau effects. In the empirical analysis, voxel-wise regression identified fewer and less spatially continuous significant regions than the surface-based SUVR analysis. This difference was most apparent for the subtype 1 age and cognitive-status effects, where the surface-based model showed broader bilateral and medial hippocampal involvement, whereas the voxel-wise maps showed more fragmented and spatially restricted findings (Figure~\ref{fig:spatial_SUVR}; Supplementary Figure~\ref{fig:sup_voxel}a--b). A similar pattern was observed for the linear SuStaIn stage effect, for which the surface-based model detected more extensive hippocampal regions than voxel-wise regression (Supplementary Figure~\ref{fig:sup_voxel}c). For the quadratic stage effect, voxel-wise regression produced isolated significant clusters near anterior boundary regions (Supplementary Figure~\ref{fig:sup_voxel}d), whereas the surface-based results were more spatially coherent with the overall hippocampal progression pattern.

The simulation provided a quantitative assessment of method performance under the prespecified subtype 1 age-effect region used to generate the simulated datasets. Across the three common-noise settings, the principal-surface IPW model showed higher average true-region recovery than voxel-wise regression, and this advantage increased as the Gaussian noise standard deviation increased (Table~\ref{tab:monte_carlo_age_simulation_metrics}). At \(\sigma=0.18\), mean sensitivity was 0.854 for the surface-based model and 0.803 for voxel-wise regression, with similar Dice overlap (0.890 versus 0.887). At higher noise levels, the difference became more pronounced (statistically significant): at \(\sigma=0.24\), sensitivity was 0.739 versus 0.582 and Dice was 0.821 versus 0.731; at \(\sigma=0.30\), sensitivity was significantly higher at 0.614 versus 0.365 and Dice was 0.735 versus 0.529.

Voxel-wise regression showed significantly lower false positive rate among true-null points and lower false discovery proportion among detected significant points across all noise settings, reflecting a more conservative detection pattern. However, this conservativeness came at the cost of substantially reduced recovery of the true age-effect region, particularly under higher measurement variation. In contrast, the principal-surface IPW model preserved greater sensitivity and spatial overlap as noise increased. Together, the empirical maps and simulation study indicate that the proposed surface-based framework has the potential to provide greater sensitivity to organized hippocampal tau effects and more robust spatial recovery than conventional voxel-wise regression, especially under noisier measurement conditions.

\begin{table}[htbp]
\centering
\caption{Simulation performance under common Gaussian noise settings.}
\label{tab:monte_carlo_age_simulation_metrics}

\vspace{0.5em}

{\small
\begin{tabular}{llcc}
\toprule
\textbf{\(\sigma\)} & \textbf{Metric} & \textbf{Surface IPW} & \textbf{Voxel-wise LM} \\
\midrule
0.18 & Dice
& 0.890 (0.845--0.923)
& 0.887 (0.825--0.934) \\
     & Sensitivity
& 0.854 (0.801--0.909)
& 0.803 (0.707--0.884) \\
     & FPR
& 0.062 (0.027--0.115)
& 0.005 (0.000--0.012) \\
     & FDP
& 0.071 (0.031--0.127)
& 0.007 (0.000--0.016) \\
\midrule
0.24 & Dice
& 0.821 (0.772--0.865)
& 0.731 (0.582--0.831) \\
     & Sensitivity
& 0.739 (0.676--0.803)
& 0.582 (0.412--0.714) \\
     & FPR
& 0.056 (0.019--0.100)
& 0.004 (0.000--0.012) \\
     & FDP
& 0.074 (0.027--0.127)
& 0.007 (0.000--0.021) \\
\midrule
0.30 & Dice
& 0.735 (0.668--0.794)
& 0.529 (0.345--0.682) \\
     & Sensitivity
& 0.614 (0.525--0.687)
& 0.365 (0.210--0.520) \\
     & FPR
& 0.051 (0.017--0.094)
& 0.002 (0.000--0.008) \\
     & FDP
& 0.080 (0.029--0.143)
& 0.007 (0.000--0.023) \\
\bottomrule
\end{tabular}
}

\vspace{0.5em}

\begin{minipage}{0.82\textwidth}
\footnotesize
\textit{Note.} Values are reported as mean (2.5\%--97.5\% simulation interval) across 100 Monte Carlo simulation replicates. Dice measures spatial overlap between the detected and true age-effect regions. Sensitivity measures recovery of the true age-effect region. FPR, false positive rate; FDP, false discovery proportion.
\end{minipage}
\end{table}

\subsection{Spatial Patterns of Choroid Plexus Contamination}

\paragraph{Subject-Level Patterns of Contamination}
To visualize spatial manifestations of CP contamination at the individual level, we present representative participants from each cognitive-status group (CN, MCI, AD), arranged in order of increasing hippocampal tau burden (Supplementary Figure~\ref{fig:supp_contam_examples}). Each row corresponds to one subject, with significance maps for both positive and negative contamination directions. Consistent with our hypothesis, CN and MCI participants — who show minimal hippocampal tau — exhibit strong \textbf{positive contamination}, where SUVR increases near the hippocampal surface adjacent to the CP. In contrast, AD participants show more pronounced \textbf{negative contamination}, suggesting that elevated hippocampal tau leads to signal mixing with adjacent non-tau regions, resulting in SUVR underestimation.

\begin{figure}[htbp]
    \centering
    \begin{tabular}{cc}

        \includegraphics[width=0.45\textwidth]{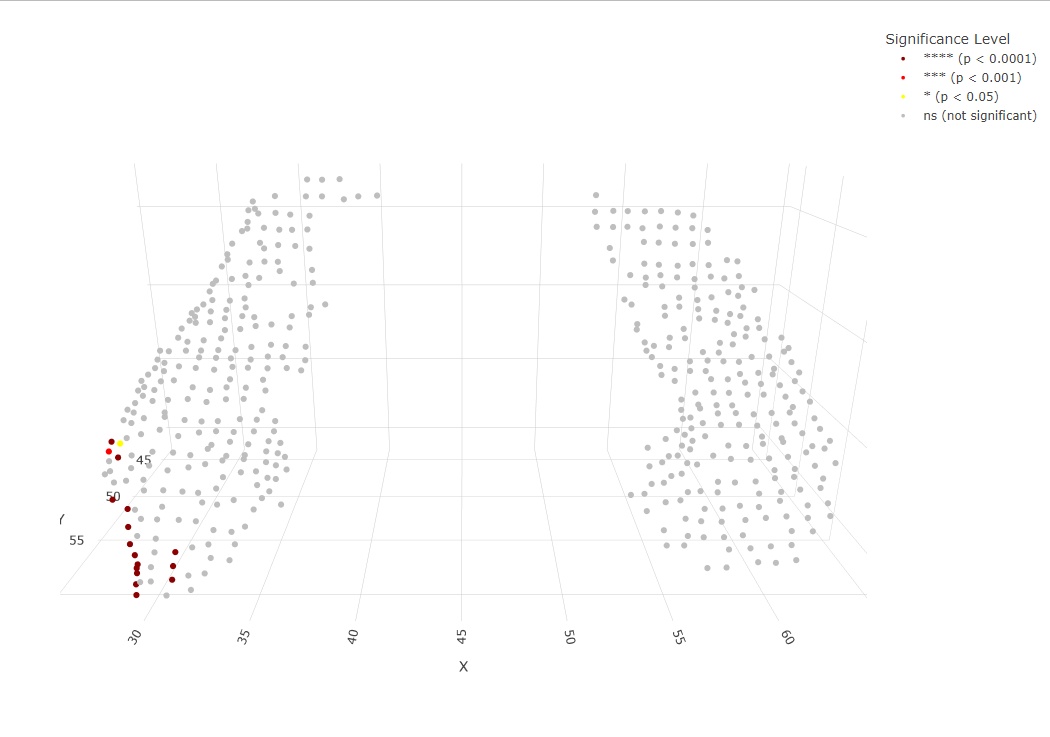} &
        \includegraphics[width=0.45\textwidth]{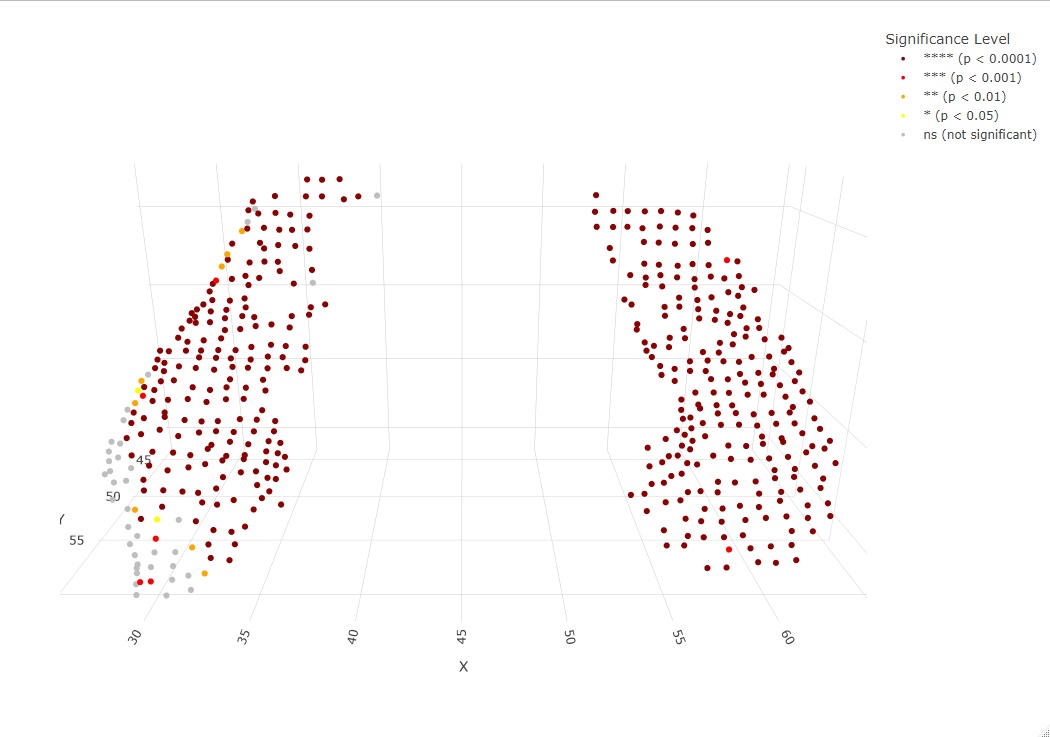} \\
        (a) ADNI CN negative & (b) ADNI CN positive \\
 
        \includegraphics[width=0.45\textwidth]{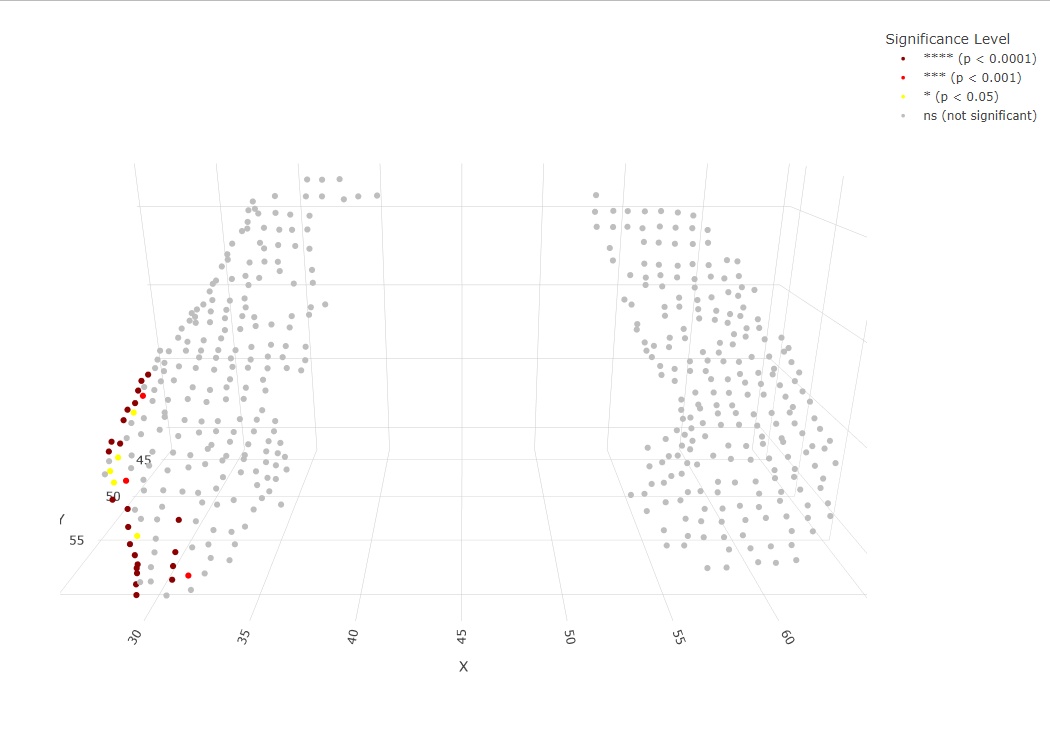} &
        \includegraphics[width=0.45\textwidth]{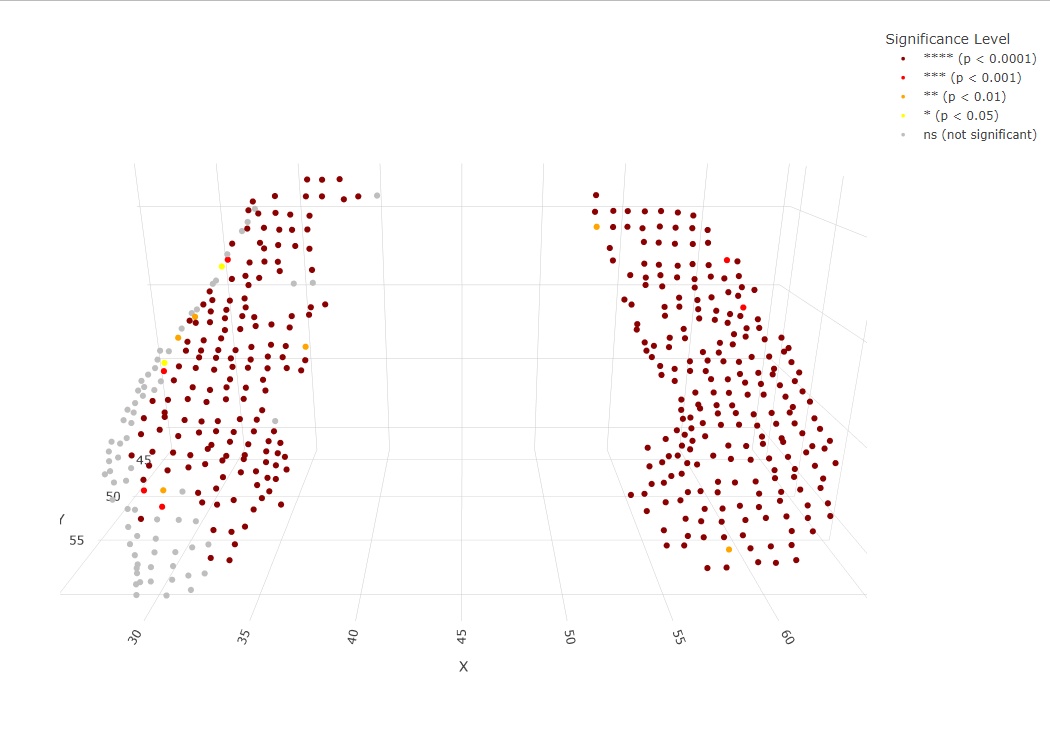} \\
        (c) ADNI MCI negative & (d) ADNI MCI positive \\

        \includegraphics[width=0.45\textwidth]{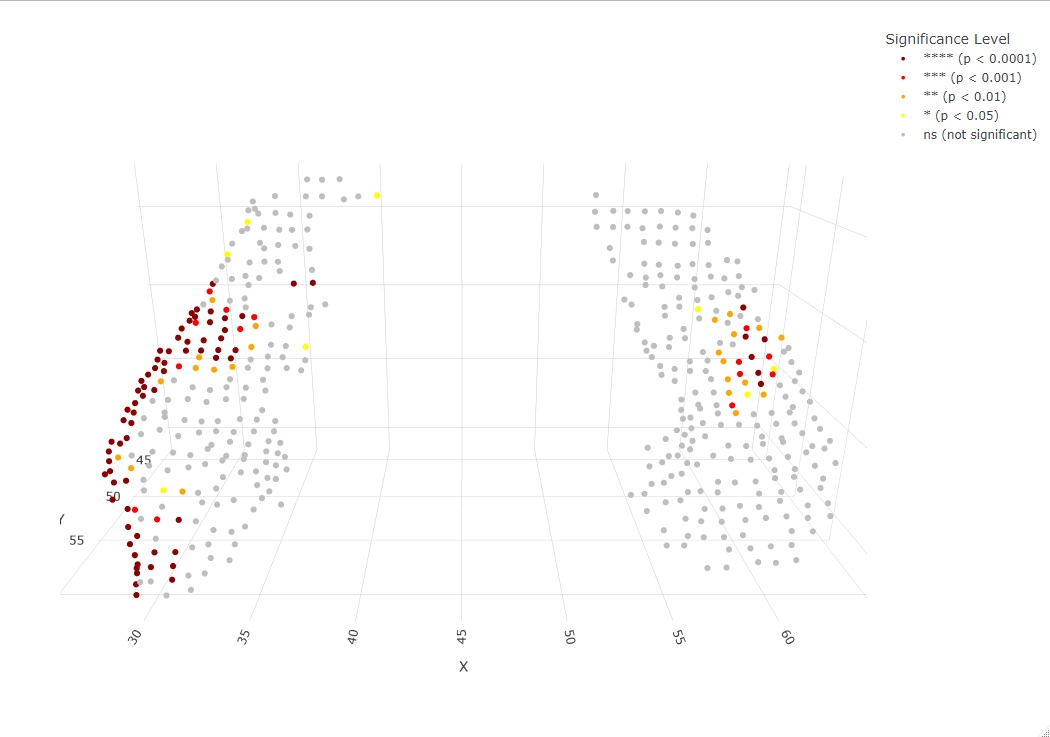} &
        \includegraphics[width=0.45\textwidth]{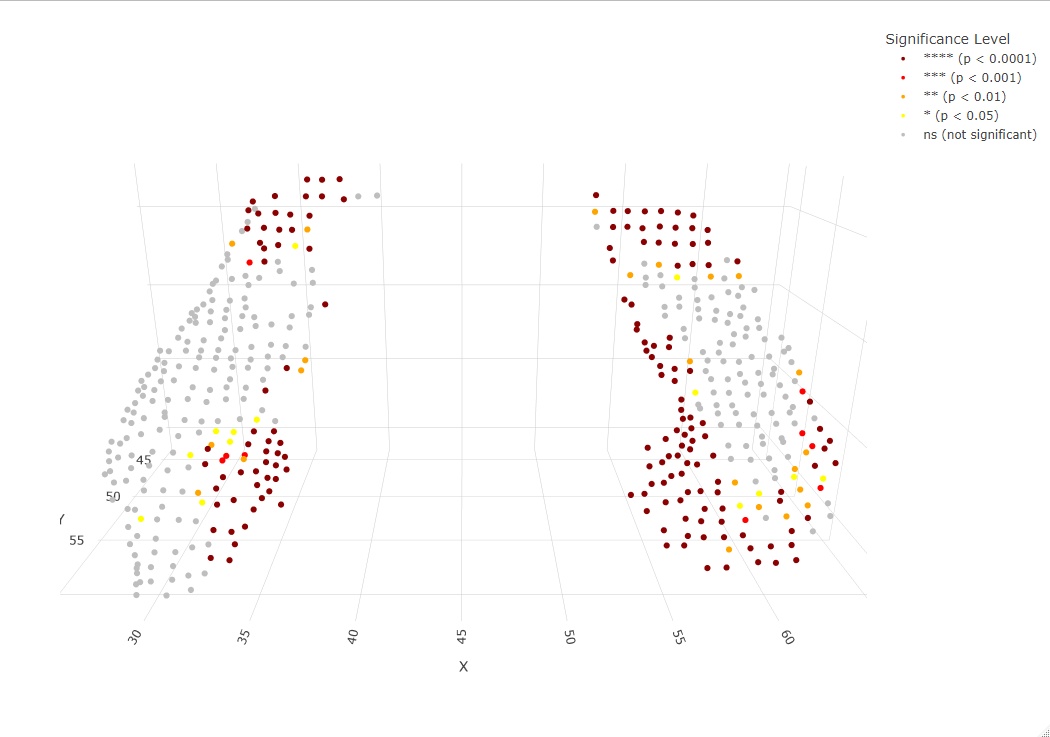} \\
        (e) ADNI AD negative & (f) ADNI AD positive \\
    \end{tabular}
    \caption{Group-level CP contamination significance maps from one-sample Right column: positive contamination (overestimation). Left column: negative contamination (underestimation). CN and MCI show widespread positive contamination, whereas the AD group exhibits central negative contamination with residual anterior/posterior positivity.}
    \label{fig:group_cp_contamination}
\end{figure}

\paragraph{Group-Level Patterns of Contamination}
Figure~\ref{fig:group_cp_contamination} shows group-level CP contamination patterns for the full cohort of 628 participants from ADNI derived from one-sample $t$-tests on $\beta_1$ coefficients across individuals. Each panel presents significant contamination effects, separated by direction (positive vs. negative) and cognitive-status group. In CN and MCI groups, where hippocampal tau signal is generally low, we observe widespread positive contamination (panels b and d). This aligns with the hypothesis that CP signal inflates SUVR when native tau is weak. In contrast, AD groups display prominent negative contamination in the central body of the hippocampus (panel e), with residual positive contamination in both anterior and posterior extremities (panel f). This spatial separation reflects the curved anatomical structure of the hippocampus, where CP contamination proximity varies along the anterior-posterior axis. These findings reinforce the directionality of CP contamination: weak tau in hippocampus → upward bias (positive $\beta_1$); strong tau in hippocampus → downward bias (negative $\beta_1$). Importantly, the contamination is not spatially uniform but follows systematic, geometry-driven patterns.

\section{Discussion}
\paragraph{Surface-based representation of hippocampal tau.}
In this study, we proposed a surface-based framework for high-resolution modeling of hippocampal tau deposition, integrating spatial coverage, SUVR intensity, and projection-based thickness features onto a medial principal surface. Using pre-defined SuStaIn subtypes and progression stages, we investigated how tau accumulation varies across subtypes within the hippocampus, quantifying both spatial distributions and covariate effects. Our analyses revealed subtype-specific trajectories of tau pathology in this region, characterized by divergent nonlinear stage effects, differential covariate associations, and anatomically consistent signal patterns. These results demonstrate the utility of our framework for capturing biologically meaningful heterogeneity in regional tau deposition. A key strength of our framework lies in its ability to represent tau pathology along multiple complementary dimensions while preserving spatial specificity. By projecting tau features onto a shared medial principal surface, we achieve geometric alignment across individuals and reduce data dimensionality, enabling localized regression with greatly reduced multiple comparison burden compared to voxelwise analysis. Unlike traditional ROI approaches, the proposed framework retains fine-grained spatial resolution, and the principal surface representation preserves sufficient geometric structure to summarize and reconstruct full hippocampal shapes when needed. This framework also facilitates a clear separation between shape-derived features (coverage and thickness) and signal intensity (SUVR), allowing distinct patterns to emerge across feature domains.

From a methodological perspective, our framework integrates anatomical structure via a medial manifold representation and restricts analysis to biologically meaningful suprathreshold signal—excluding low-SUVR regions that are unlikely to reflect true tau deposition. This over-threshold constraint provides greater biological interpretability than regional SUVR averages and ensures that regression targets reflect true signal distribution rather than background variability. While the optimal threshold remains an open question, our approach provides a powerful and interpretable model for investigating spatial tau accumulation patterns. The comparison with voxel-wise regression demonstrates that the proposed surface-based framework offers improved statistical power, anatomical alignment, and biological interpretability relative to voxel-wise regression, making it a preferable choice for analyzing sparse, spatially structured PET data.

\paragraph{Statistical modeling and threshold robustness.}

Our surface-based representation of hippocampal tau deposition allows decoupling structural features into three interpretable components: suprathreshold coverage, local thickness, and signal intensity. To model covariate effects on these features, we employed a two-stage regression strategy with IPW, correcting for selection bias induced by signal-dependent inclusion. This IPW-adjusted method was especially effective in improving the stability of SUVR-based regression. As shown in Supplementary Figure~\ref{fig: age_diagnosis_SUVR_no_IPW}, IPW correction revealed cognitive status effects in subtype 2 that were otherwise obscured, bringing them into spatial agreement with the more robust patterns observed in subtype 1. Similar gains were observed for age effects and across thickness and coverage models. These improvements highlight the utility of IPW-enhanced modeling for detecting spatially nuanced effects in sparsely expressed tau signals.

Importantly, we found that the covariate modeling results were robust to variations in the threshold used to define tau positivity. Although we adopted a fixed SUVR threshold of 2.0 in our analysis, alternative thresholds (e.g., 1.8, 2.2) produced qualitatively similar regression coefficients despite changes in surface coverage patterns. The robustness to threshold choice means that downstream modeling results remain stable across reasonable cutoffs, allowing flexible yet reliable use in practice. In future applications, e.g. in other brain regions, the SUVR threshold can be redefined as a task-specific, tunable parameter.

\paragraph{Subtype-specific tau dynamics.}

Our results reveal consistent spatial and temporal differences in tau accumulation across SuStaIn-derived subtypes. The limbic-predominant subtype (subtype 1) exhibited early and widespread tau deposition in the medial temporal lobe, followed by a plateau in both spatial coverage and projection-based thickness—evidenced by strong negative quadratic effects of SuStaIn stage. In contrast, the posterior subtype (subtype 2) showed minimal nonlinearity in these shape-related features, but exhibited accelerating SUVR increases in posterior and lateral hippocampal regions at later stages. These divergent trajectories suggest that subtype 1 follows a rapid early accumulation with subsequent saturation, while subtype 2 progresses more gradually but with sustained expansion in signal intensity and anatomical extent.

Beyond staging effects, we also observed subtype-specific associations with age. Only subtype 1 showed significant age-related increases in hippocampal SUVR after adjustment for sex, cognitive status, and SuStaIn stage, suggesting that tau intensity in the limbic-predominant subtype may be more closely coupled to aging-related processes within the hippocampus. In contrast, subtype 2 showed no significant age-related hippocampal SUVR effect despite having an age-at-scan distribution comparable to subtype 1 and little correlation between age at scan and SuStaIn stage. This suggests that the subtype difference in age associations is unlikely to be driven simply by participant age distribution or by age acting as a proxy for inferred disease stage.

This finding should be distinguished from clinical symptom-onset timing. Prior studies have reported that posterior or atypical AD phenotypes, including posterior cortical atrophy and hippocampal-sparing/posterior tau patterns, are often associated with earlier clinical onset or greater cortical tau burden in younger individuals~\citep{RN325,RN320,RN277,Whitwell2019AgeAtypicalAD}. Our findings are broadly compatible with this literature in that subtype 2 reflects a posterior occipitotemporal phenotype with less direct age dependence within the hippocampus. In our cohort, ADNI-reported cognitive symptom-onset year suggested numerically earlier symptom onset and longer symptom duration in subtype 2, but these differences were not statistically significant. Therefore, while clinical onset timing may still be relevant to subtype-specific AD heterogeneity, the absence of significant age-related hippocampal SUVR effects in subtype 2 does not appear to be explained solely by participant age, symptom-onset timing, or symptom duration.

In addition to subtype-level differences in overall signal and shape metrics, our predictive modeling revealed that tau deposition within the hippocampus also follows a subtype-specific spatial expansion pattern. Using the two-stage regression models, we simulated tau progression across disease stages under fixed covariates, and found that in subtype 1, tau consistently originated in the middle segment of the hippocampus and extended along both anterior–posterior and medial–lateral axes, accompanied by increasing deposition thickness. Subtype 2, by contrast, exhibited an initial signal peak in the posterior hippocampus, aligning with its broader posterior cortical phenotype. These results suggest that even within a region traditionally regarded as the starting point of tau accumulation, the spatial dynamics of pathology may differ depending on subtype. The observation that tau does not emerge diffusely, but instead follows a systematic expansion trajectory, reinforces the biological coherence of SuStaIn-derived staging, and provides a fine-grained view of intra-hippocampal propagation that may inform future models of tau spread.

\paragraph{Tracer-specific artifacts and off-target signal.}

Beyond regression-based phenotyping, our framework provides additional analytical capabilities for spatially localized imaging artifact identification. Motivated by the known relevance of choroid plexus off-target signal for hippocampal \textsuperscript{18}F-AV1451/flortaucipir PET quantification \citep{RN334,RN335}, we used directional patterns of SUVR signal on the principal surface to evaluate potential choroid plexus spill-in. In our results, we identified asymmetric anterior or inferior signal distributions that were misaligned with expected hippocampal anatomy. This observation served as indicator of potential spill-in. With this application, we illustrate how the surface representation can support artifact detection within an anatomically organized coordinate system. The results and interpretation of this analysis are tracer-specific: other tau PET tracers may have different off-target binding profiles. For example, \textsuperscript{18}F-MK6240 has been reported to show extracerebral or meningeal off-target signal, with less prominent choroid plexus binding than \textsuperscript{18}F-AV1451/flortaucipir \citep{Gogola2022DirectComparison,Smith2021OffTargetTauPET,Tissot2023MK6240OffTarget}. Thus, while the surface-based representation and regression framework may be adapted to other tau PET tracers, these implementations would require tracer-specific choices of reference region, SUVR thresholding approaches, partial-volume correction strategy, and off-target artifact model.

\paragraph{Model assumptions and limitations.}

The principal-surface representation provides a compact approach to analyze hippocampal tau deposition, but it is not a lossless transformation of the original PET image. By mapping suprathreshold voxels onto a shared medial surface, the framework emphasizes the surface-aligned organization of tau signal and summarizes each scan through projected surface location, local SUVR intensity, suprathreshold coverage, and positive/negative projection-distance features. This design is well suited to the hippocampus, which has a coherent elongated geometry and where tau signal is expected to accumulate along anatomical structure rather than appear as highly fragmented 3D clusters. However, the representation does not retain the full voxel-wise 3D intensity field or the exact 3D location of each voxel. Highly focal, disconnected, or strongly depth-specific uptake patterns could therefore be smoothed by LOESS interpolation or simplified by the projection-distance summary. In case of high noise, this smoothing can have a positive impact on further analysis. For tau-positive regions with more complex geometry, future extensions of the surface-fitting procedure may be needed to represent disconnected, highly nonconvex, or strongly depth-specific patterns more faithfully. Finally, the dimension reduction reduces the number of tests performed in 3D voxel-level analyses thereby reducing the effect of multiple comparisons.

The interpretation of the surface features also depends on image alignment and tracer-specific sources of signal bias. PET--MRI registration and template alignment must provide sufficient anatomical correspondence for projected surface locations to be comparable across participants; misalignment could blur spatial effects or assign signal to an incorrect surface location. Although this is a strong assumption, it is shared by most population-level neuroimaging analyses that rely on image registration and template-space comparison, and is therefore not specific to the proposed surface-based framework. Partial volume effects and spillover from adjacent structures remain additional limitations, particularly for \textsuperscript{18}F-AV1451/flortaucipir in the hippocampus because of choroid plexus off-target binding. Residual spill-in could overestimate local hippocampal SUVR and derived coverage, whereas spill-out from small or atrophic hippocampal regions could underestimate them. Our directional contamination analysis showed that CP-related signal patterns were detectable and spatially structured, illustrating how the proposed surface framework can be used to localize and assess this tracer-specific artifact.

Additional assumptions arise from thresholding and statistical modeling. The suprathreshold coverage measure depends on the SUVR cutoff used to define tau-positive voxels; this threshold-related assumption was evaluated in the threshold-sensitivity analysis described above. The two-stage regression model also assumes that the first-stage coverage model provides adequate inverse-probability weights for the observed data. If the coverage model is misspecified or yields unstable weights, downstream SUVR and projection-distance effects may be over- or under-corrected, potentially biasing covariate-effect estimates or increasing uncertainty. This is a common consideration in regression modeling, and future work could evaluate weight diagnostics, alternative coverage models, weight-stabilization strategies, and nonlinear or semiparametric regression approaches. 

\paragraph{Future directions.}
The generalizability of our approach opens avenues for future work. While this study focused on the hippocampus in AD, the framework is readily extensible to other brain regions and disease models. For example, projection-based tau thickness could be adapted for neocortical surfaces, or applied to other tracers such as amyloid PET or FDG. The regression modeling scheme can accommodate additional biomarkers or longitudinal trajectories, offering a flexible platform for capturing disease heterogeneity. We anticipate that this framework can contribute to more precise biomarker analysis in both clinical and research settings, especially in subtype-aware modeling of neurodegeneration. The Centiloid and amyloid-status summaries presented in Table~\ref{table:demographics} followed the expected clinical gradient from CN to MCI to AD, and future analyses could incorporate amyloid status or continuous Centiloid burden directly as covariates or effect modifiers in this surface-based regression framework.

\section*{Ethics}
This research used data from the Alzheimer's Disease Neuroimaging Initiative (ADNI) database. All ADNI participants provided written informed consent, and study protocols were approved by the institutional review boards at each participating site. No additional human subjects were recruited for this study.

\section*{Data and Code Availability}

\paragraph{ADNI data set}  
Data used in the preparation of this article were obtained from the Alzheimer’s Disease Neuroimaging Initiative (ADNI) database (\url{http://adni.loni.usc.edu}).

\paragraph{Code Availability}

All code for preprocessing, geometric modeling, regression analysis, and figure generation 
is openly available at \url{https://github.com/LaoWang-123/PS-analysis}. 
The repository includes a general pipeline, scripts for reproducing the analyses and figures in this paper, 
and supporting utility functions. 
Due to ADNI data use restrictions, no subject-level imaging or demographic data are included. 
Researchers can request access to the raw data directly from ADNI at \url{http://adni.loni.usc.edu}.

\section*{Author Contributions}
L.W. designed the study, adapted and implemented the computational framework, performed all analyses, and drafted the manuscript. A.K. performed data preprocessing. A.E. supervised the study, provided methodological guidance, and contributed to manuscript revision. All authors reviewed and approved the final version of the manuscript.

\section*{Funding}
This research was supported by the National Institutes of Aging grant 5R01AG075511. The content is solely the responsibility of the authors and does not necessarily represent the official views of the NIH.

\section*{Declaration of Competing Interests}
The authors declare no competing interests.

\section*{Acknowledgements}

Data collection and sharing for this project was funded by the Alzheimer's Disease Neuroimaging Initiative
(ADNI) (National Institutes of Health Grant U01 AG024904) and DOD ADNI (Department of Defense award
number W81XWH-12-2-0012). ADNI is funded by the National Institute on Aging, the National Institute of
Biomedical Imaging and Bioengineering, and through generous contributions from the following: AbbVie,
Alzheimer’s Association; Alzheimer’s Drug Discovery Foundation; Araclon Biotech; BioClinica, Inc.; Biogen;
Bristol-Myers Squibb Company; CereSpir, Inc.; Cogstate; Eisai Inc.; Elan Pharmaceuticals, Inc.; Eli Lilly and
Company; EuroImmun; F. Hoffmann-La Roche Ltd and its affiliated company Genentech, Inc.; Fujirebio; GE
Healthcare; IXICO Ltd.; Janssen Alzheimer Immunotherapy Research \& Development, LLC.; Johnson \&
Johnson Pharmaceutical Research \& Development LLC.; Lumosity; Lundbeck; Merck \& Co., Inc.; Meso
Scale Diagnostics, LLC.; NeuroRx Research; Neurotrack Technologies; Novartis Pharmaceuticals
Corporation; Pfizer Inc.; Piramal Imaging; Servier; Takeda Pharmaceutical Company; and Transition
Therapeutics. The Canadian Institutes of Health Research is providing funds to support ADNI clinical sites
in Canada. Private sector contributions are facilitated by the Foundation for the National Institutes of Health
(www.fnih.org). The grantee organization is the Northern California Institute for Research and Education,
and the study is coordinated by the Alzheimer’s Therapeutic Research Institute at the University of Southern
California. ADNI data are disseminated by the Laboratory for Neuro Imaging at the University of Southern
California.

\printbibliography
\clearpage
\section*{Supplementary Materials}
\setcounter{section}{0}
\setcounter{figure}{0}
\setcounter{table}{0}
\renewcommand{\thesection}{S\arabic{section}}
\renewcommand{\thesubsection}{S\arabic{section}.\arabic{subsection}}
\renewcommand{\thefigure}{S\arabic{figure}}
\renewcommand{\thetable}{S\arabic{table}}
\section{SuStaIn Output}

\begin{figure}[h!]
    \centering
    \includegraphics[width=14cm]{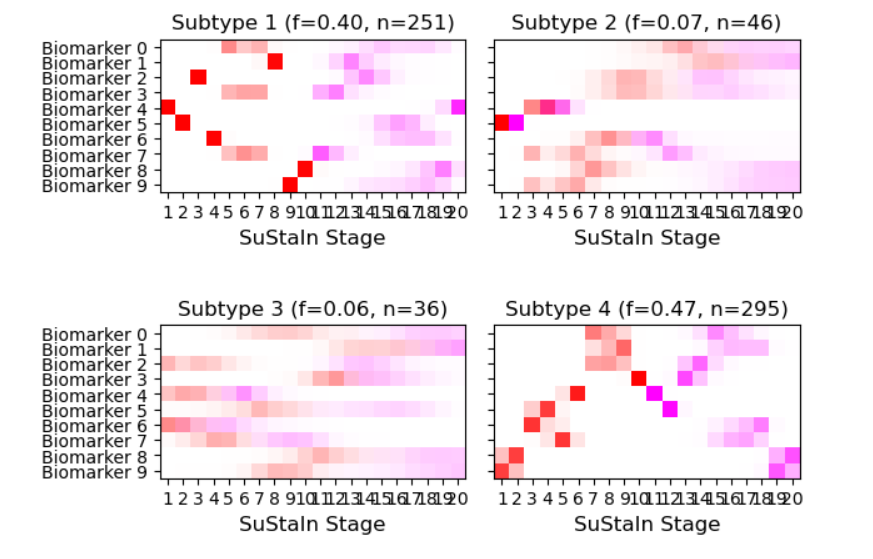}
    \caption{SuStaIn output matched with PVD plot}
    \label{fig: sustain_pvd}
\end{figure}

\section{Directional Consistency Validation}
\subsection{Principal Surface Selection} \label{sec: Principal Surface Selection}

To ensure that the principal surface (PS) provides an accurate and stable representation of tau deposition without overfitting, we evaluate its smoothness and geometric consistency using Gaussian curvature (\( K_i \)). High curvature regions indicate potential overfitting, where the surface captures noise rather than meaningful structural variations. To mitigate this, we apply curvature-based filtering to prevent self-intersections and excessive flexibility in the PS fitting process, ensuring a stable yet representative surface.

Gaussian curvature is computed at each surface point \( \mathbf{q}_i = (x_i, y_i, z_i) = f(t_{1i}, t_{2i}) \) using the first and second fundamental forms:
\[
\mathbf{K}_i = \frac{\mathbf{L}_i \mathbf{N}_i - \mathbf{M}_i^2}{\mathbf{E}_i \mathbf{G}_i - \mathbf{F}_i^2}.
\]
where \( \mathbf{E}_i, \mathbf{F}_i, \mathbf{G}_i \) define local metric properties:
\[
  \mathbf{E}_i = \left\| \frac{\partial f}{\partial t_{1i}} \right\|^2, \quad 
  \mathbf{F}_i = \frac{\partial f}{\partial t_{1i}} \cdot \frac{\partial f}{\partial t_{2i}}, \quad 
  \mathbf{G}_i = \left\| \frac{\partial f}{\partial t_{2i}} \right\|^2,
\]
and \( \mathbf{L}_i, \mathbf{M}_i, \mathbf{N}_i \) capture curvature variations:
\[
  \mathbf{L}_i = \mathcal{N}_i \cdot \frac{\partial^2 f}{\partial t_{1i}^2}, \quad 
  \mathbf{M}_i = \mathcal{N}_i \cdot \frac{\partial^2 f}{\partial t_{1i} \partial t_{2i}}, \quad 
  \mathbf{N}_i = \mathcal{N}_i \cdot \frac{\partial^2 f}{\partial t_{2i}^2}.
\]

The unit normal vector \( \mathcal{N}_i \) at each point is computed as the cross product of the tangent vectors:
\[
\mathcal{N}_i = \frac{\partial f}{\partial t_{1i}} \times \frac{\partial f}{\partial t_{2i}}.
\]
To ensure numerical stability, derivatives are approximated using finite differences:
\[
\frac{\partial f}{\partial t_{1i}} \approx \frac{f(t_{1i}+h, t_{2i}) - f(t_{1i}-h, t_{2i})}{2h},
\]
\[
\frac{\partial^2 f}{\partial t_{1i} \partial t_{2i}} \approx \frac{f(t_{1i}+h, t_{2i}+h) - f(t_{1i}+h, t_{2i}-h) - f(t_{1i}-h, t_{2i}+h) + f(t_{1i}-h, t_{2i}-h)}{4h^2}.
\]

By analyzing the distribution of curvature values, we discard PS candidates with excessive local curvature that could lead to artifacts. The final selected surface balances smoothness and structural representation, ensuring a robust framework for subsequent tau deposition analysis.

\subsection{Projection Vectors Difference}
\label{sec:supp_projection_vectors}

\begin{figure}[htbp]
    \centering
    \begin{tabular}{cc}
        \includegraphics[width=0.45\textwidth]{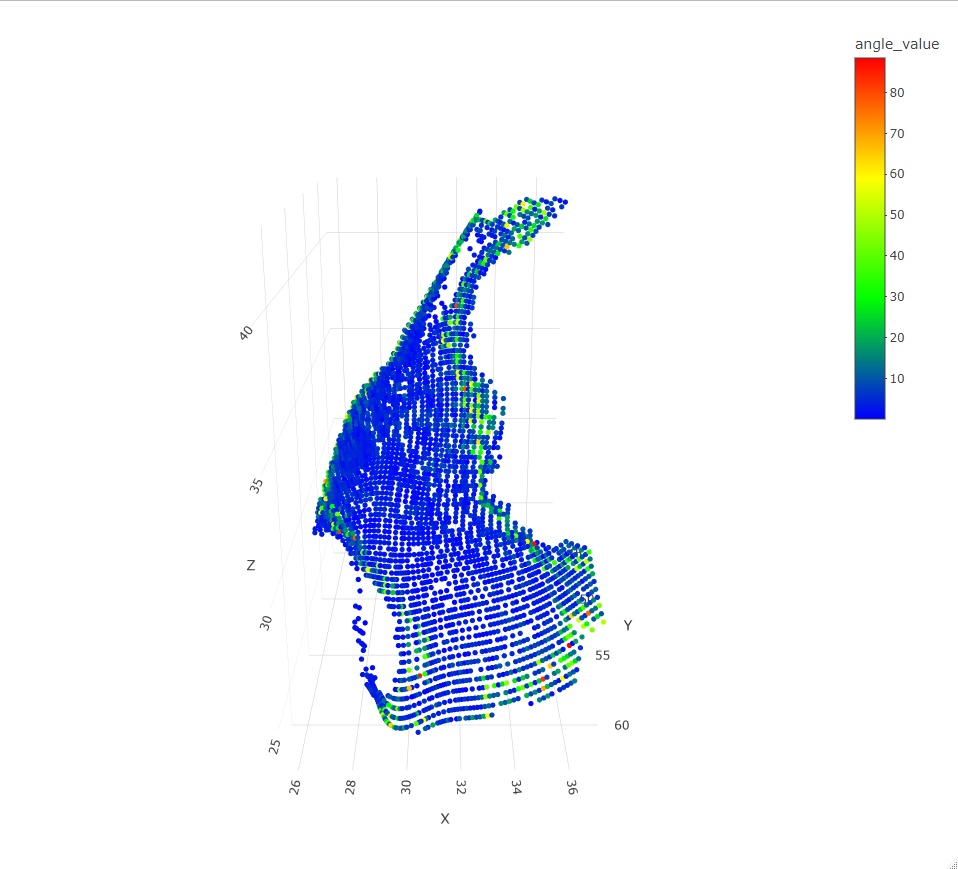} &
        \includegraphics[width=0.45\textwidth]{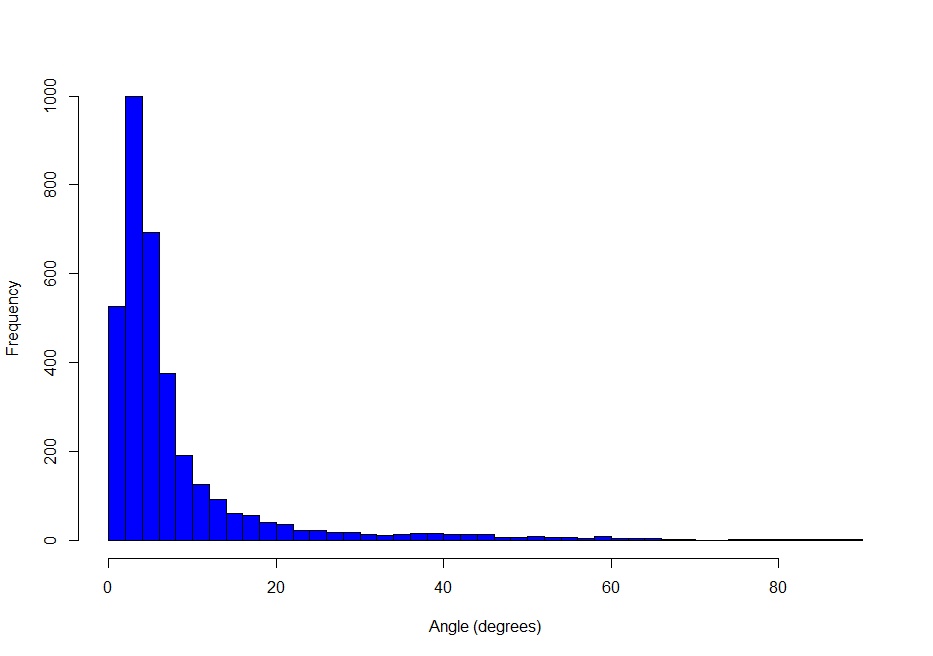} \\
        (a) 3D Angle Distribution Map & (b) Angle Distribution Histogram \\

    \end{tabular}
    \caption{
Directional consistency validation 
(a) Spatial map of angular differences between projection and normal vectors, showing localized misalignment. 
(b) Histogram of angle deviations confirms strong directional consistency across surface points. 
}
    \label{fig:validation_reconstruction}
\end{figure}

This representation assumes that the intrinsic normal vectors of the PS are well aligned with the projection directions from hippocampal boundaries. That is, when projecting a voxel from the boundary onto the PS (for distance computation or SUVR interpolation), the resulting vector should approximate the PS normal at the target point. While this holds in most regions, local misalignment may occur in areas of high curvature due to numerical approximation of normals (via finite differences) or the shortest-path projection strategy. The misalignment of two projection vectors is shown in Figure~\ref{fig:validation_reconstruction}.

\section{Prediction Model Diagnostics}

\begin{figure}[h]
    \centering
    \begin{tabular}{cc}
        \includegraphics[width=0.45\textwidth]{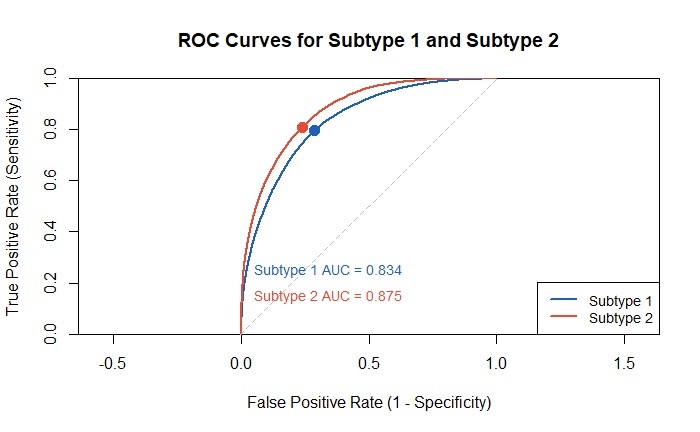} \\ 
        (a) ROC curves for tau coverage 
    \end{tabular}
    \caption{
ROC curves for pointwise logistic models of tau coverage used in model-based prediction of hippocampal tau deposition. 
The models were evaluated across all selected surface points and subjects for subtype 1 and subtype 2. 
The models achieved high discriminative performance, with AUC = 0.834 for subtype 1 and AUC = 0.875 for subtype 2.
}
\label{fig:prediction_ROC}
\end{figure}

\section{Additional Results for Regression}

\begin{figure}[h]
    \centering
    \begin{tabular}{cc}
        \includegraphics[width=0.45\textwidth]{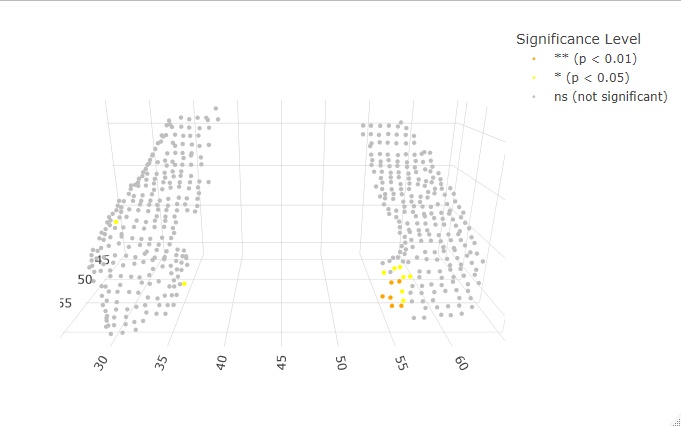} &
        \includegraphics[width=0.45\textwidth]{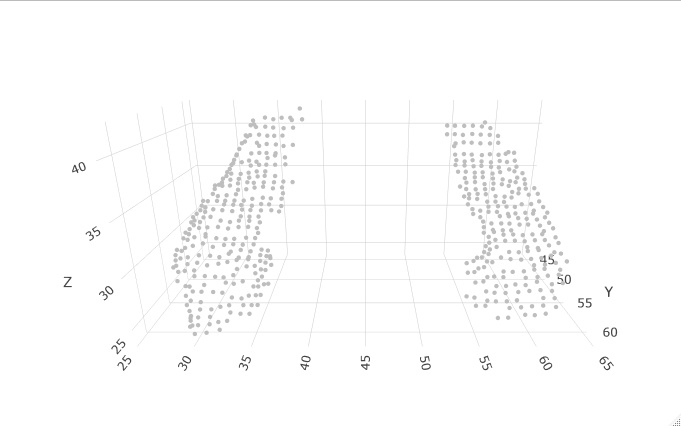} 
        \\
        (a) Age effect significance for subtype 1 & (b) Age effect significance for subtype 2
        \\
        \includegraphics[width=0.45\textwidth]{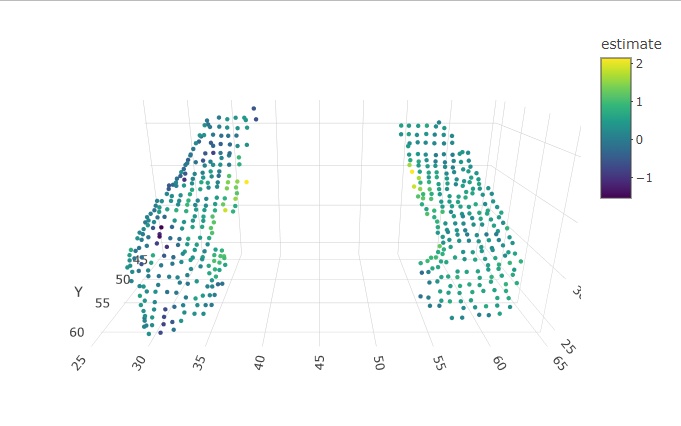} &
        \includegraphics[width=0.45\textwidth]{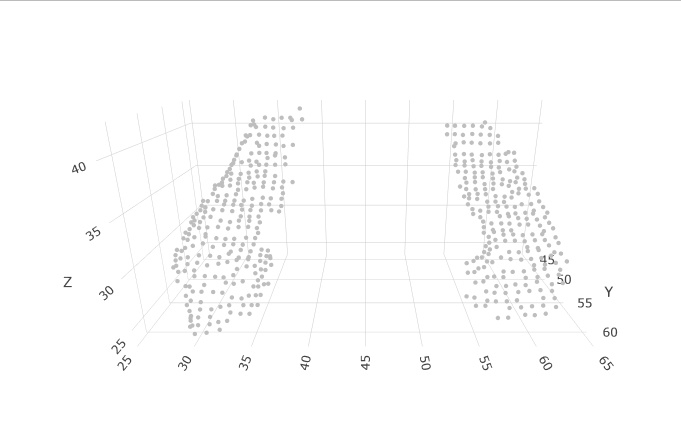} 
        \\
        (c) Sex effect significance for subtype 1 & (d) Sex effect significance for subtype 2
    \end{tabular}
    \caption{Spatial distribution of age and sex effects. }
    \label{fig:spatial_coverage_age_sex}
\end{figure}

\begin{figure}[h]
    \centering
    \begin{tabular}{cc}
        \includegraphics[width=0.45\textwidth]{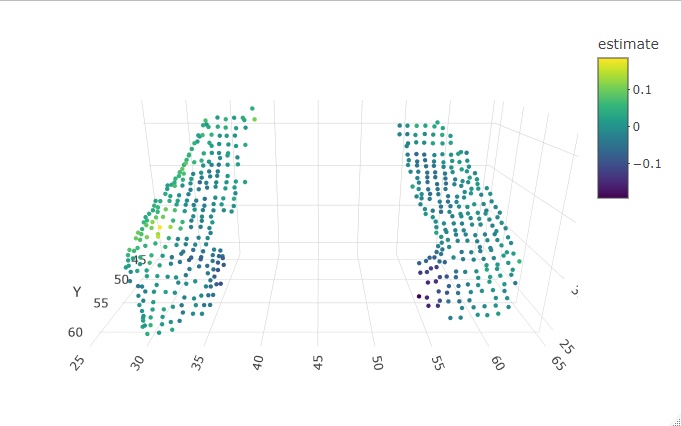}&
        \includegraphics[width=0.45\textwidth]{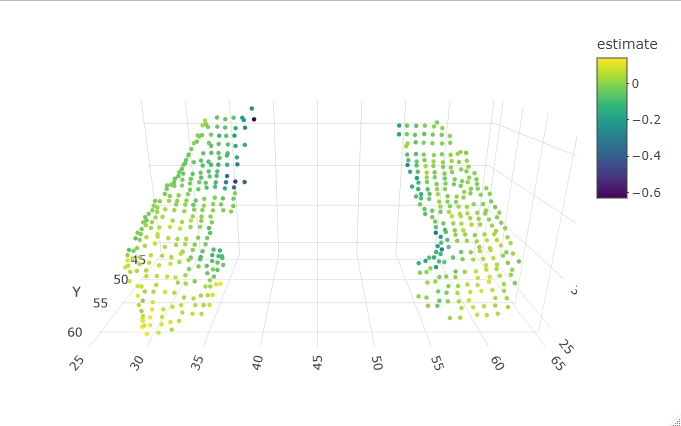} \\
        (a)Age estimate effect for subtype 1  & (b) Age estimate effect for subtype 2 \\
        \includegraphics[width=0.45\textwidth]{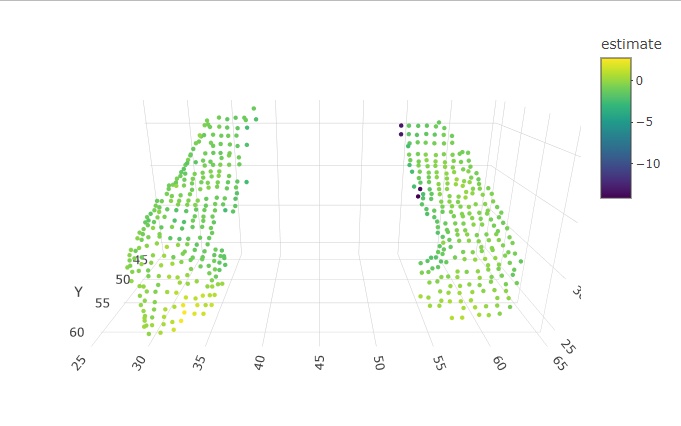}&
        \includegraphics[width=0.45\textwidth]{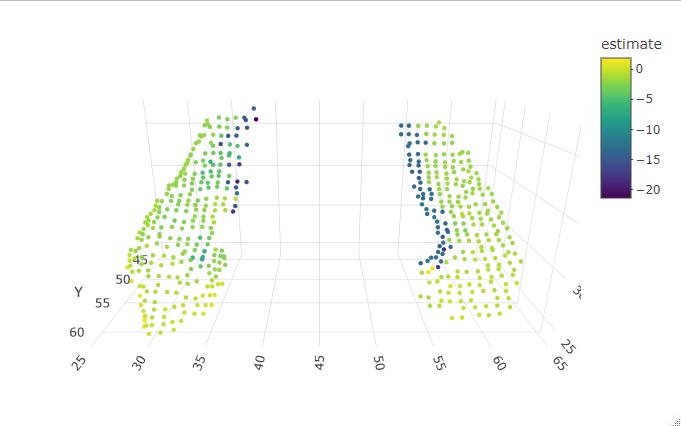} \\
        (c) Cognitive status effect estimate for subtype 1  & (d) Cognitive status effect estimate for subtype 2 \\
        \includegraphics[width=0.45\textwidth]{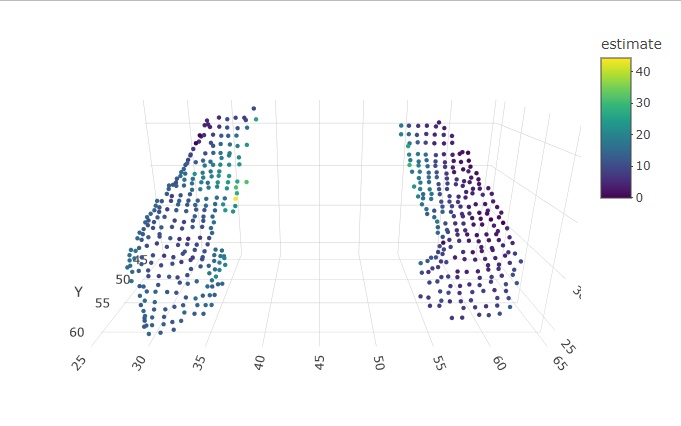} &
        \includegraphics[width=0.45\textwidth]{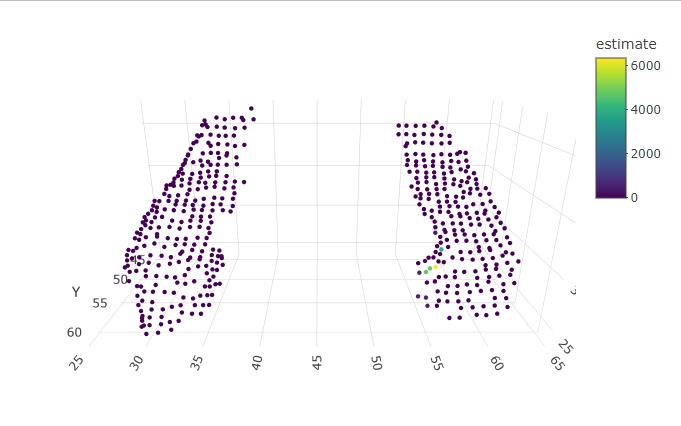} 
        \\
        (e) Linear Stage effect estimate for subtype 1 & (f)  Linear Stage effect for subtype 2
    \end{tabular}
    \caption{Spatial distribution of covariate-associated effects on hippocampal tau coverage across SuStaIn subtypes.
Each panel displays estimated voxel-wise effects (standardized beta coefficients) of age, cognitive status, and SuStaIn stage on hippocampal tau coverage, modeled separately within subtype 1 (left column) and subtype 2 (right column). Notably, age and stage effects show consistent spatial gradients within each subtype, while cognitive-status effects differ substantially between subtypes. Subtype 2 exhibits stronger and more localized cognitive-status-related effects compared to subtype 1, where cognitive status appears to have limited influence on hippocampal tau distribution. These findings highlight subtype-specific spatial patterns and covariate sensitivities in tau deposition.}
    \label{fig:spatial_coverage_estimate}
\end{figure}

\begin{figure}[h]
    \centering
    \begin{tabular}{cc}
        \includegraphics[width=0.45\textwidth]{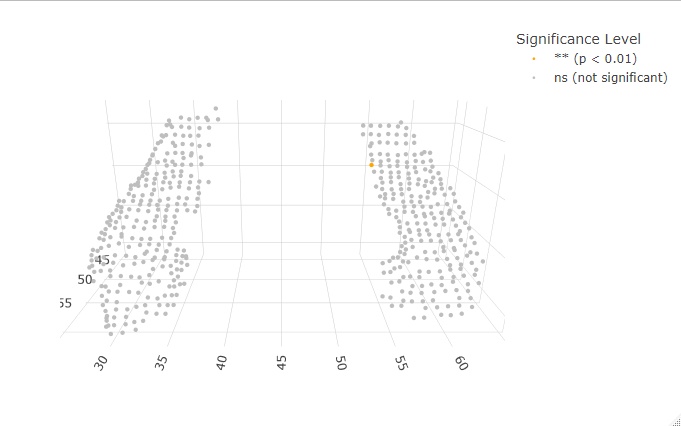}&
        \includegraphics[width=0.45\textwidth]{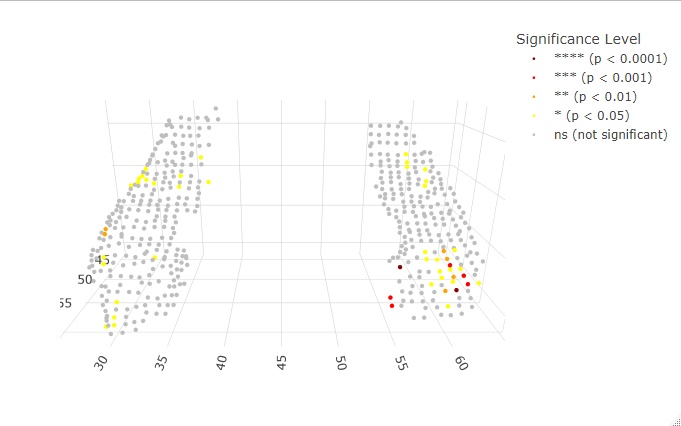} \\ 
        (a)Sex effect significance for subtype 1  & (b)Sex effect significance for subtype 2\\
        \includegraphics[width=0.45\textwidth]{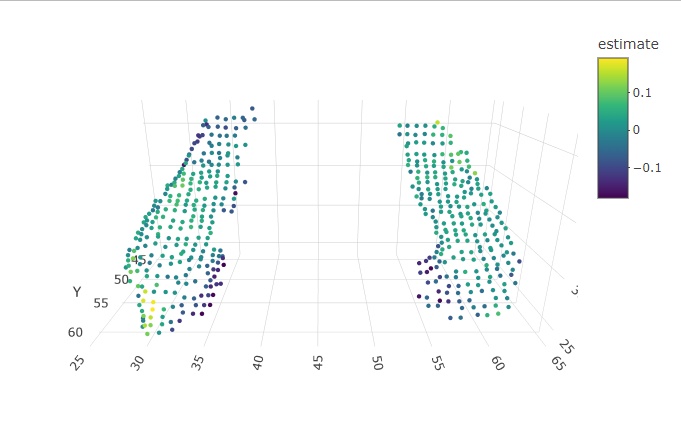}&
        \includegraphics[width=0.45\textwidth]{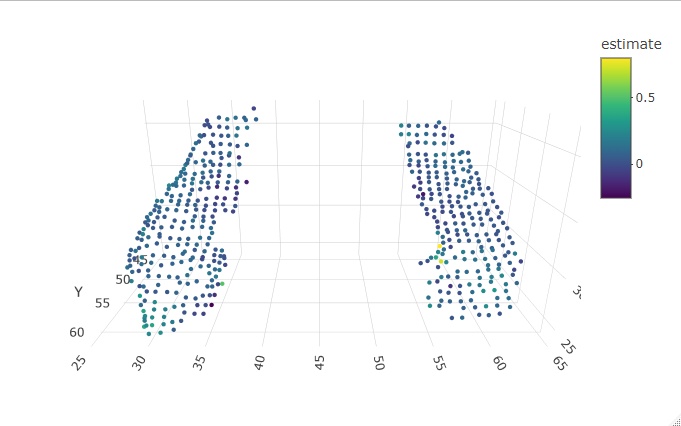} \\
        (c)Sex effect estimate for subtype 1 & (d)Sex effect estimate for subtype 2\\
    \end{tabular}
    \caption{
    Supplementary Figure. Sex-related effects on hippocampal tau SUVR. 
    (a–b) Significance maps for the effect of sex in subtypes 1 and 2. 
    (c–d) Corresponding regression coefficient estimates for each subtype.
    }
    \label{fig:sex_SUVR}
\end{figure}

\begin{figure}[h]
    \centering
    \begin{tabular}{cc}
        \includegraphics[width=0.45\textwidth]{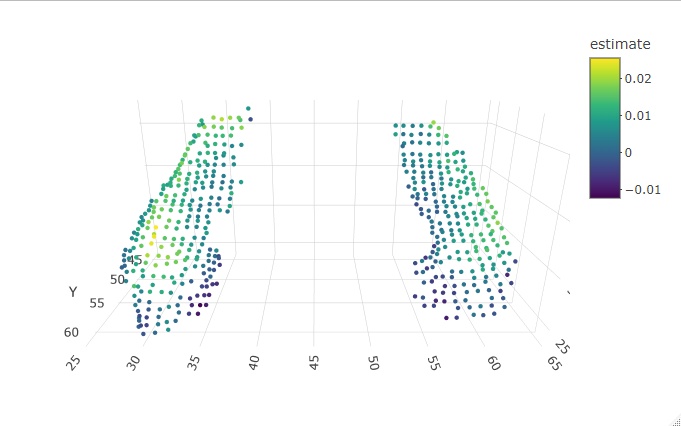}&
        \includegraphics[width=0.45\textwidth]{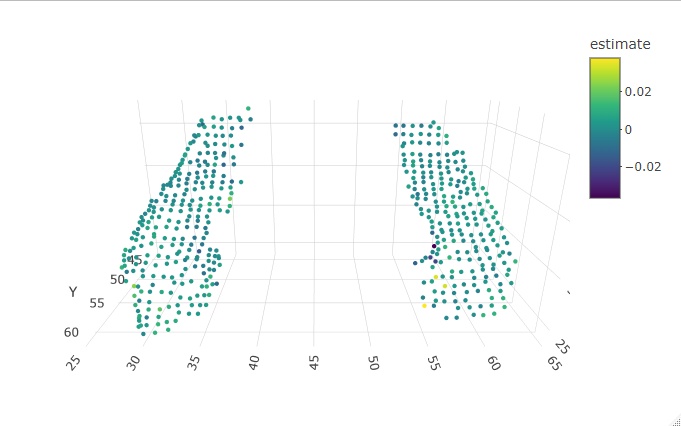} \\ 
        (a) Age estimate for subtype 1  & (b) Age estimate for subtype 2\\
        \includegraphics[width=0.45\textwidth]{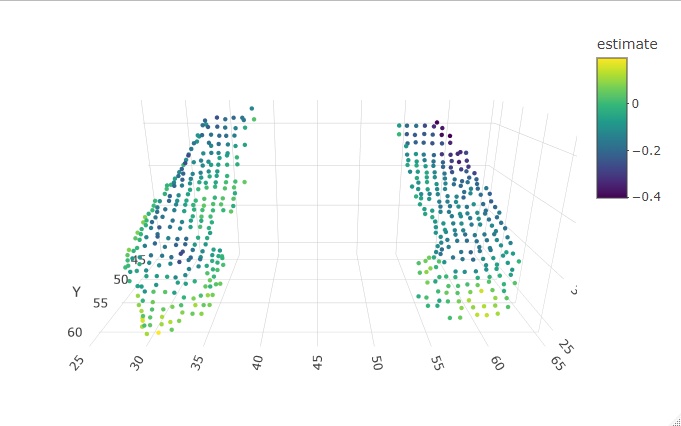}&
        \includegraphics[width=0.45\textwidth]{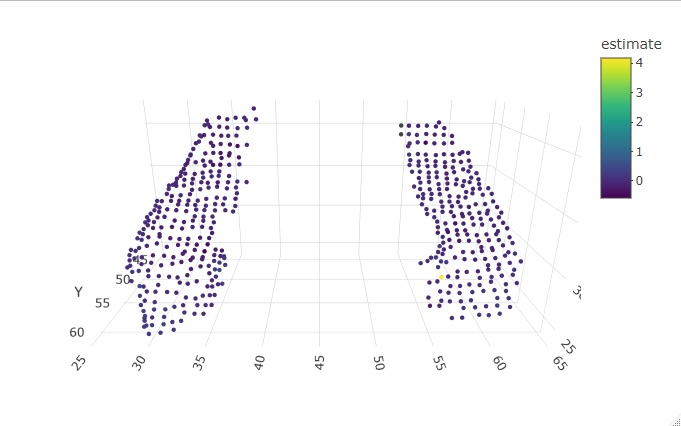} \\
        (c) Cognitive status estimate for subtype 1 & (d) Cognitive status estimate for subtype 2\\
        \includegraphics[width=0.45\textwidth]{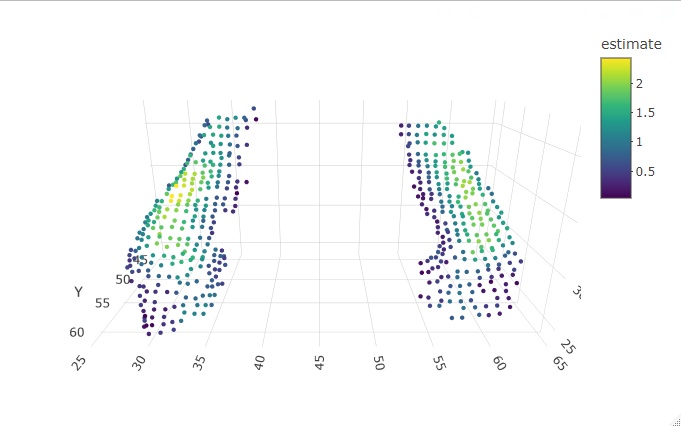}&
        \includegraphics[width=0.45\textwidth]{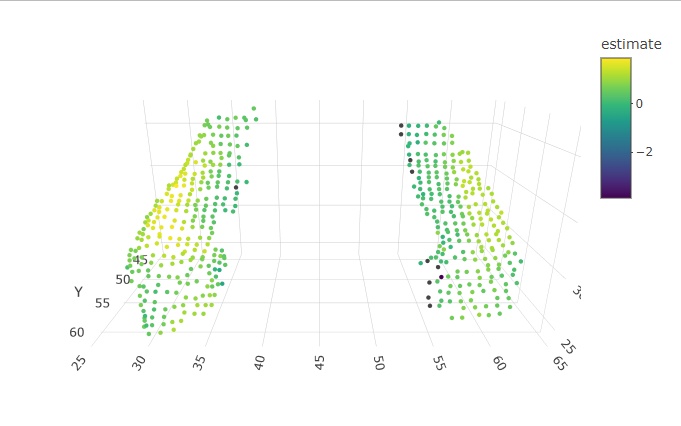} \\
        (e) Linear stage estimate for subtype 1 & (f) Linear stage estimate for subtype 2
    \end{tabular}
    \caption{
    Supplementary Figure. Regression coefficient estimates for covariate effects on hippocampal tau SUVR. 
    (a–b) Age effect estimates. 
    (c–d) Cognitive status effect estimates. 
    (e–f) Linear stage effect estimates, shown separately for subtypes 1 and 2.
    }    
    \label{fig:spatial_SUVR_supp}
\end{figure}

\begin{figure}[h]
    \centering
    \begin{tabular}{cc}
        \includegraphics[width=0.45\textwidth]{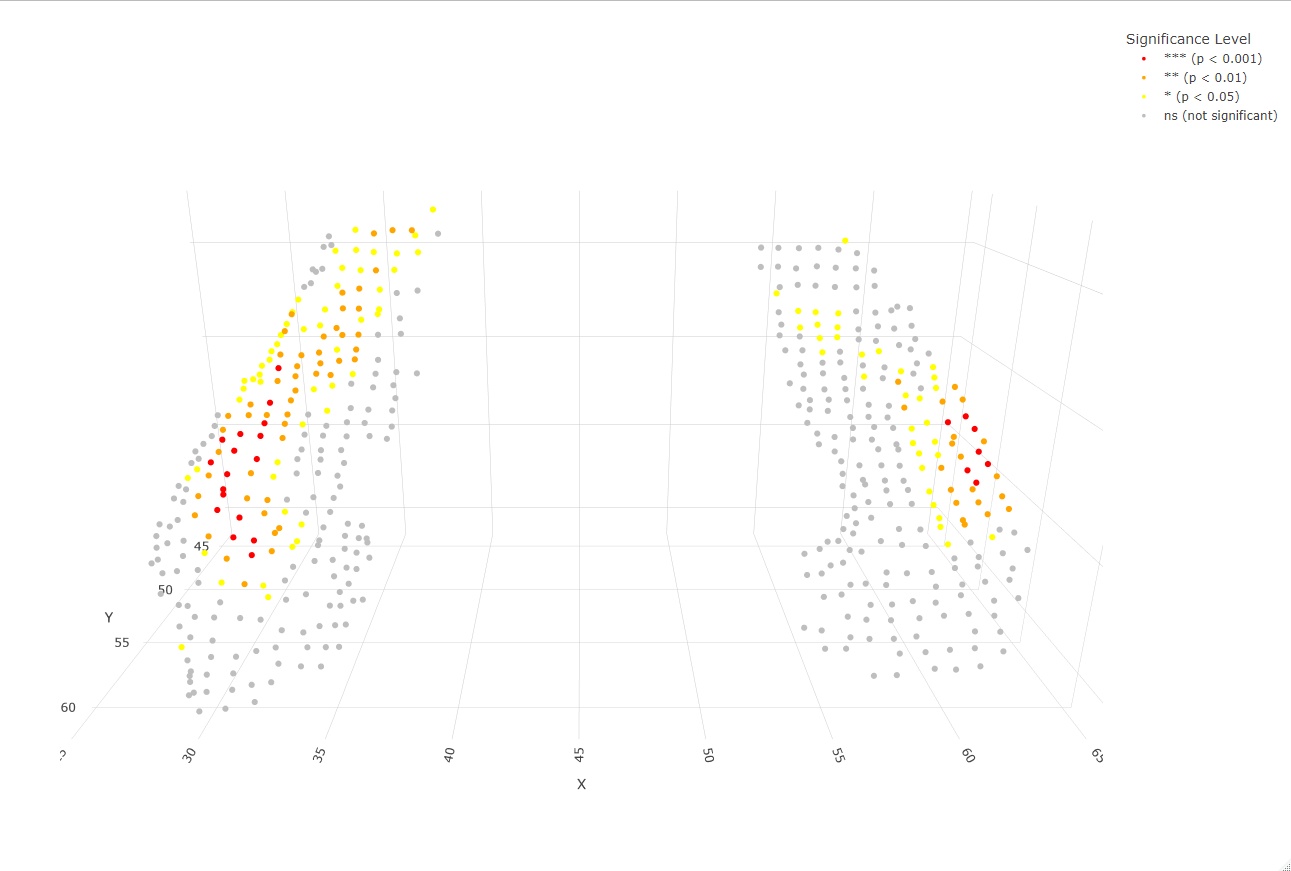}&
        \includegraphics[width=0.45\textwidth]{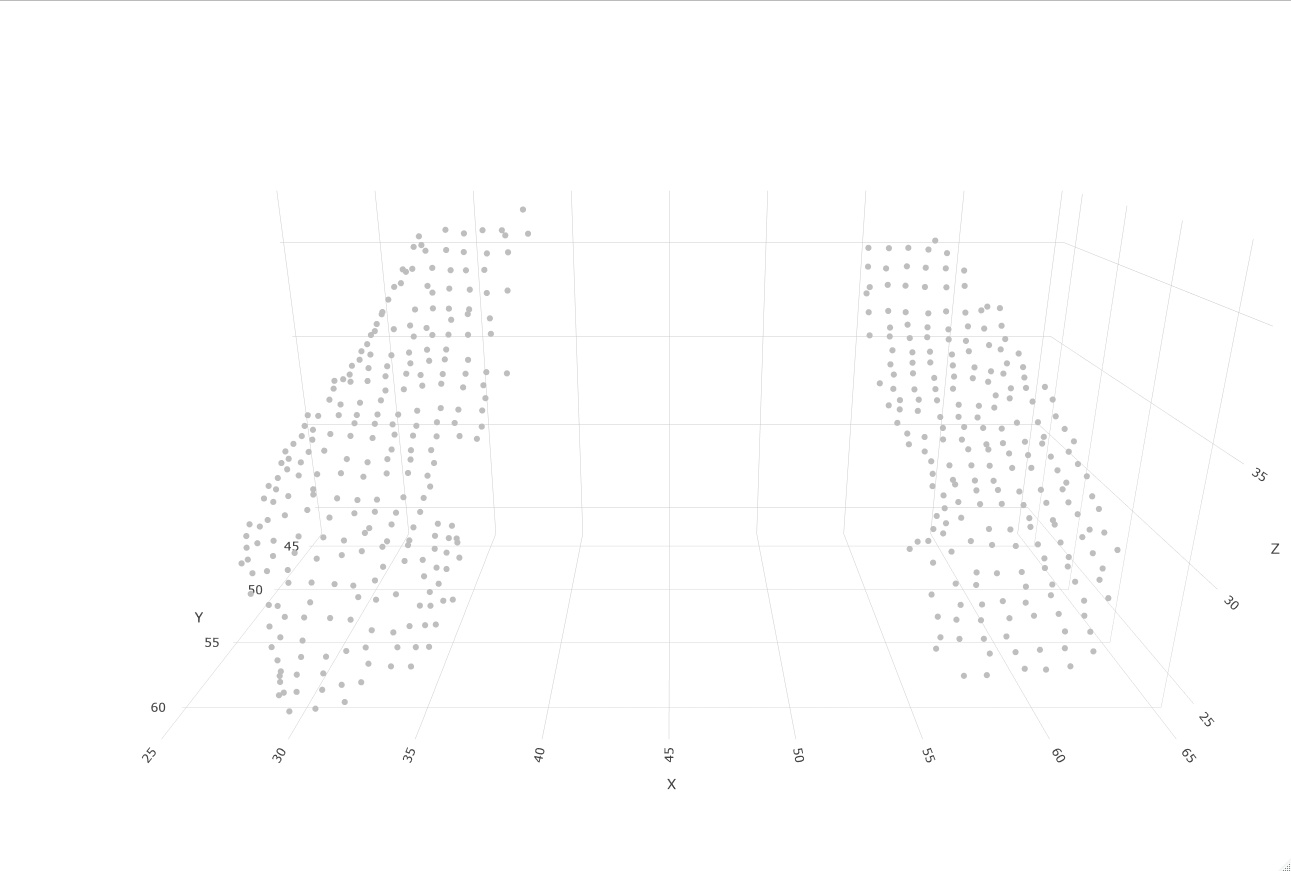} \\ 
        (a)Age significance subtype 1 no IPW & (b)Age significance subtype 2 no IPW\\
        \includegraphics[width=0.45\textwidth]{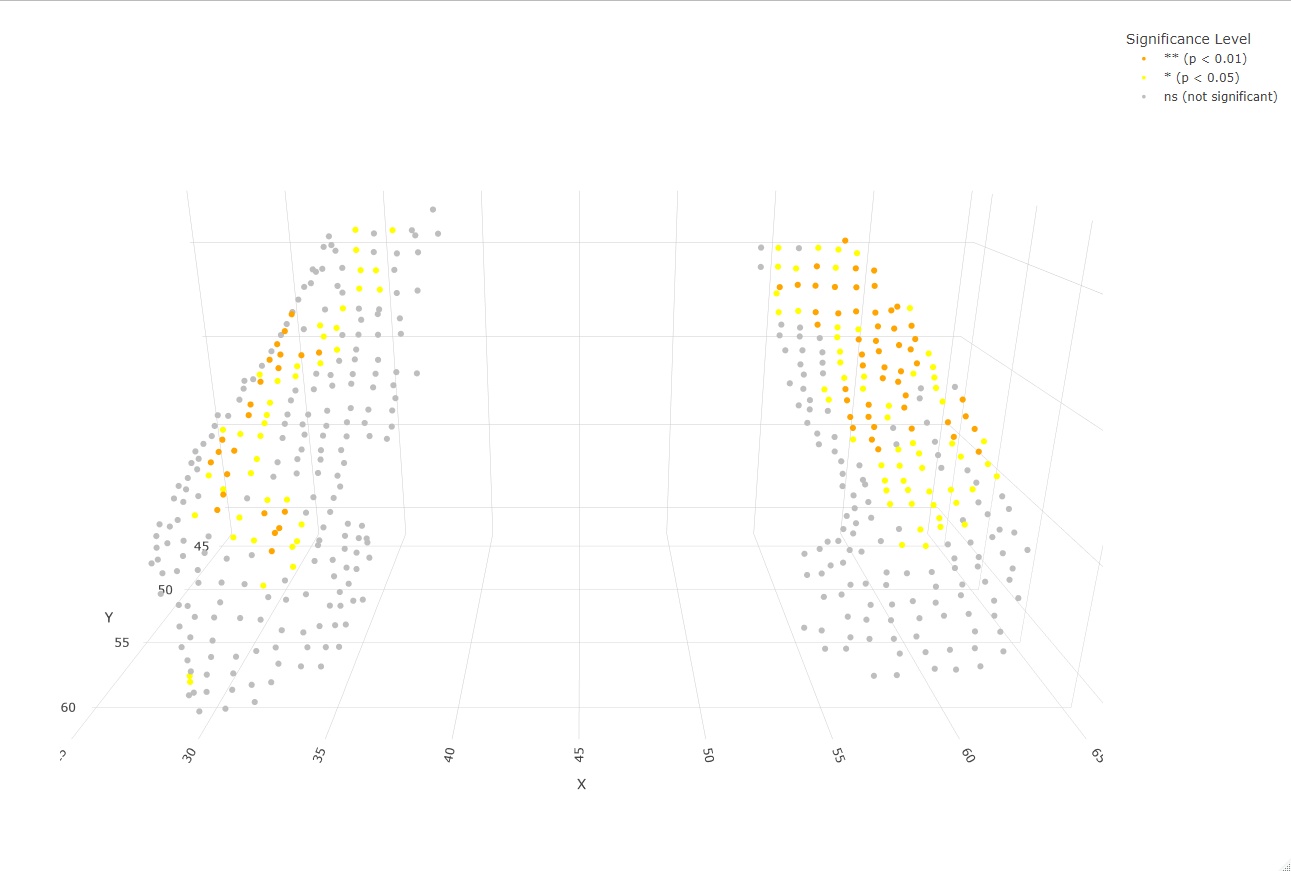}&
        \includegraphics[width=0.45\textwidth]{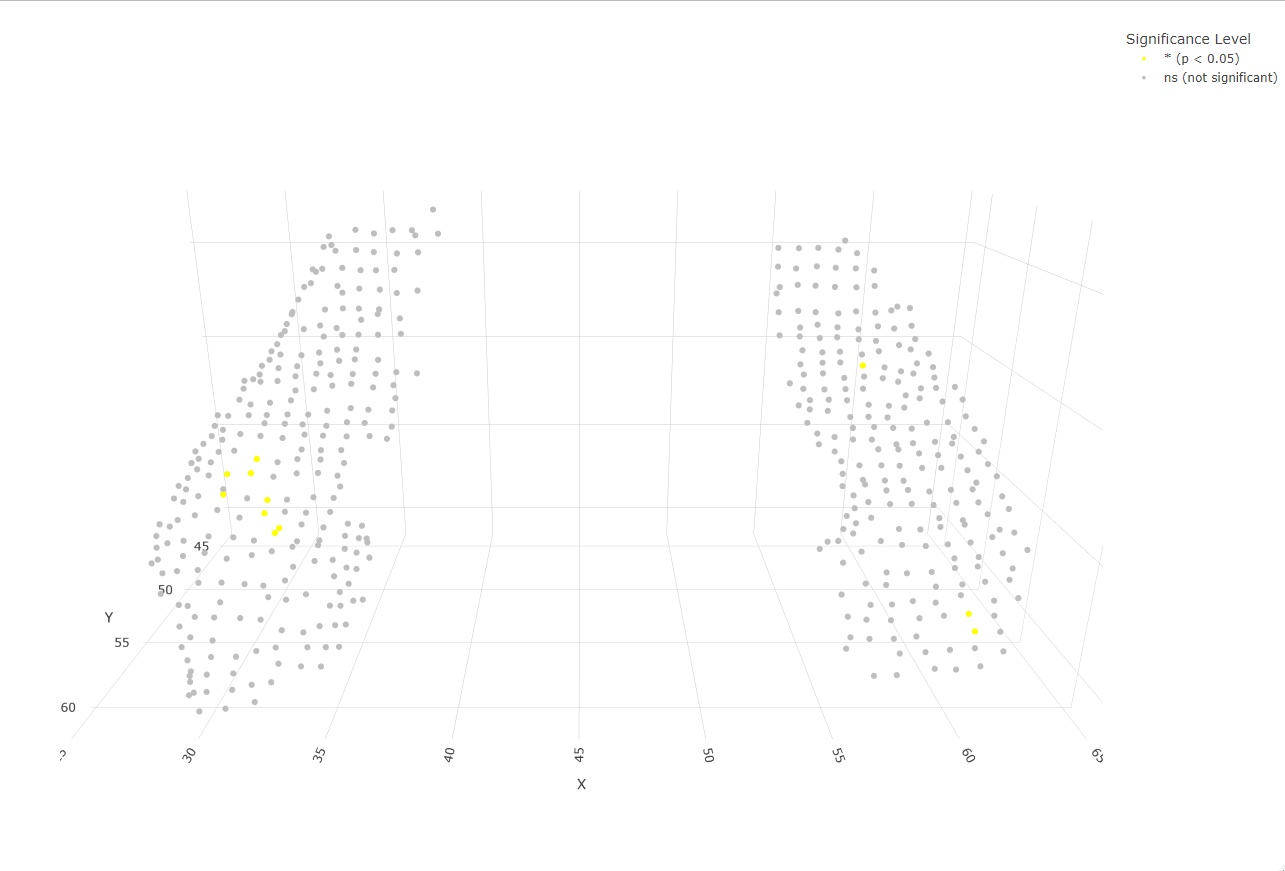} \\
        (c) Cognitive status significance subtype 1 no IPW & (d) Cognitive status significance subtype 2 no IPW\\
    \end{tabular}
    
    \caption{
    Supplementary Figure. Age- and cognitive-status-related effects on hippocampal tau SUVR. 
    (a–b) Significance maps for the effect of Age in subtypes 1 and 2 without IPW. 
    (c–d) Significance maps for the effect of cognitive status in subtypes 1 and 2 without IPW. 
    }
    \label{fig: age_diagnosis_SUVR_no_IPW}
\end{figure}

\begin{figure}[h]
    \centering
    \begin{tabular}{cc}
        \includegraphics[width=0.45\textwidth]{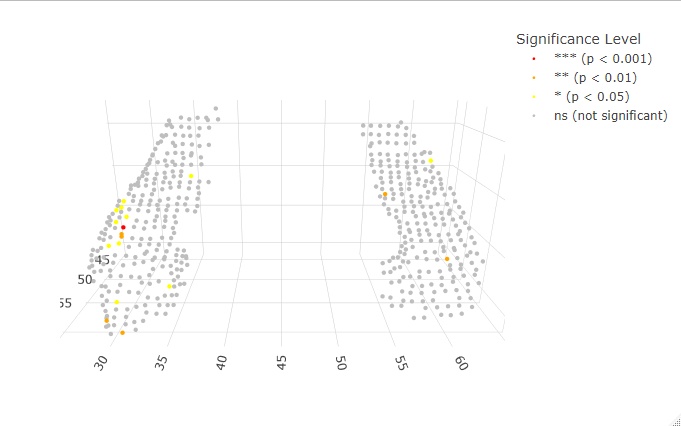} &
        \includegraphics[width=0.45\textwidth]{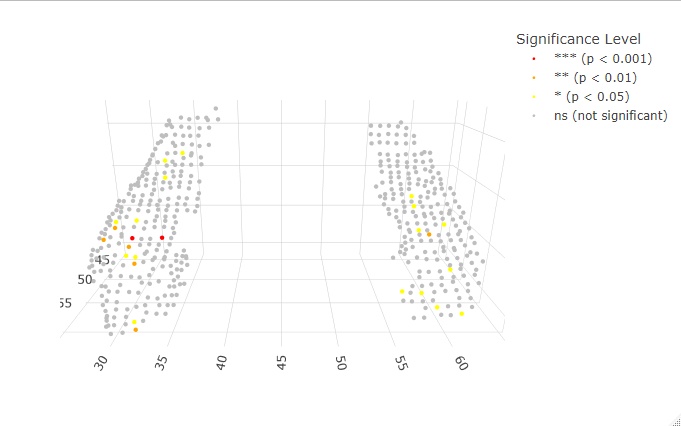} 
        \\
        (a) Age negative distance subtype 1 & (b) Age negative distance for subtype 2
        \\
        \includegraphics[width=0.45\textwidth]{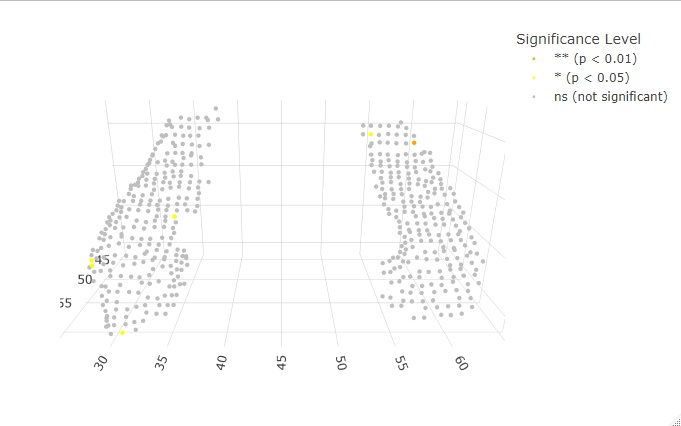} &
        \includegraphics[width=0.45\textwidth]{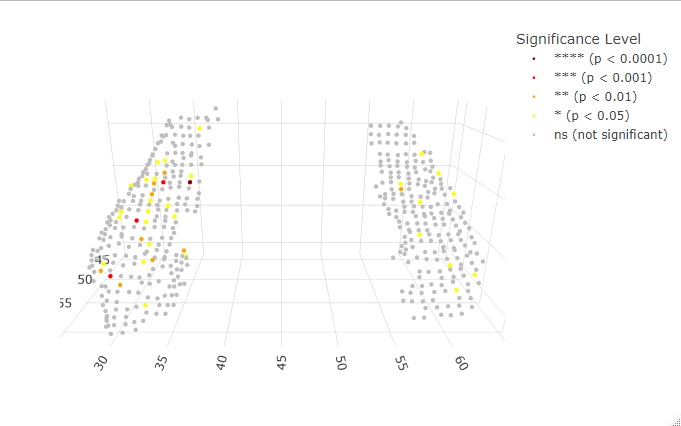} 
        \\
        (c) Cognitive status negative distance subtype 1 & (d) Cognitive status negative distance for subtype 2\\
        \includegraphics[width=0.45\textwidth]{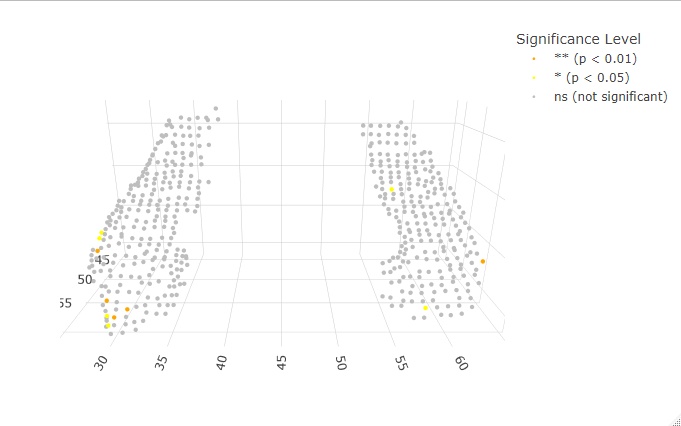} &
        \includegraphics[width=0.45\textwidth]{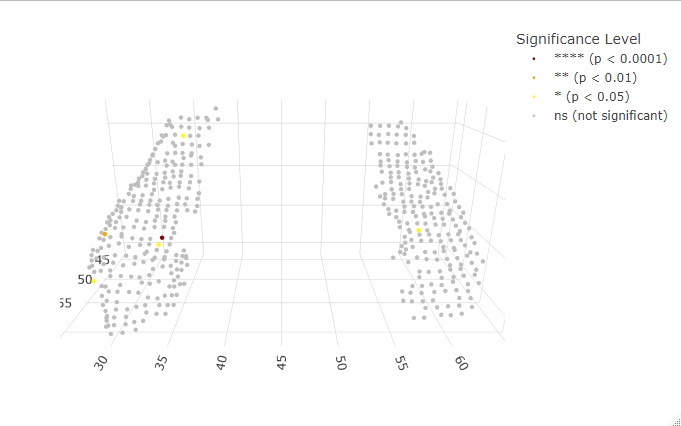} 
        \\
        (e) Sex negative distance subtype 1 & (f) Sex negative distance for subtype 2
    \end{tabular}
    \caption{Thickness of age and cognitive status effects. }
    \label{fig:thickness_age_sex_diagnosis}
\end{figure}

\begin{figure}[h]
    \centering
    \begin{tabular}{cc}
        \includegraphics[width=0.45\textwidth]{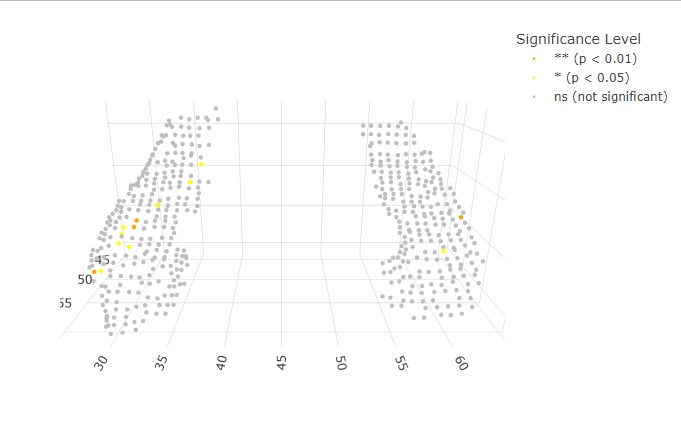} &
        \includegraphics[width=0.45\textwidth]{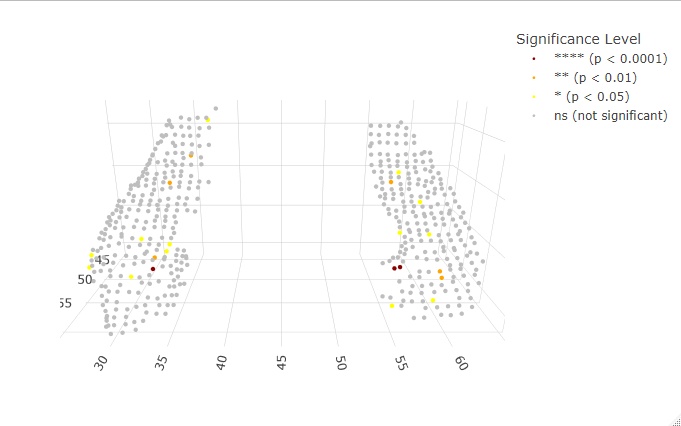} 
        \\
        (a) Age positive distance subtype 1 & (b) Age positive distance for subtype 2
        \\
        \includegraphics[width=0.45\textwidth]{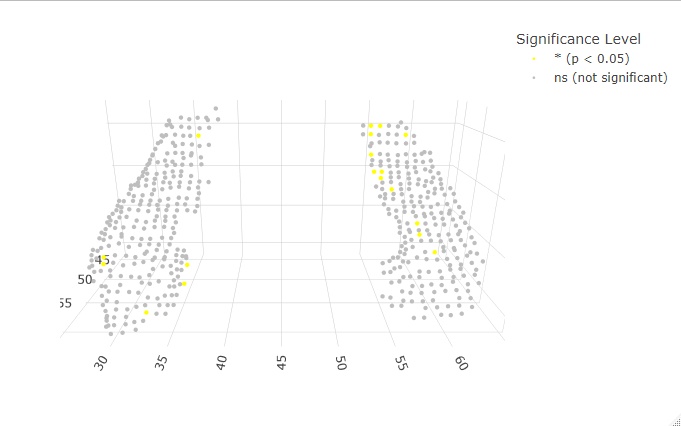} &
        \includegraphics[width=0.45\textwidth]{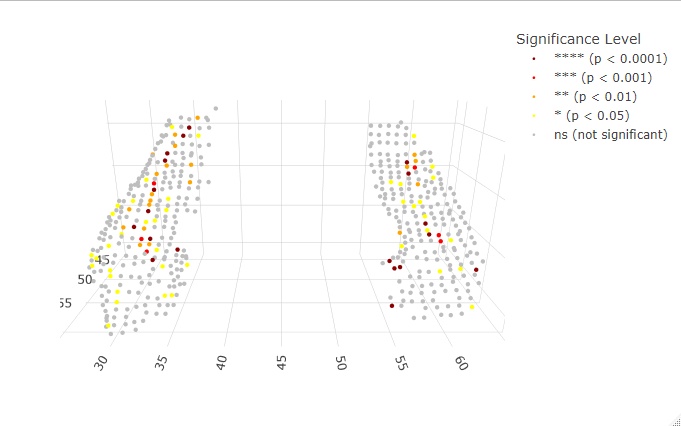} 
        \\
        (c) Cognitive status positive distance subtype 1 & (d) Cognitive status positive distance for subtype 2\\
        \includegraphics[width=0.45\textwidth]{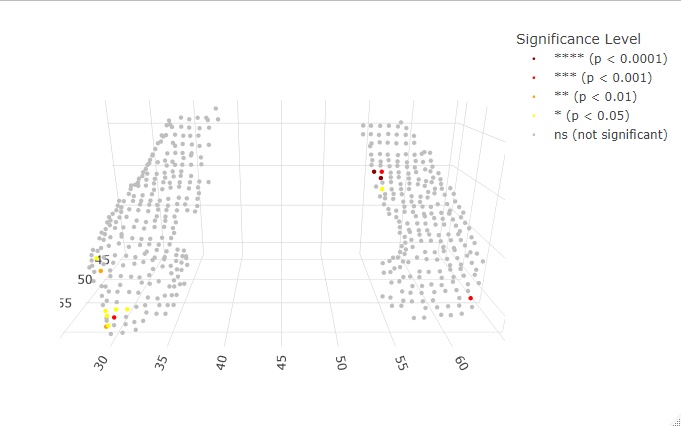} &
        \includegraphics[width=0.45\textwidth]{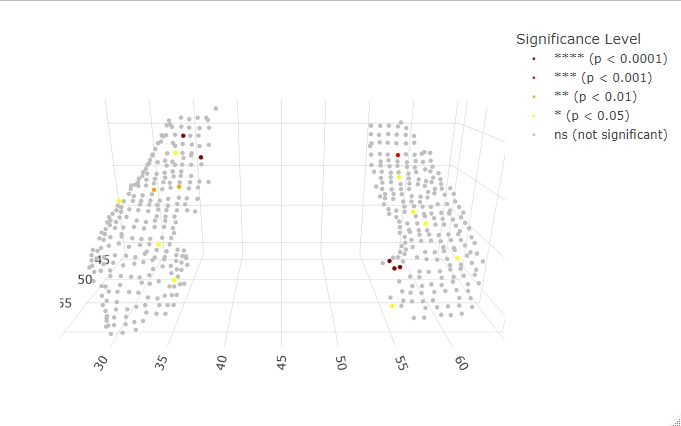} 
        \\
        (e) Cognitive status positive distance subtype 1 & (f) Cognitive status positive distance for subtype 2
    \end{tabular}
    \caption{Thickness of age and cognitive status effects. }
    \label{fig:positive_thickness_age_sex_diagnosis}
\end{figure}

\begin{figure}[h]
    \centering
    \begin{tabular}{cc}
        \includegraphics[width=0.45\textwidth]{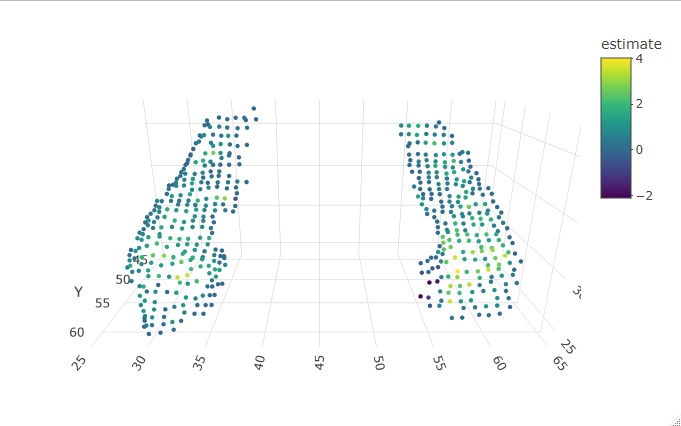}&
        \includegraphics[width=0.45\textwidth]{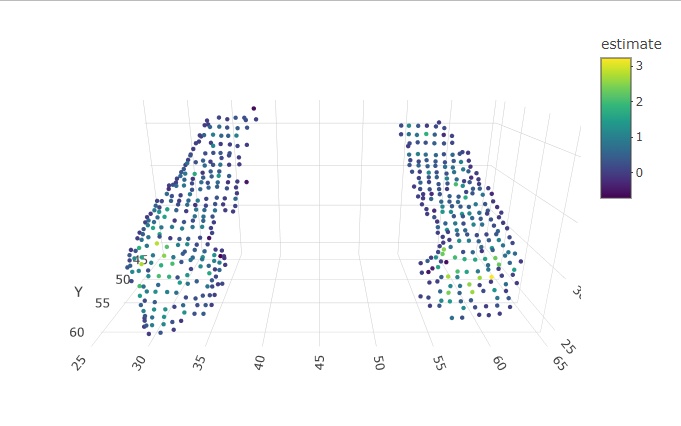} \\ 
        (a) Linear stage estimate, subtype 1, negative direction & (b) Linear stage estimate, subtype 1, positive direction\\
        \includegraphics[width=0.45\textwidth]{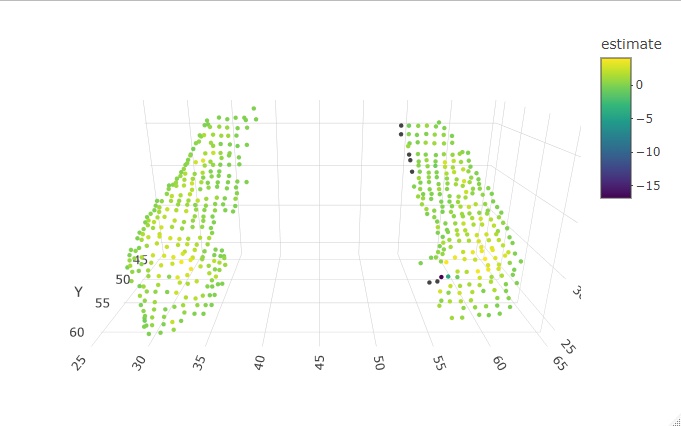}&
        \includegraphics[width=0.45\textwidth]{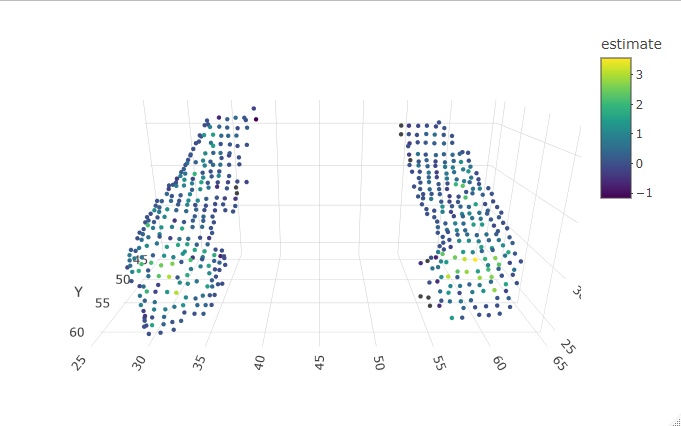} \\
        (c) Linear stage estimate, subtype 2, negative direction & (d) Linear stage estimate, subtype 2, positive direction\\
        \includegraphics[width=0.45\textwidth]{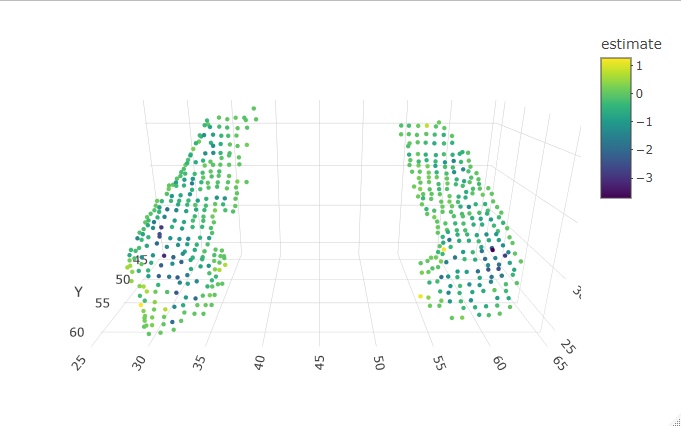}&
        \includegraphics[width=0.45\textwidth]{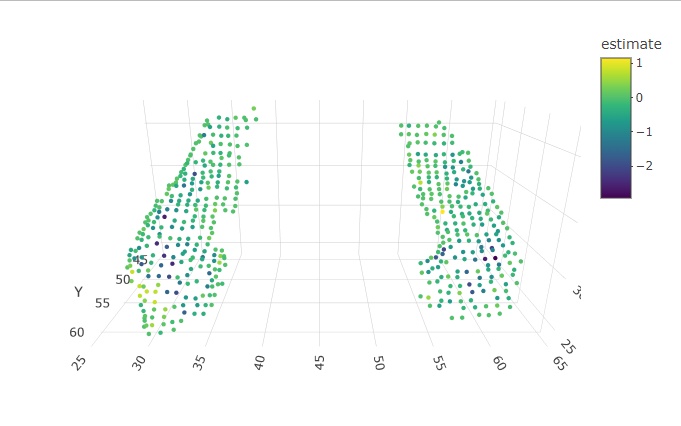} \\ 
        (e) Quadratic stage estimate, subtype 1, negative direction & (f) Quadratic stage estimate, subtype 1, positive direction\\
        \includegraphics[width=0.45\textwidth]{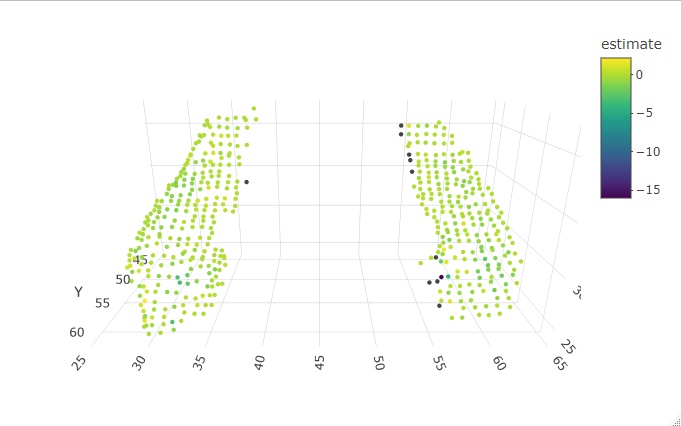}&
        \includegraphics[width=0.45\textwidth]{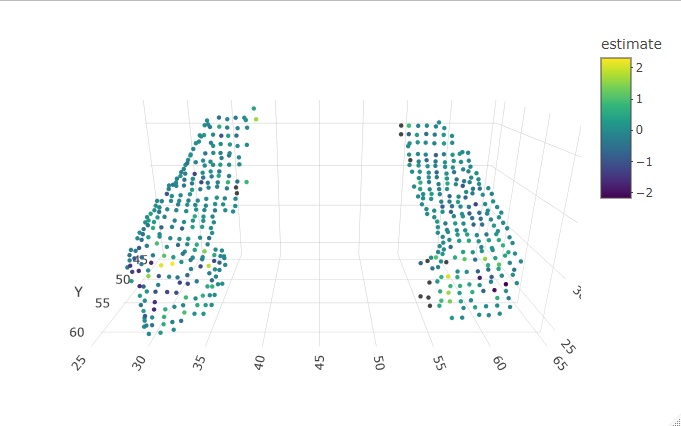} \\
        (g) Quadratic stage estimate, subtype 2, negative direction & (h) Quadratic stage estimate, subtype 2, positive direction\\
    \end{tabular}
    \caption{
    Supplementary Figure. SuStaIn stage effects on hippocampal tau deposition thickness, modeled via projection distance features. 
    (a--d) Linear stage coefficient estimates for subtypes 1 and 2, shown separately for negative and positive projection directions. 
    (e--h) Quadratic stage coefficient estimates for subtypes 1 and 2, shown separately for negative and positive projection directions.
    }
\label{fig:thickness_sup}
\end{figure}

\begin{figure}[h]
    \centering
    \begin{tabular}{cc}
        \includegraphics[width=0.45\textwidth]{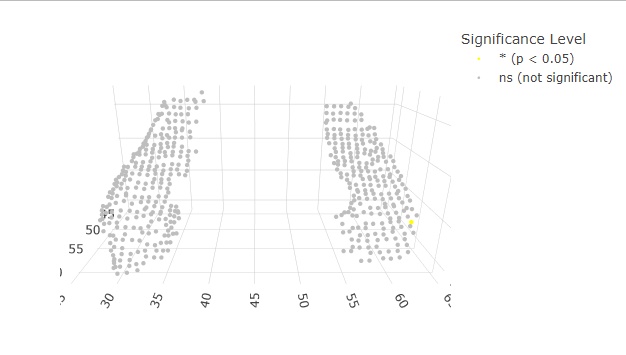}&
        \includegraphics[width=0.45\textwidth]{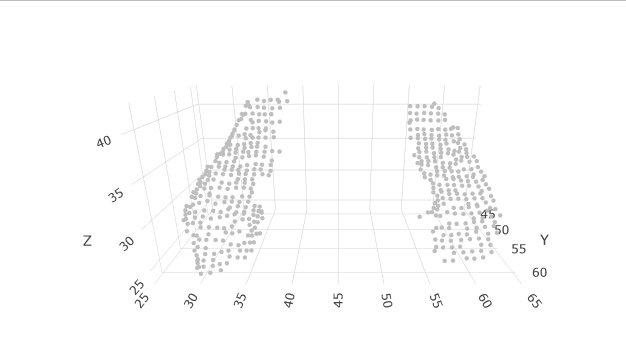} \\ 
        (a) Cognitive status × Subtype (SUVR)   & (b) Quadratic stage × Subtype (SUVR)\\
        \includegraphics[width=0.45\textwidth]{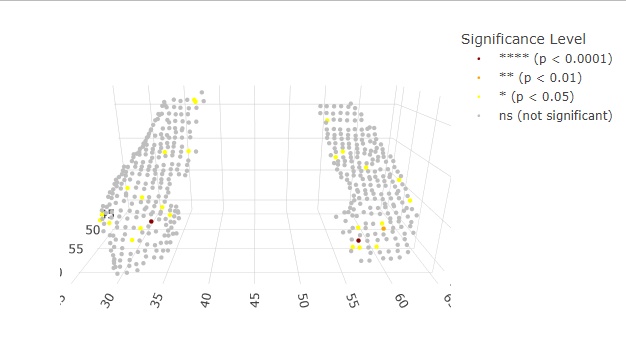}&
        \includegraphics[width=0.45\textwidth]{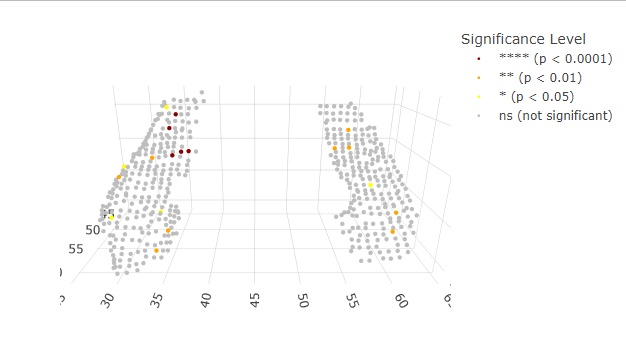} \\
        (c) Age × Subtype (Thickness)  & (d) Sex × Subtype (Thickness)\\
        \includegraphics[width=0.45\textwidth]{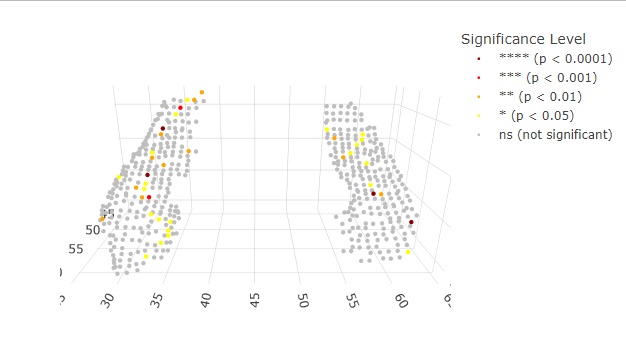}&
        \includegraphics[width=0.45\textwidth]{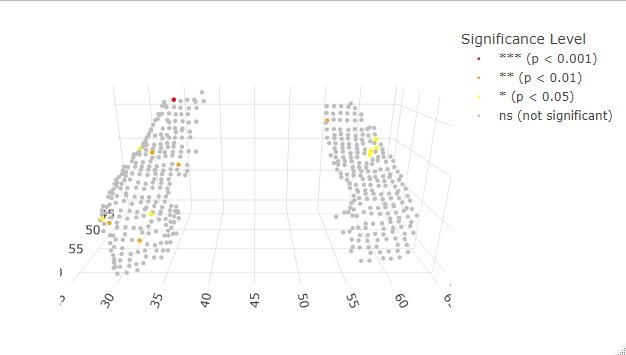} \\
        (e) Cognitive status × Subtype (Thickness)  & (f) Linear stage × Subtype (Thickness)\\
    \end{tabular}
\caption{
Significance maps for covariate-by-subtype interaction terms in the unified regression model. 
Each map displays surface regions where the effect of a covariate on tau deposition differs significantly between subtypes. 
(a)~Significance of the cognitive-status effect interaction on SUVR intensity; 
(b)~Interaction significance for SuStaIn stage (quadratic term) on SUVR intensity; 
(c)~Interaction significance of age on positive projection thickness; 
(d)~Interaction significance of sex on positive projection thickness; 
(e)~Interaction significance of cognitive status on positive projection thickness; 
(f)~Interaction significance of SuStaIn stage (linear term) on positive projection thickness. All maps display $-{\log_{10}}(p)$ values from IPW-adjusted regression models, with multiple significance levels indicated by color intensity. 
Benjamini–Hochberg correction was applied across all vertices to control for multiple comparisons.}
\label{fig:sup_interaction}
\end{figure}

\section{Threshold Sensitivity Analysis}
\label{sec:supp_threshold_sensitivity}

\begin{figure}[htbp]
    \centering
    \begin{tabular}{c}
        \includegraphics[width=0.88\textwidth]{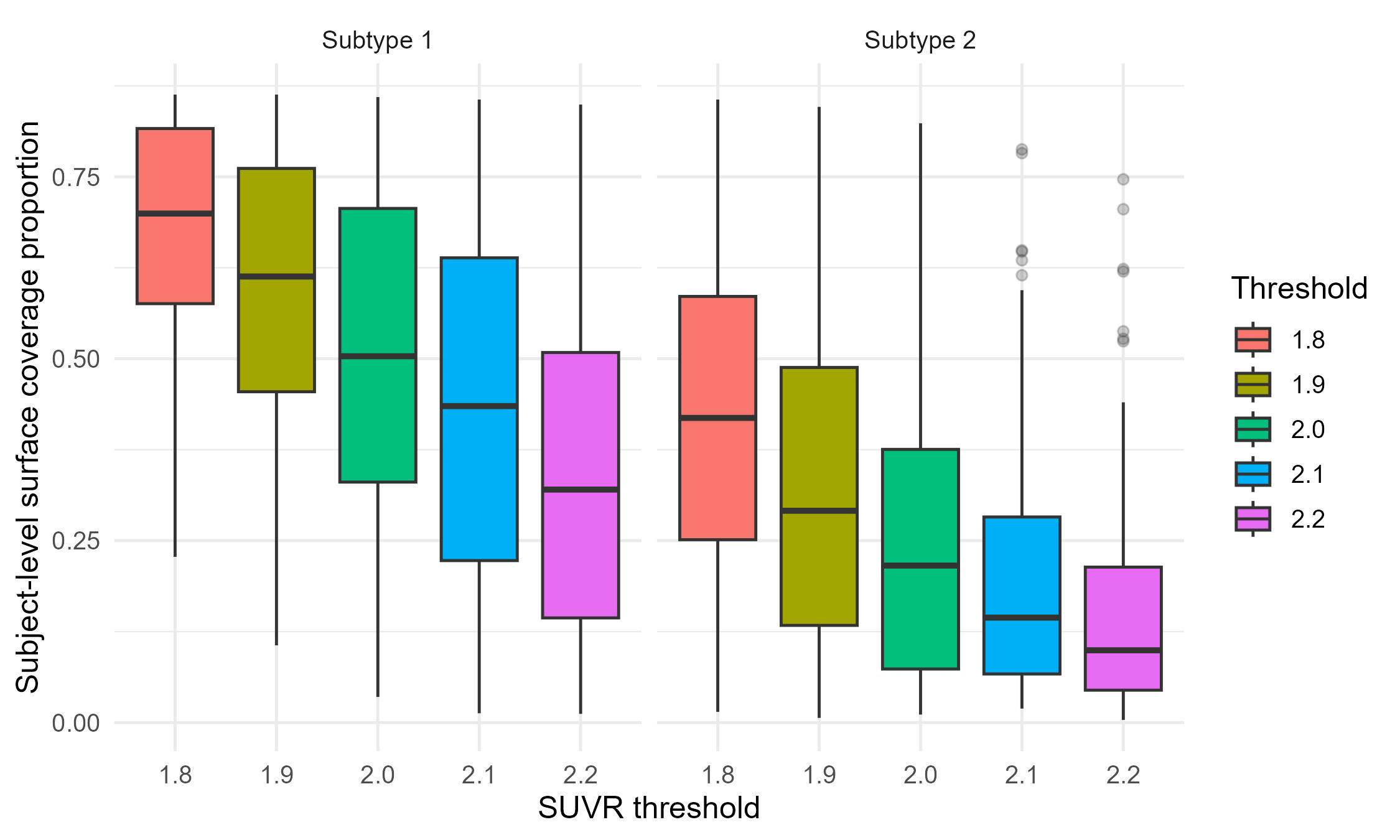} \\
        (a) Subject-level surface coverage burden across thresholds \\[1em]
        \includegraphics[width=0.98\textwidth]{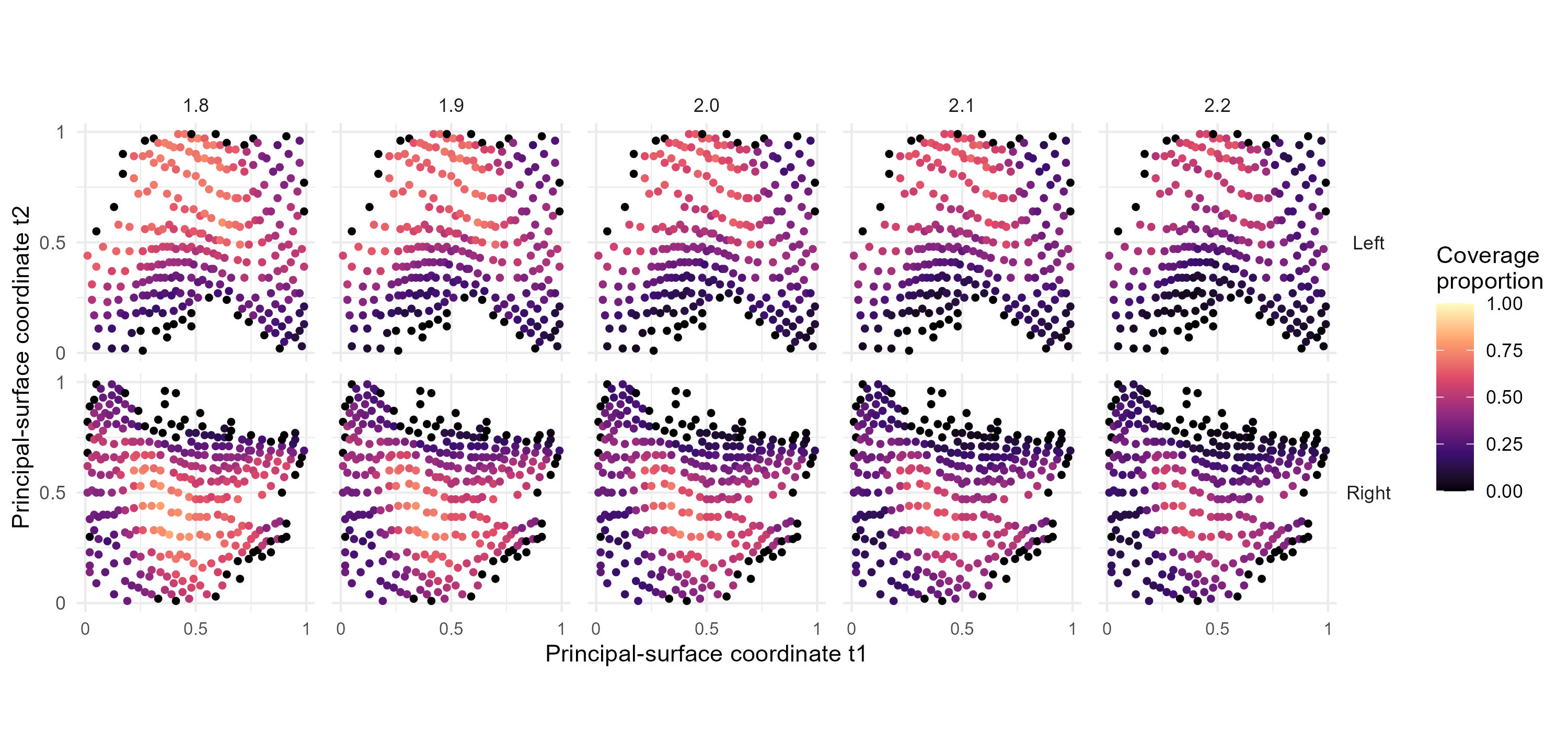} \\
        (b) Surface coverage frequency across thresholds
    \end{tabular}
    \caption{\rev{Sensitivity of hippocampal surface coverage to the SUVR threshold. 
    (a) Subject-level surface coverage burden across thresholds from 1.8 to 2.2, defined as the proportion of principal-surface points covered by suprathreshold hippocampal tau signal within each scan. Results are shown for SuStaIn subtypes 1 and 2. 
    (b) Surface coverage frequency across thresholds, defined at each principal-surface point as the proportion of scans with an available surface projection in which suprathreshold hippocampal tau signal was mapped to that point. Spatial coverage-frequency patterns were highly similar across thresholds in both hemispheres.}}
    \label{fig:threshold_sensitivity_coverage}
\end{figure}

\begin{table}[htbp]
\centering
\caption{\rev{Pearson and Spearman correlations of IPW-adjusted smoothed SUVR coefficient maps across SUVR thresholds.}}
\label{tab:threshold_ipw_correlation}

\vspace{0.8em}

{\scriptsize
\begin{tabular}{llcccc}
\toprule
\textbf{Subtype} & \textbf{Regression term} & \textbf{1.8 vs 2.0} & \textbf{1.9 vs 2.0} & \textbf{2.1 vs 2.0} & \textbf{2.2 vs 2.0} \\
\midrule
1 & Age              & 0.853 / 0.849 & 0.885 / 0.895 & 0.885 / 0.881 & 0.721 / 0.797 \\
1 & Cognitive status & 0.909 / 0.909 & 0.934 / 0.936 & 0.887 / 0.885 & 0.624 / 0.715 \\
1 & Sex              & 0.674 / 0.598 & 0.752 / 0.707 & 0.763 / 0.731 & 0.536 / 0.587 \\
1 & Linear stage     & 0.934 / 0.942 & 0.956 / 0.959 & 0.962 / 0.958 & 0.828 / 0.925 \\
1 & Quadratic stage  & 0.209 / 0.164 & 0.568 / 0.491 & 0.679 / 0.606 & 0.580 / 0.475 \\
\midrule
2 & Age              & 0.029 / 0.431 & -0.376 / 0.475 & 0.114 / 0.474 & 0.068 / 0.224 \\
2 & Cognitive status & 0.217 / 0.605 & 0.167 / 0.734 & 0.141 / 0.677 & 0.117 / 0.520 \\
2 & Sex              & -0.012 / 0.423 & -0.015 / 0.505 & 0.179 / 0.509 & 0.081 / 0.265 \\
2 & Linear stage     & 0.290 / 0.830 & 0.328 / 0.896 & 0.436 / 0.867 & 0.189 / 0.776 \\
2 & Quadratic stage  & 0.195 / 0.192 & 0.413 / 0.416 & 0.118 / 0.402 & 0.043 / 0.291 \\
\bottomrule
\end{tabular}
}

\vspace{0.5em}

\begin{minipage}{0.90\textwidth}
\footnotesize
\textit{Note.} Each cell reports Pearson / Spearman correlation. Correlations compare point-wise IPW-adjusted smoothed SUVR regression coefficient maps from each alternative threshold with the corresponding coefficient map from the primary \(\mathrm{SUVR}>2.0\) analysis. Values are shown for pooled left-and-right hippocampal surface points.
\end{minipage}
\end{table}

\section{Voxel-wise Regression Comparison}

\begin{figure}[h]
    \centering
    \begin{tabular}{cc}
        \includegraphics[width=0.45\textwidth]{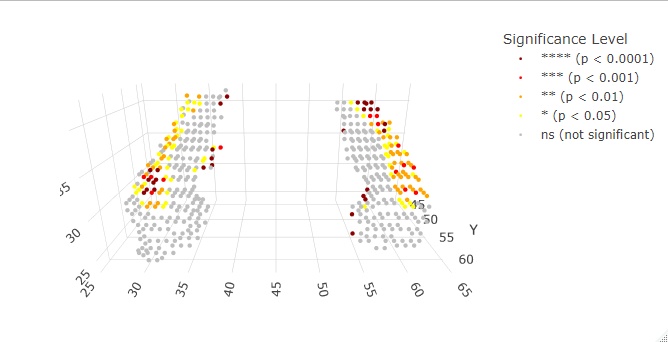}&
        \includegraphics[width=0.45\textwidth]{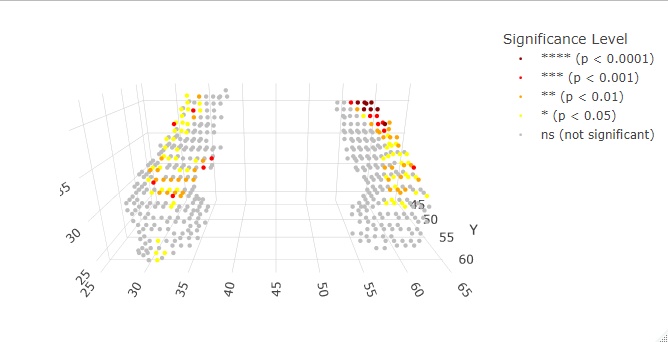} \\ 
        (a) Age effect (subtype 1)   & (b) Cognitive status effect (subtype 1)\\
        \includegraphics[width=0.45\textwidth]{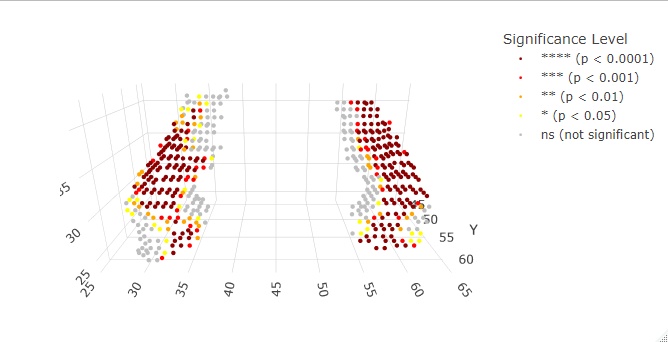}&
        \includegraphics[width=0.45\textwidth]{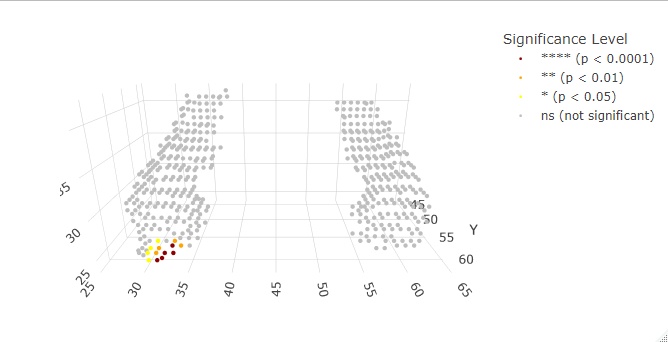} \\
        (c) Linear stage effect (subtype 1)  & (d) Quadratic stage effect (subtype 1)\\
    \end{tabular}
\caption{
Voxel-wise regression significance maps for subtype 1, displayed in MNI coordinate space. Each map shows the spatial distribution of covariate effects on SUVR at individual voxel locations without thresholding or projection onto surface geometry. Panels (a--d) depict the significance of regression coefficients for age, cognitive status, linear SuStaIn stage, and quadratic SuStaIn stage, respectively. Statistical testing was performed at each voxel using linear regression, and $p$-values were corrected for multiple comparisons using the Benjamini--Hochberg (BH) procedure. These maps provide a conventional voxel-wise baseline for comparison with the surface-based modeling framework.
}
\label{fig:sup_voxel}
\end{figure}

\section{Contamination Analysis}

\begin{figure}[h]
    \centering
    \begin{tabular}{cc}

        \includegraphics[width=0.45\textwidth]{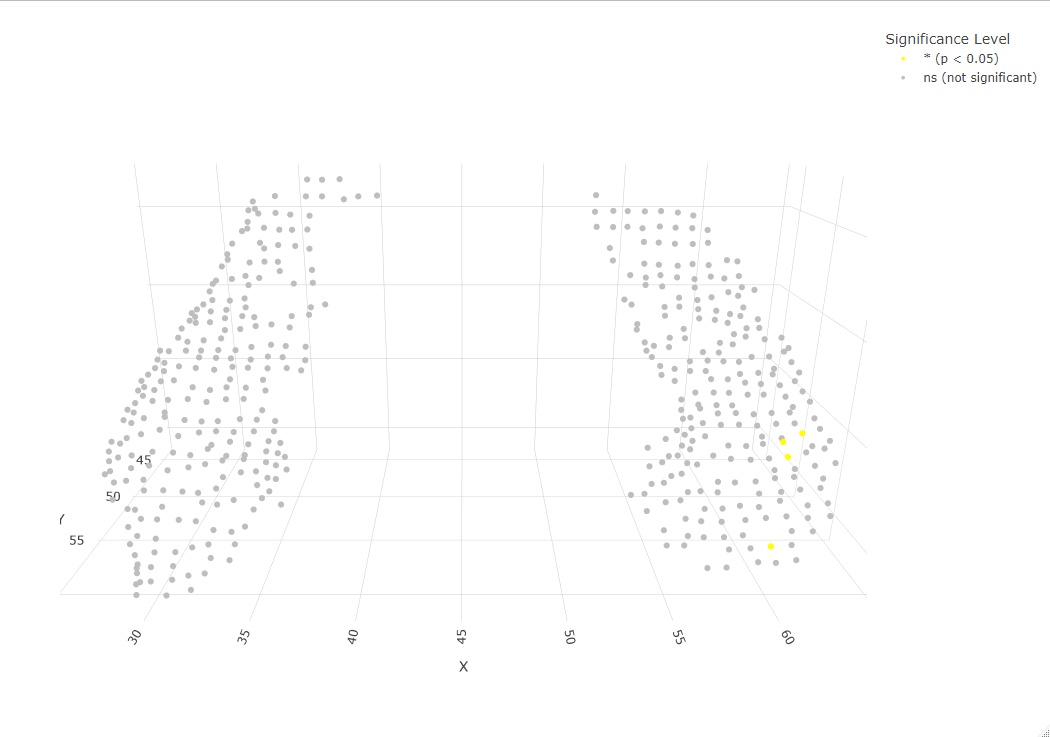} &
        \includegraphics[width=0.45\textwidth]{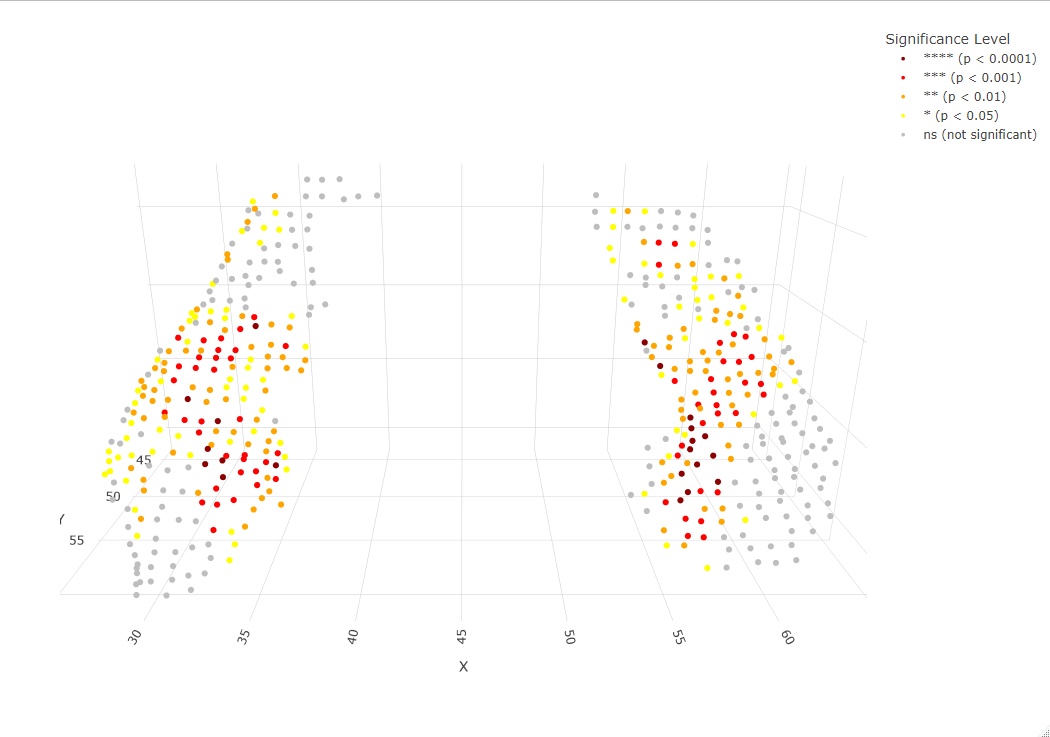} \\
        (a) Subject A (CN, negative) & (b) Subject A (CN, positive) \\
 
        \includegraphics[width=0.45\textwidth]{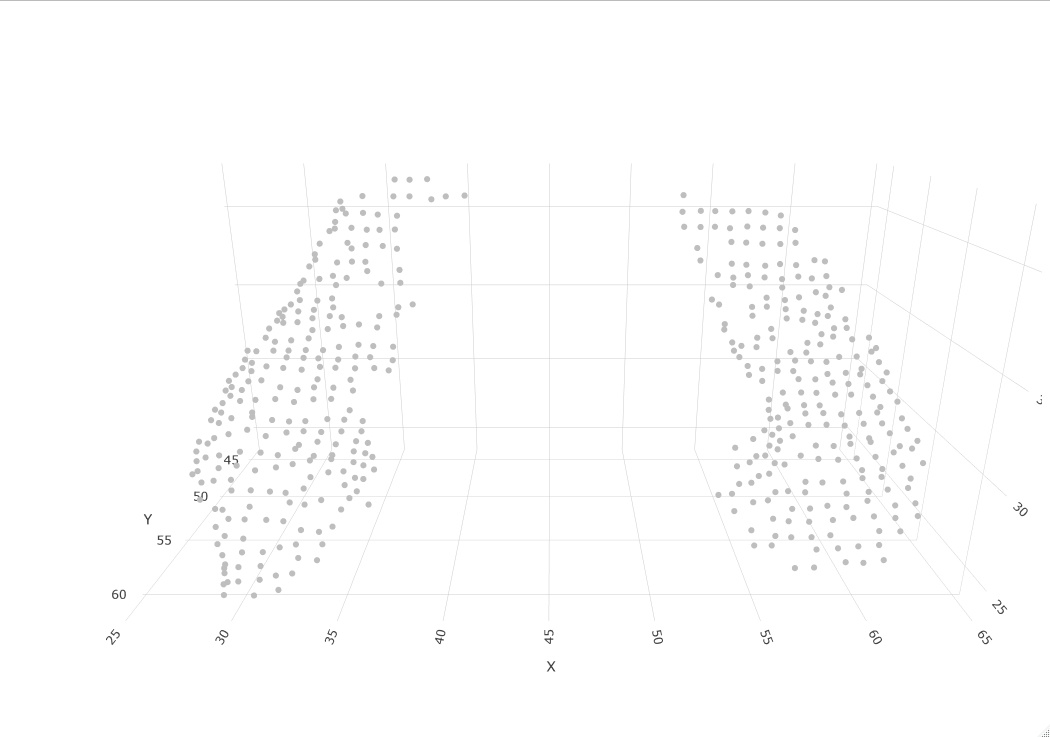} &
        \includegraphics[width=0.45\textwidth]{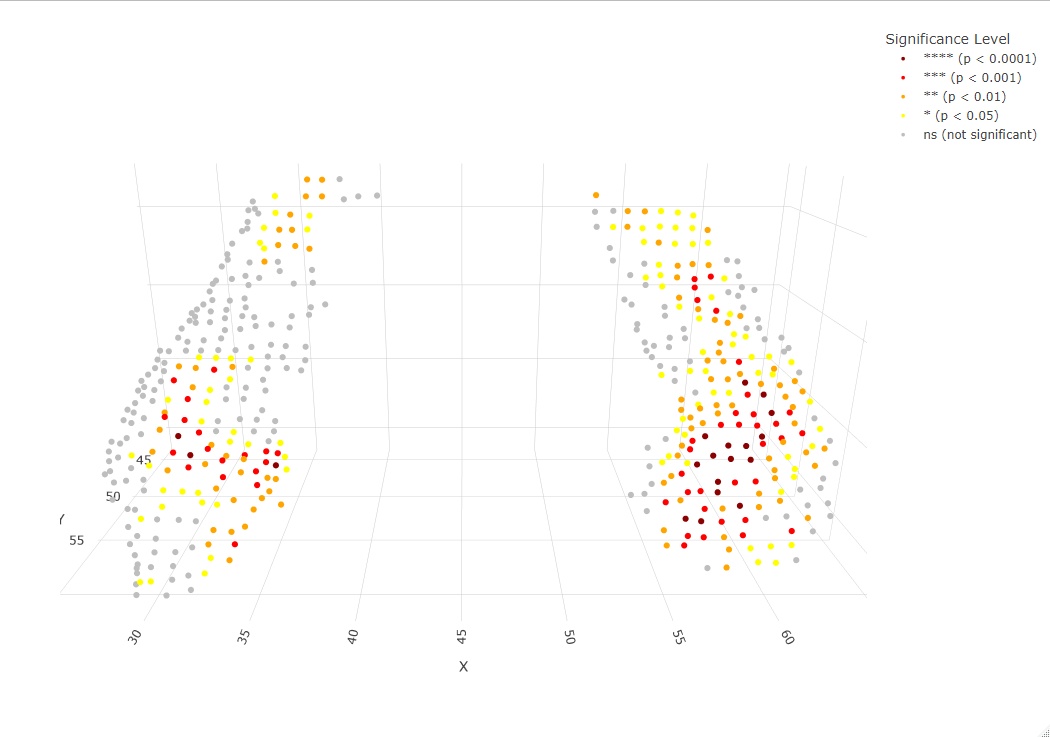} \\
        (c) Subject B (MCI, negative) & (d) Subject B (MCI, positive) \\

        \includegraphics[width=0.45\textwidth]{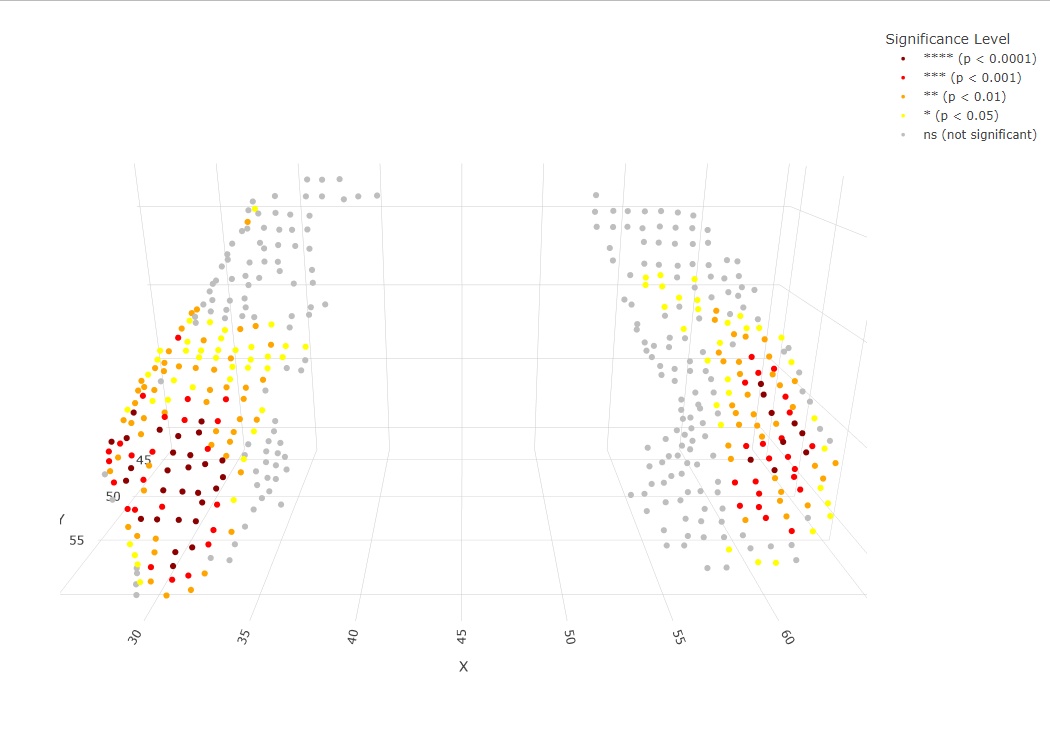} &
        \includegraphics[width=0.45\textwidth]{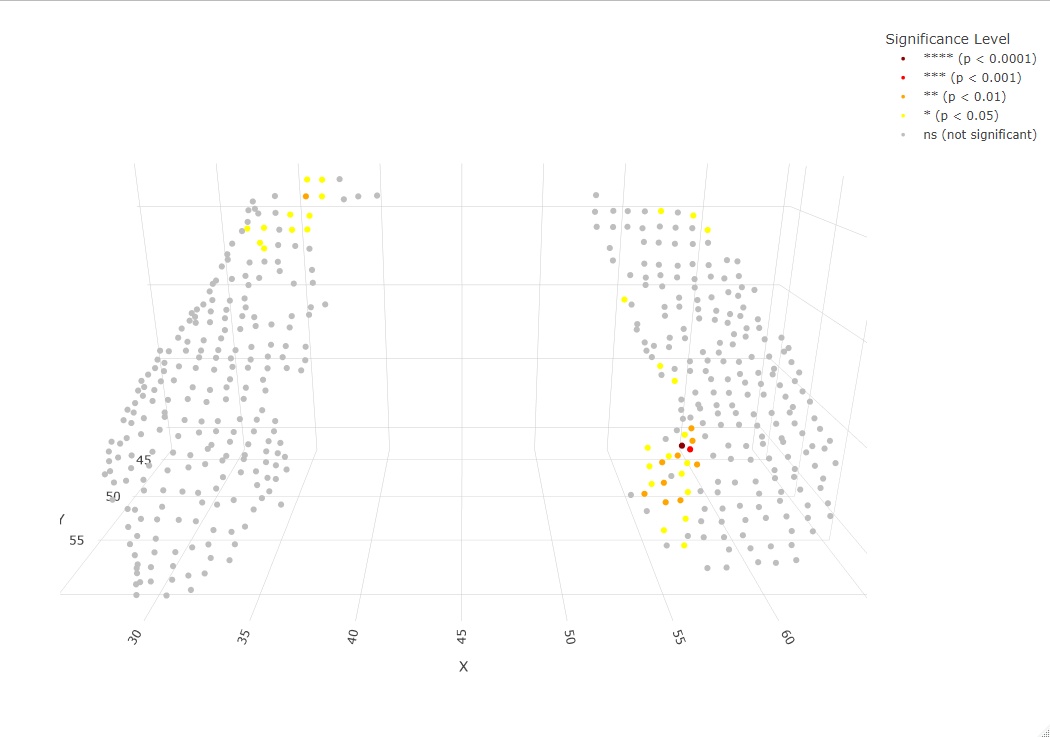} \\
        (e) Subject C (AD, negative) & (f) Subject C (AD, positive) \\
    
    \end{tabular}
     \caption{\textbf{Subject-Level Contamination Significance Maps.}
    Left: significance of negative contamination; Right: significance of positive contamination. 
    Participants are shown in order of increasing hippocampal tau burden. 
    Positive contamination is prominent in CN and MCI, whereas negative contamination emerges in AD.
    }
    \label{fig:supp_contam_examples}
\end{figure}

\end{document}